\documentclass[prd,twocolumn,nofootinbib,showpacs,superscriptaddress]{revtex4-1}
\pdfoutput=1
\usepackage{amsfonts}
\usepackage{amsmath}
\usepackage{amssymb}
\usepackage{bm}
\usepackage{dcolumn}
\usepackage{multirow}
\usepackage{graphicx}
\usepackage{graphics}
\usepackage[latin1]{inputenc}
\usepackage{latexsym}
\usepackage{rotating}
\usepackage[colorlinks=true]{hyperref}
\usepackage[all]{hypcap} %ky: I need to comment "\usepackage{hyperref}" and "\usepackage[all]{hypcap}" out since I cannot somehow compile with these two on. If I don't forget, I will try to put these back on before pushing this text into svn.
\usepackage{xspace} % Sensible space treatment at end of simple macros
\usepackage[usenames]{color}
\usepackage[T1]{fontenc}
\usepackage{microtype}
\usepackage{mathrsfs}
\usepackage{ulem}
\definecolor {darkgreen}{rgb}{0.2,0.7,0.2}

\newcommand\be{\begin{equation}}
\newcommand\ba{\begin{eqnarray}}
\newcommand\ee{\end{equation}}
\newcommand\ea{\end{eqnarray}}

\newcommand\bw{\begin{widetext}}
\newcommand\ew{\end{widetext}}

\newcommand{\nn}{\nonumber}

\newcommand{\hGR}{h}
\newcommand{\hDef}{\mathfrak{h}}

\newcommand{\GR}{{\mbox{\tiny GR}}}

\newcommand{\FZ}{{\mbox{\tiny FZ}}}
\newcommand{\TT}{{\mbox{\tiny TT}}}
\newcommand{\MAT}{{\mbox{\tiny mat}}}

\newcommand{\CS}{{\mbox{\tiny CS}}}

\newcommand{\pd}{\partial}

\newcommand{\plusonetotwo}{+(1\leftrightarrow 2)}

\newcommand{\NS}{*}

\newcommand{\ext}{\mathrm{ext}}
\newcommand{\inter}{\mathrm{int}}
\newcommand{\RNS}{\mathcal{R}_*}

\newcommand{\mrm}{\mathrm}

\newcommand{\accR}{\mathscr{R}}
\newcommand{\accS}{\mathscr{S}}
\newcommand{\accW}{\mathscr{W}}

\newcommand{\ave}[1]{\left< #1 \right>}

\begin{document}
\title{Isolated and Binary Neutron Stars\\* in Dynamical Chern-Simons Gravity}

\author{Kent Yagi}
%\email{kyagi@physics.montana.edu}
\affiliation{Department of Physics, Montana State University, Bozeman, MT 59717, USA.}

\author{Leo C. Stein}
\affiliation{Center for Radiophysics and Space Research, Cornell University, Ithaca, NY 14853}

\author{Nicol\'as Yunes}
\affiliation{Department of Physics, Montana State University, Bozeman, MT 59717, USA.}

\author{Takahiro Tanaka}
\affiliation{Yukawa Institute for Theoretical Physics, Kyoto University, Kyoto, 606-8502, Japan.}

\date{\today}

%%%%%%%%%%%%%%%%%%%%%%%%%%%%%%%%%%%%%%%%%%%%%%%%%
\begin{abstract} 

We study isolated and binary neutron stars in dynamical Chern-Simons gravity. This theory modifies the Einstein-Hilbert action through the introduction of a dynamical scalar field coupled to the Pontryagin density. We here treat this theory as an effective model, working to leading order in the Chern-Simons coupling. We first construct isolated neutron star solutions in the slow-rotation expansion to quadratic order in spin. We find that isolated neutron stars acquire a scalar dipole charge that corrects its spin angular momentum to linear order in spin and corrects its mass and quadrupole moment to quadratic order in spin, as measured by an observer at spatial infinity. We then consider neutron stars binaries that are widely separated and solve for their orbital evolution in this modified theory. We find that the evolution of post-Keplerian parameters is modified, with the rate of periastron advance being the dominant correction at first post-Newtonian order. We conclude by applying these results to observed pulsars with the aim to place constraints on dynamical Chern-Simons gravity. We find that the modifications to the observed mass are degenerate with the neutron star equation of state, which prevents us from testing the theory with the inferred mass of the millisecond pulsar J1614-2230. We also find that the corrections to the post-Keplerian parameters are too small to be observable today even with data from the double binary pulsar J0737-3039. Our results suggest that pulsar observations are not currently capable of constraining dynamical Chern-Simons gravity, and thus, gravitational-wave observations may be the only path to a stringent constraint of this theory in the imminent future. 

\end{abstract}


% 04.30.-w Gravitational waves
% 04.50.Kd Modified theories of gravity
% 04.25.-g Approximation methods; equations of motion
%97.60.Jd Neutron stars

\maketitle

%%%%%%%%%%%%%%%%%%%%%%%%%%%%%%%%%%%
\section{Introduction}

Current astrophysical observations suggest that General Relativity (GR) may have to be modified on large scales. Dark energy, dark matter and even inflation have been suggested as natural consequences of certain modified gravity theories (see e.g. Ref.~\cite{fR}). Perhaps similar modifications will be necessary in the non-linear, dynamical regime, the so-called {\emph{strong-field}}, such as when black holes (BHs) merge and neutron stars (NSs) collide. In this regime, physical phenomena is not well-described by a leading-order weak-field and slow-motion expansion of the Einstein equations. Instead, one must either retain high-orders in such perturbative expansion, or solve the full set of Einstein equations numerically. The only way to determine whether modifications to GR in the strong field are necessary is to make observations and test GR in this regime. 

Modified gravity theories have been tested accurately in the Solar System and with binary pulsar observations~\cite{TEGP,will-living,stairs}. While the former allow for tests in the weak-field regime only, the latter has led to tests when gravitational fields are moderately strong. Binary pulsar observations, however, are still not capable of probing the non-linear regime of GR. For example, the orbital velocity of the double binary pulsar J0737-3039~\cite{burgay,lyne,kramer-double-pulsar} is roughly $10^{-3}$ smaller than the speed of light, and thus, its orbital behavior can be well-approximated by a weak-field, slow-velocity expansion of the Einstein equations, retaining only the leading and first subleading terms. On the other hand, GR will soon be tested during compact binary late coalescence through gravitational wave (GW) observations~\cite{TEGP,will-living,gair-living}, which will allow for a probe of the fully non-linear and dynamical, strong-field regime. 

Recently, there has been a suggestion that binary pulsar observations may constrain deviations from GR to such a degree that GW tests will not be necessary in the future~\cite{damour-GW-vs-pulsar}. Such a suggestion emerged from studies of certain scalar-tensor theories~\cite{fujii}, the constraints on which indeed cannot be improved with second-generation GW detectors, such as Advanced LIGO~\cite{eardley,will1977,zaglauer,will1994,arun-dipole}. But of course, this suggestion depends strongly on the particular modified gravity model considered. For example, the same type of GW detectors may be able to place constraints on massive graviton propagation that are three orders of magnitude stronger than current binary pulsar constraints~\cite{will1998,arunwill,keppel,delpozzo,cornishsampson,mirshekari,sutton}. 

By studying NSs in dynamical Chern-Simons (CS) gravity~\cite{CSreview}, we here find another counterexample to this suggestion. Currently, the only constraints on this theory come from Solar System 
observations~\cite{alihaimoud-chen}, by comparing
the CS correction to gravitomagnetic precession to observations with 
Gravity Probe B, and from table top~\cite{kent-CSBH} experiments,
by requiring that no CS corrections are present above the smallest gravitational
length scales sampled experimentally on Earth. These tests lead to
comparable constraints, namely $\sqrt{\alpha} < 10^8 \; {\rm{km}}$, 
where $\alpha$ is the dimensional CS coupling constant. Such an 
incredibly weak
constraint is perhaps not surprising, given the weakness of the 
gravitational field in the Solar System. Recent work has shown that future GW observations of BH binary inspirals could improve upon Solar System constraints by as much as seven orders of magnitude~\cite{sopuerta-yunes-DCS-EMRI,pani-DCS-EMRI,canizares,kent-CSGW}. In this paper, we study whether current isolated and binary pulsar observations can already constrain dynamical CS gravity. We will see that indeed this is not possible and that only future GW observations of compact binary coalescence can place stringent constraints on this theory. 

Dynamical CS modified gravity is a parity-violating, quadratic-curvature theory that is defined by modifying the Einstein-Hilbert action with a term that is the product of a dynamical scalar field and the Pontryagin density (contraction of the Riemann tensor and its dual). The scalar field has dynamics through the addition of a standard kinetic term and a potential to the action. The Pontryagin term in the action is involved in anomaly cancellation~\cite{polchinski2}. Such a term also appears naturally in heterotic superstring theory~\cite{polchinski1,polchinski2,alexandergates} and loop quantum gravity~\cite{taveras,calcagni,Mercuri:2009zt}. Dynamical CS gravity also arises naturally in effective field theories of inflation~\cite{weinberg-CS}. Historically, CS gravity was introduced without scalar field dynamics, assuming the field was given {\emph{a priori}}~\cite{jackiw}. Such a formulation was plagued with problems, such as metric instability~\cite{Alexander:2007kv} and overconstrained field equations. The modern (dynamical) incarnation of the theory restores the dynamics of the scalar field through a standard kinetic term in the action~\cite{Smith:2007jm}, while treating the model as an effective theory, thus avoiding instabilities~\cite{yunespretorius}. 

Dynamical CS gravity has only recently begun to be studied in detail. Non-rotating BHs are described by the Schwarzschild metric, but rotating ones are not. Analytic slowly-rotating BH solutions have been constructed to linear order in spin~\cite{yunespretorius,konnoBH} and to quadratic order in spin~\cite{kent-CSBH} within the small-coupling approximation, i.e. linearizing all expressions in the CS coupling. Slowly-rotating NS solutions were first constructed in~\cite{yunes-CSNS} to linear order in spin and within the small-coupling approximation. This work was extended in~\cite{alihaimoud-chen}, where NS solutions were obtained still to linear order in spin, but without imposing the small-coupling approximation. 

The purpose of this paper is to investigate whether any meaningful constraints on dynamical CS gravity can be obtained from isolated~\cite{1.97NS} and binary pulsar observations~\cite{stairs}. In order to achieve this goal, one must first study isolated NSs solutions to quadratic-order in spin, since the latter will introduce modifications to the NS mass as measured by an observer at spatial infinity. One must then place such NS solutions in binary systems and study the conservative and dissipative corrections to the evolution of the binary. The former come from the CS deformation of the quadrupole moment of each individual NS, as well as from the scalar dipole-dipole interaction of the binary components. The latter are caused by the modification to the emitted gravitational and scalar radiation. These corrections can be calculated once one finds the scalar dipole charge induced by each individual NS and its associated CS quadrupole moment deformation. The charge can be extracted by studying the asymptotic behavior of the scalar field at spatial infinity to linear order in spin. The quadrupole moment deformation appears at quadratic-order in spin. All throughout this paper, we treat dynamical CS gravity as an 
effective field theory. Such a treatment is definitely valid for the
systems considered here, as explained in detail in~\cite{kent-CSBH}. In turn, this implies that we can use the weak-coupling
approximation to linearize all expression in the CS coupling constant. Such a treatment guarantees that the field 
equations are second order, and the theory is ghost-free~\cite{kent-CSBH}. 

\subsection{Executive Summary}

The first half of this paper focuses on finding isolated NS solutions to quadratic order in spin in the small-coupling approximation. We follow Hartle's approach~\cite{hartle1967}, in which one treats rotation perturbatively, i.e.~one expands in powers of the product of the NS mass and the NS rotational angular frequency. As expected, we find that CS corrections at quadratic order in spin modify the NS mass monopole and quadrupole moments, where the former can be absorbed by a redefinition of mass. At linear order in spin, the CS corrections appear at much higher multipole order than quadrupole, since the correction to the current dipole moment can be absorbed by a redefinition of spin. Therefore, CS corrections at quadratic order in spin can be larger than those at linear order if NSs are spinning moderately fast, yet sufficiently slowly for the small-rotation expansion to hold. 

\begin{figure}[b]
\begin{center}
\includegraphics[width=8.5cm,clip=true]{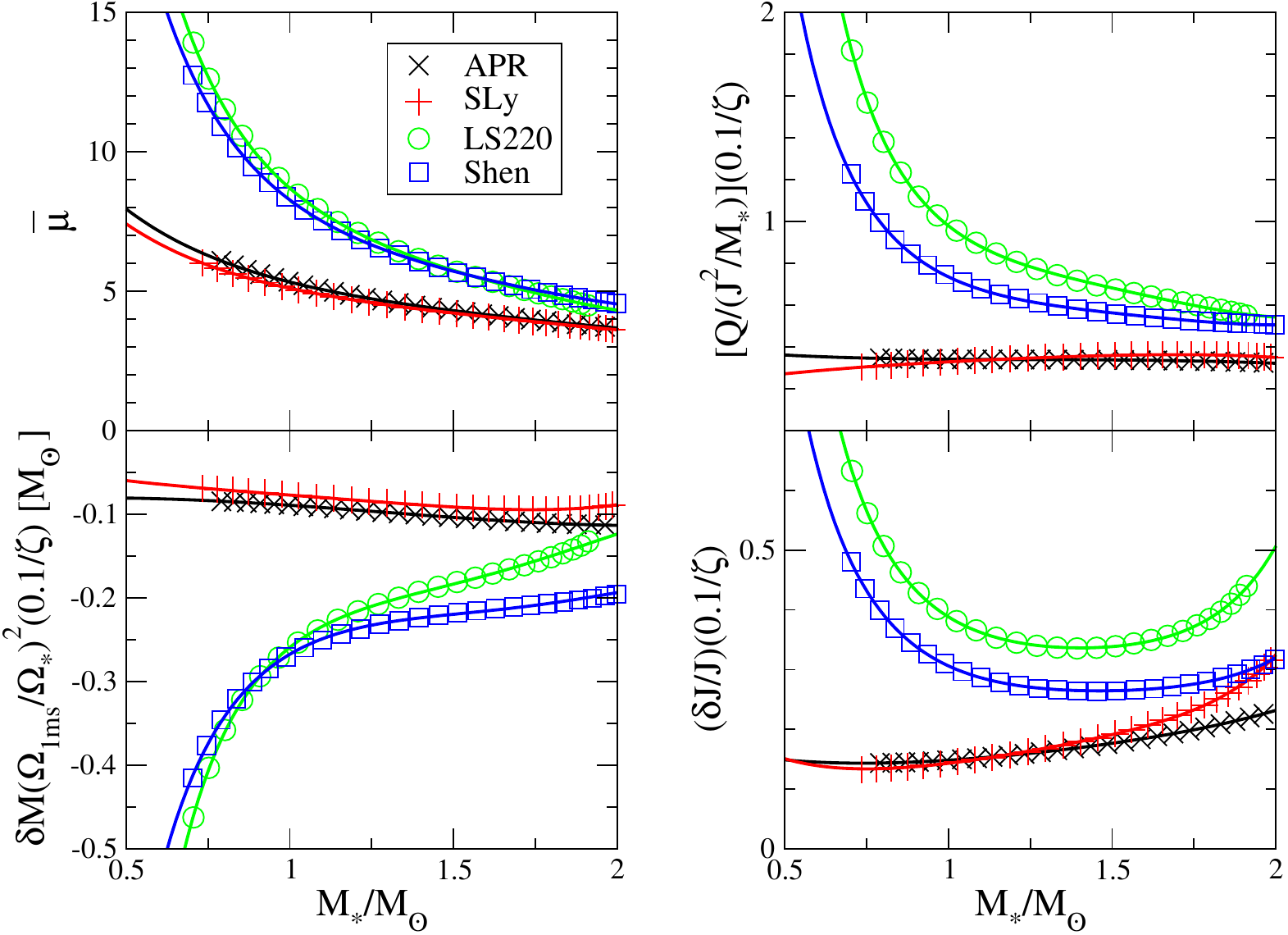}  
\caption{\label{fig:mu-M} 
(Color online)
Numerical (symbols) and fitted (curve) results as functions of NS mass with various EoSs. 
We plot the dimensionless scalar dipole susceptibility $\bar{\mu}$ (top left), the
quadrupole correction $Q$ (top right), the mass shift $\delta M$ (bottom
left), and the angular momentum shift $\delta J$ (bottom right), which are 
defined in Eqs.~\eqref{eq:mu-mubar-relation},
\eqref{eq:QA}, and \eqref{eq:asymptotic-deltaM-deltaJ} respectively. 
The y-axis in the top-right panel is normalized by $J^2/M_\NS$, the absolute value of the NS quadrupole moment 
in the point particle limit in GR, while the y-axis of the bottom-left panel is normalized by $\Omega_\mrm{1ms}$, 
the angular frequency of a NS with a period of 1ms defined by ($\Omega_\mrm{1ms} \equiv 2 \pi / \mrm{1ms}$). 
Moreover, $Q$, $\delta M$ and $\delta J$ are also re-scaled to $\zeta = 0.1$. Observe how the mass shift is always
negative, while the quadrupole moment deformation is always positive.}
\end{center}
\end{figure}
Figure~\ref{fig:mu-M} shows the scalar dipole charge $\bar{\mu}$ (top left) and the CS corrections to the mass $\delta M$ (bottom left), spin angular momentum $\delta J$ (bottom right) and quadrupole moment $Q$ (top right) as a function of the GR mass parameter $M_{\NS}$ in solar masses $M_{\odot}$. In this figure, $\zeta$ is the dimensionless coupling constant of dynamical CS gravity\footnote{This quantity is proportional to the fourth power of the Chern-Simons natural length scale and it is such that as $\zeta \to 0$ one recovers GR. See Eq.~\eqref{zeta-def} for a more precise definition.}, $\Omega_\NS$ is the NS angular velocity, $\Omega_\mrm{1ms}$ is the angular frequency of a NS with a period of 1ms, $J$ is the NS spin angular momentum in GR and we used the APR~\cite{APR}, SLy~\cite{SLy,shibata-fitting}, Lattimer-Swesty (LS220)~\cite{LS, ott-EOS} and Shen~\cite{Shen1,Shen2,ott-EOS} equations of state (EoSs). Observe that the CS corrections \emph{reduce} the observed mass, but they \emph{increase} the observed angular momentum and quadrupole moment with increasing $M_{\NS}$.

\begin{figure}[tb]
\begin{center}
%\begin{tabular}{l}
\includegraphics[width=8.5cm,clip=true]{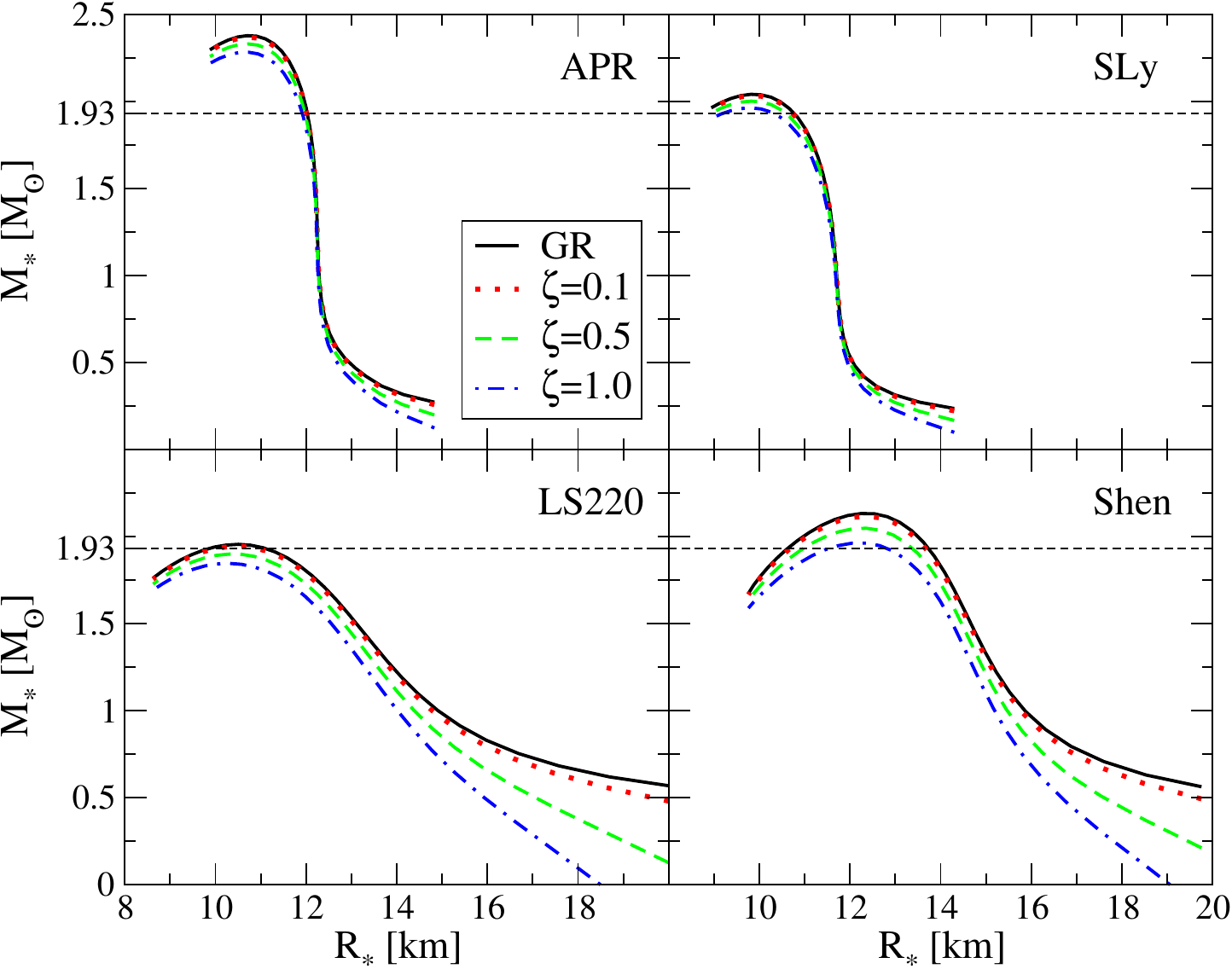}  
%\end{tabular}
\caption{\label{fig:MR-CS}
(Color online)
Mass-radius relations for GR (solid curves) and CS with different
coupling strengths ($\zeta=0.1,0.5,1$ curves are respectively dotted,
dashed, and dot-dashed). Each panel uses a different EoS: 
APR (top left), SLy (top right),  LS220 (bottom left) and Shen
(bottom right). The horizontal dashed line corresponds to the lower bound on the NS mass, 
$1.93M_\odot$, provided by observations of PSR J1614-2230. The
mass-radius relations depend on the spin of the NS, which we have set
to match that for PSR J1614-2230. Since APR,  SLy and Shen EoSs can each
produce NSs above the mass bound for all $\zeta < 1$, there is a clear degeneracy between
EoS and the CS correction to the mass-radius relation. Thus, one cannot constrain the theory
from observations of PSR J1614-2230.}
\end{center}
\end{figure}

Such corrected NS observables can then be contrasted with NS observations to constrain dynamical CS gravity. Observations of the massive millisecond pulsar J1614-2230~\cite{1.97NS} require that one be able to construct NSs with observed masses larger than 1.93$M_\odot$. Unfortunately, but perhaps not surprisingly, the allowed maximum NS mass in dynamical CS gravity depends not only on the EoS but also on the magnitude of the dimensionless CS coupling parameter $\zeta \propto \xi_{\CS} M_{\NS}^{2}/\RNS^{6}$. Figure~\ref{fig:MR-CS} shows the mass-radius relation for different EoSs with different choices of $\zeta$. 
Therefore, such a test of dynamical CS gravity cannot be performed due to degeneracies with the EoS.

The CS corrections to the NS quadrupole moment and the moment of inertia induce modifications to the orbital evolution of binary pulsars, which then can be contrasted with binary pulsar observations to constrain the modified theory. Observations of the double binary pulsar J0737-3039~\cite{burgay,lyne,kramer-double-pulsar} require that the evolution of certain Keplerian elements~\cite{damour-taylor,TEGP,stairs} agree with GR to within a certain observational uncertainty. The CS modification to the quadrupole moment leads to corrections to these evolution equations, which we explicitly calculate in this paper. We find that the rate of periastron advance $\dot{\omega}$ is the best observable to constrain CS gravity, because its CS modification enters at first PN order relative to the GR behavior, i.e. this modification is suppressed by a single factor of ${\cal{O}}(m/a)$, where $m$ is the total mass of the binary and $a$ is the semi-major axis of the binary. However, for J0737-3039, this still implies a suppression of the CS correction of $\mathcal{O}(10^{-8})$ relative to the measured GR $\dot{\omega}$. Moreover, this CS effect is even smaller than the GR spin-orbit correction to $\dot{\omega}$, which enters at 0.5PN order and depends on the NS's moments of inertia. Therefore, current binary pulsar observations are not accurate enough to allow for tests of dynamical CS gravity.

Unlike for scalar-tensor theories~\cite{damour-GW-vs-pulsar}, our results suggest that dynamical CS gravity cannot be constrained well with binary pulsar observations using current data. Instead, GW observations may be the only way to place stringent constraints. We have therefore found a concrete example of a modified gravity theory that cannot be stringently constrained with current electromagnetic observations.  

\subsection{Organization}

The organization of the paper is as follows. In Sec.~\ref{sec:DCS}, we introduce the basics of dynamical CS gravity and explain the approximations that we use throughout the paper. In Sec.~\ref{sec:decomposition}, we explain the coordinate systems that we use and impose an ansatz on the metric and the stress-energy momentum tensor of the matter field. In Sec.~\ref{sec:field-equations}, we derive both GR and CS field equations at each order in spin and derive the exterior solutions, modulo integration constants, which are discussed in Sec.~\ref{sec:exterior}. In Sec.~\ref{sec:interior}, we explain how to construct the interior solution. This amounts to finding the matching conditions at the NS surface and the initial conditions at the center. Then, we explain the numerical procedure that we use to solve these sets of differential equations. In Sec.~\ref{sec:solutions}, we present the numerical results obtained by solving the interior equations and matching the solutions to the exterior ones. We derive the evolution of the NS binary in Sec.~\ref{sec:binary-evolution}, and in Sec.~\ref{sec:applications} we apply the results to the massive millisecond pulsar J1614-2230 and to the double binary pulsar PSR J0737-3039. We conclude in Sec.~\ref{sec:conclusions} with a discussion of possible avenues for future work. 

All throughout the paper, we mostly follow the conventions of Misner, Thorne and Wheeler~\cite{MTW}. We use the Greek letters $(\alpha, \beta, \cdots)$ to denote spacetime indices. The metric is denoted by $g_{\mu \nu}$ and it has signature $(-,+,+,+)$. We use geometric units, with $G=c=1$.

%%%%%%%%%%%%%%%%%%%%%%%%%%%%%%%%%%%
%\newpage
\section{Dynamical Chern-Simons Gravity}
\label{sec:DCS}

%------------------------------------------------------------
\subsection{Basics}

The action in dynamical CS gravity is given by~\cite{CSreview}
\ba
S &\equiv & \int d^4x \sqrt{-g} \Big[ \kappa_g R + \frac{\alpha}{4} \vartheta R_{\nu\mu \rho \sigma} {}^* R^{\mu\nu\rho\sigma}  \nn \\
& &  - \frac{\beta}{2} \nabla_\mu \vartheta \nabla^{\mu} \vartheta + \mathcal{L}_{\MAT} \Big]\,.
\label{action}
\ea
Here, $\kappa_g \equiv (16\pi)^{-1}$, $g$ denotes the determinant of the metric $g_{\mu\nu}$ and $R_{\mu\nu \rho \sigma}$ is the Riemann tensor. ${}^* R^{\mu\nu\rho\sigma}$ is the dual of the Riemann tensor which is defined by~\cite{CSreview}
\be
{}^* R^{\mu\nu\rho\sigma} \equiv \frac{1}{2} \varepsilon^{\rho \sigma \alpha \beta} R^{\mu\nu}{}_{\alpha \beta}\,,
\ee
where $\varepsilon^{\rho \sigma \alpha \beta}$ is the Levi-Civita tensor. $\vartheta$ is a scalar field while $\alpha$ and $\beta$ are coupling constants. $\mathcal{L}_{\MAT}$ denotes the matter Lagrangian density. Following Refs.~\cite{yunespretorius,quadratic,kent-CSBH}, we have neglected the potential of the scalar field for simplicity. We take $\vartheta$ and $\beta$ to be dimensionless while we set $\alpha$ to have a dimension of (length)$^2$~\cite{yunespretorius}.
%For convenience, we define a dimensionless parameter $\zeta$ as
%
%\be
%\zeta \equiv \frac{\xi}{M^4}, \qquad \xi \equiv \frac{\alpha^2}{\kappa_g \beta}\,,
%\label{zeta}
%\ee
%
%where $M$ is the total mass of the system.

The field equations in dynamical CS gravity are given by~\cite{CSreview}
\be
G_{\mu\nu} + \frac{\alpha}{\kappa_g} C_{\mu\nu} =\frac{1}{2\kappa_g} (T_{\mu\nu}^\mrm{mat} + T_{\mu\nu}^\vartheta)\,,
\label{field-eq}
\ee
where $G_{\mu\nu}$ is the Einstein tensor and $T_{\mu\nu}^\mrm{mat}$ is the matter stress-energy tensor. The C-tensor and the stress-energy tensor for the scalar field are defined by
\begin{align}
C^{\mu\nu} & \equiv  (\nabla_\sigma \vartheta) \epsilon^{\sigma\delta\alpha(\mu} \nabla_\alpha R^{\nu)}{}_\delta + (\nabla_\sigma \nabla_\delta \vartheta) {}^* R^{\delta (\mu\nu) \sigma}\,, \\
\label{eq:Tab-theta}
T_{\mu\nu}^\vartheta & \equiv  \beta (\nabla_\mu \vartheta) (\nabla_\nu \vartheta) -\frac{\beta}{2} g_{\mu\nu} \nabla_\delta \vartheta \nabla^\delta \vartheta\,.
\end{align}

The evolution equation of the scalar field is given by
\be
\square \vartheta = -\frac{\alpha}{4 \beta} R_{\nu\mu \rho \sigma} {}^*R^{\mu\nu\rho\sigma}\,.
\label{scalar-wave-eq}
\ee
By using this equation, one can show that
\ba
\nabla_\nu C^{\mu \nu} &=& -\frac{1}{8} (\nabla^\mu \vartheta) R_{\alpha \beta \rho \sigma} {}^*R^{\beta \alpha\rho\sigma} \nn \\
 &=& \frac{1}{2\kappa} \nabla_{\nu} T_\vartheta^{\mu\nu}\,.
\ea
Together with the Bianchi identity, if we take the divergence of Eq.~\eqref{field-eq}, we end up with 
\be
\nabla_{\nu} T_\MAT^{\mu\nu}=0\,.
\label{mat-cons}
\ee
Thus, the equation of motion for a test particle is not modified.

The evolution equation for the scalar field, Eq.~\eqref{scalar-wave-eq}, admits a flat background solution of the form $\vartheta = C_{\mu} x^{\mu}$, for $C_{\mu} = {\rm{const}}$, i.e.~a solution to the homogeneous equation in a flat background. The constants $C_{\mu}$ control the strength of the CS modification, and thus, one would expect them to be proportional to $\alpha/\beta$. Although such a solution is in principle allowed, $C_{\mu}$ would have to be prescribed {\emph{a priori}}, as some form of cosmological term. Moreover, such a term would ruins the $\vartheta$ shift-invariance of the equations of motion, which is a desirable property that allows us to treat $\vartheta$ as massless. Also, such a scalar field would have an infinite energy density, when its stress-energy tensor is integrated over the entire manifold. Finally, dynamical CS gravity with such a homogeneous scalar field is very similar to the non-dynamical theory, and thus, it is already severely constrained by cosmological observations~\cite{Dyda:2012rj}. For the rest of this paper, we will ignore the homogeneous solution for $\vartheta$, and instead, concentrate on dynamically generated scalar fields.

%------------------------------------------------------------
\subsection{Small-Coupling, Slow-Rotation Approximations}
\label{sec:approx}

We here work in the small-coupling and slow-rotation approximations. The former implies that we treat the action given in Eq.~\eqref{action} as defining an effective theory, which requires the CS quadratic term (the second term) to be much smaller than the Einstein-Hilbert term. For isolated NSs, the small-coupling approximation is valid if the following inequality holds: 
\be
\label{zeta-def}
\zeta \equiv \frac{\xi_\CS  M_\NS^2}{\RNS^6} \ll 1\,, \quad \xi_\CS \equiv \frac{\alpha^2}{\beta \kappa_g}\,,
\ee
where $M_\NS$ and $\RNS$ are the mass and radius of the NS respectively. $\xi^{1/4}$ corresponds to the characteristic length scale of the theory while $\sqrt{\RNS^3/M_\NS}$ corresponds to the curvature length scale of the NS. This definition of the dimensionless small coupling constant $\zeta$ is the same as $\zeta^2$ defined in~\cite{alihaimoud-chen} modulo an $\mathcal{O}(1)$ numerical factor. Figure~7 of~\cite{alihaimoud-chen} shows that the small coupling approximation is valid when $\zeta \ll 1$. Solar System experiments require $\xi_\CS^{1/4} \leq \mathcal{O}(10^8)$km~\cite{alihaimoud-chen}. In this paper, we work to linear order in $\zeta$.

Dynamical CS gravity must thus have cut-off scale beyond which one cannot treat it as an effective model. Ref.~\cite{kent-CSBH} estimated this scale by calculating the critical length scale at which loop corrections to the second term in Eq.~\eqref{action} due to $n$-point interactions become of order unity. Requiring that this length scale be smaller than the smallest scales probed by table-top experiments led to the requirement that $\sqrt{\alpha} < {\cal{O}}(10^{8} \; {\rm{km}})$. This requirement is of the same order as the current constraints from Solar System experiments. For the NS systems we will consider here, both this requirement and the small coupling condition $\zeta \ll 1$ are satisfied. 

We further assume that NSs are slowly rotating (i.e. $|\chi | \ll 1$ where $\chi =J/M^{2}$ is the dimensionless spin parameter, with $J$ the magnitude of the NS spin angular momentum and $M$ is its mass) and consider terms up to quadratic order. 

All physical quantities, $A$, can be expanded bivariately as
\be
A = \sum_{m,n} \chi'^m \alpha'^n A_{(m,n)}\,,
\ee
where $\chi'$ and $\alpha'$ are book keeping parameters to label the order of the slow-rotation and the small-coupling approximations, respectively. Notice that $A_{(m,n)} \propto \chi^m \alpha^n$. For the spherically symmetric case, $R_{\nu\mu \rho \sigma} {}^* R^{\mu\nu\rho\sigma} = 0$, and thus, there is no source to the inhomogeneous equation of motion for $\vartheta$. Therefore, for the scalar field, $\vartheta_{(0,n)}=0$ and 
\be
\vartheta = \alpha' \chi' \vartheta_{(1,1)} + \mathcal{O}(\alpha' \chi'^3)\,.
\ee
Notice that $\vartheta_{(2m,1)}=0$ due to parity.

%%%%%%%%%%%%%%%%%%%%%%%%%%%%%%%%%%%
%\newpage
\section{Spacetime and Matter Decomposition}
\label{sec:decomposition}

%------------------------------------------------------------
\subsection{Metric}

Following Hartle~\cite{hartle1967}, we start with the metric ansatz given by
\begin{align}
ds^2 &= -e^{\bar{\nu}(r)} \left(1+2 \bar{h}(r,\theta) \right) dt^2 \nn \\
&  + e^{\bar{\lambda}(r)} \left( 1+\frac{2\bar{m}(r,\theta)}{r-2\bar{M}(r)} \right) dr^2 \\
&  + r^2 \left( 1+2\bar{k}(r,\theta) \right) \left[ d\theta^2 + \sin^2 \theta \left( d\phi - \bar{\omega}(r,\theta) dt \right)^2 \right]\,, \nn
\label{metric-ansatz-rth}
\end{align}
where $\bar{\nu}$ and $\bar{\lambda}$ are $\mathcal{O}(\alpha'^0 \chi'^0)$ quantities that only depend on $r$ while $\bar{h}$, $\bar{k}$, $\bar{m}$ and $\bar{\omega}$ refer to perturbations that depend\footnote{These quantities do not depend on $t$ and $\phi$ because we are searching for stationary and axisymmetric NS solutions.} on both $r$ and $\theta$. $\bar{M}$ is related to $\bar{\lambda}$ by
\be
\bar{M}(r) \equiv \frac{\left( 1-e^{-\bar{\lambda}(r)} \right)r}{2}\,.
\ee
The coordinates $(t,r,\theta,\phi)$ are Hartle-Thorne coordinates. In particular, $(r,\theta)$ denote ordinary polar coordinates. 

Since we are treating rotation perturbatively, one must be careful with the choice of polar coordinates~\cite{hartle1967}. A perturbative approach is valid only when the fractional change in a quantity between rotating and non-rotating cases is small. If one were to use $(r,\theta)$ coordinates, this condition could not be met for the density $\rho$ and the pressure $p$ near the surface. This is because the rotation changes the shape of a star, and hence there are points on the NS surface where these quantities vanish in the non-rotating case while they acquire finite values in the rotating case, leading to infinite fractional changes. Instead, we adopt the coordinates $(R,\Theta)$ as proposed by Hartle~\cite{hartle1967}, which are defined via
\be
\rho \left[ r(R,\Theta ), \Theta \right] = \rho (R) = \rho_{(0,0)}(R), \quad \Theta = \theta\,.
\ee
In other words, the new radial coordinate $R$ is chosen such that the density at $\rho \left[ r(R,\Theta ), \Theta \right]$ in the rotating configuration is the same as $\rho_{(0,0)}(R)$ in the non-rotating configuration. By construction, the density and the pressure in the new coordinates contain the non-rotating part only:
\be
\rho(R) = \rho_{(0,0)}(R), \quad p(R)=p_{(0,0)}(R)\,.
\ee 
We expand $r(R,\Theta)$ as
\be
r(R,\Theta) = R + \xi(R,\Theta)\,,
\ee
where
\be
\xi(R,\Theta) = \alpha'^2 \chi'^2 \xi_{(2,2)}(R,\Theta) + \mathcal{O}(\alpha'^0 \chi'^2, \alpha'^2 \chi'^4)\,.
\label{Eq:xi}
\ee
Notice that we have neglected $\mathcal{O}(\alpha'^0 \chi'^2)$
quantities (pure GR, quadratic in spin effects) in
Eq.~\eqref{Eq:xi}. Since in this work we are interested in the
detectability of the CS corrections, and the mentioned terms would
simply add linearly but not modify the CS corrections at
$\mathcal{O}(\alpha'^2 \chi'^2)$, we ignore the $\mathcal{O}(\alpha'^0
\chi'^2)$ term henceforth.

After the coordinate transformation, the new metric is given by
\ba
ds^2 &=& - \left[\left( 1+2h+\xi \frac{d \nu}{dR} \right)e^\nu - R^2 \omega^2 \sin^2 \Theta \right]dt^2 \nn \\
& & -2 R^2 \omega \sin^2 \Theta dt d\phi + \left[ R^2 (1+2k) + 2 R \xi \right] \sin^2\Theta d\phi^2 \nn \\
& & + e^\lambda \left(1 +\frac{2m}{R-2M} + \xi \frac{d\lambda}{dR} + 2 \frac{\partial \xi}{\partial R} \right) dR^2 \nn \\
& & + 2e^\lambda \frac{\partial \xi}{\partial \Theta} dRd\Theta + \left[ R^2 (1+2k) + 2R\xi \right]d\Theta^2 \nn \\
& & + \mathcal{O}(\alpha'^0 \chi'^2, \alpha'^2 \chi'^4)\,,
\label{metric-ansatz-RTh}
\ea
where
\be
M(R) \equiv \frac{\left( 1-e^{-\lambda (R)} \right)R}{2}\,.
\label{M-lambda}
\ee
Each quantity in the above equation is bivariately expanded as
\begin{align}
\nu (R) &= \nu_{(0,0)} (R)\,, \nn \\
\lambda (R) &= \lambda_{(0,0)} (R)\,, \nn \\ 
\omega (R,\Theta) &= \chi' \omega_{(1,0)} (R,\Theta)+\alpha'^2 \chi' \omega_{(1,2)} (R,\Theta) \nn \\
& \quad + \mathcal{O}(\alpha'^0 \chi'^3, \alpha'^2 \chi'^3) \,, \nn \\
h (R,\Theta) &= \alpha'^2 \chi'^2 h_{(2,2)} (R,\Theta)+ \mathcal{O}(\alpha'^0 \chi'^2, \alpha'^2 \chi'^4)\,, \nn \\
m (R,\Theta) &= \alpha'^2 \chi'^2 m_{(2,2)} (R,\Theta)+ \mathcal{O}(\alpha'^0 \chi'^2, \alpha'^2 \chi'^4)\,, \nn \\
k (R,\Theta) &= \alpha'^2 \chi'^2 k_{(2,2)} (R,\Theta)+ \mathcal{O}(\alpha'^0 \chi'^2, \alpha'^2 \chi'^4)\,.
\end{align}
Due to parity, the $(t,t)$, $(R,R)$, $(\Theta,\Theta)$, $(\phi,\phi)$ and $(R,\Theta)$ metric components only have terms proportional to even powers of $\chi'$ while $(t,\phi)$ component only contains terms that are odd powers in $\chi'$. Also, $\mathcal{O}(\alpha')$ terms do not appear in the metric because of the structure of the field equation and the fact that $\vartheta$ is proportional to $\alpha$ in the small-coupling approximation.

Notice that the coordinate deformation $\xi(R,\Theta)$ is well defined only inside the star. We take it to be constant outside the star. This means that the exterior metric in $(t,r,\theta,\phi)$ coordinates can be obtained simply by replacing $R \to r$ and $\Theta \to \theta$ in the exterior metric in $(t,R,\Theta,\phi)$ coordinates.

%------------------------------------------------------------
\subsection{Neutron Star Stress-Energy Tensor}

In this paper, we assume that the matter field inside a NS is a perfect fluid and that the NS is rotating uniformly. The stress-energy tensor for the matter field $T_{\mu\nu}^\MAT$ is given by
\be
T_{\mu\nu}^\MAT = (\rho + p ) u_\mu u_\nu + p g_{\mu\nu}\,,
\ee
where the four-velocity $u^\mu$ is given by
\be
u^\mu = (u^0, 0,0,\Omega_\NS u^0)\,.
\ee
$\Omega_\NS$ is the constant angular velocity of the NS. By using the normalization condition $u_\mu u^\mu = -1$, we obtain the time component of the four-velocity $u^0$ as
\ba
e^{\nu/2} u^0 &=& 1 + \alpha'^2 \chi'^2 \left\{ e^{-\nu} R^2 \omega_{(1,2)} [ \omega_{(1,0)}-\Omega_\NS ] \sin^2(\Theta) \right. \nn \\
& & \left. - h_{(2,2)} - \frac{1}{2} \xi_{(2,2)} \frac{d\nu}{dR} \right\} + \mathcal{O}(\alpha'^0 \chi'^2, \alpha'^2 \chi'^4)\,. \nn \\
\ea

As mentioned in Eq.~\eqref{mat-cons}, the stress-energy tensor of this
minimally-coupled fluid is divergence free, which contributes another
equation to the system. In order to close the system, one needs a
relationship between density $\rho$ and pressure $P$, an EoS. 
In this paper we only consider one-parameter EoSs,
i.e.~$P=P(\rho)$ without any entropy dependence. The EoSs we
use are APR~\cite{APR}, SLy~\cite{SLy,shibata-fitting},
Lattimer-Swesty with nuclear incompressibility of 220MeV
(LS220)~\cite{LS, ott-EOS} and Shen~\cite{Shen1,Shen2,ott-EOS}. 
For the latter two, we use a temperature of 0.01MeV and an electron 
fraction of 30$\%$. The
EoS and conservation of matter stress-energy tensor close the
system. We discuss the matter equations of motion order-by-order
below.

%%%%%%%%%%%%%%%%%%%%%%%%%%%%%%%%%%%
%\newpage
\section{Modified Field Equations}
\label{sec:field-equations}

In this section, we derive the modified field equations  expanded in $\chi'$ and $\alpha'$.

%------------------------------------------------------------
\subsection{Zeroth Order in Spin}

First, we consider equations at $\mathcal{O}(\chi'^0)$. As mentioned previously, in the spherically symmetric case there is no CS correction, and hence the field equations reduce to the Einstein equations for a non-rotating star. The only non-vanishing components are the $(t,t)$, $(R,R)$, $(\Theta,\Theta)$ and $(\phi,\phi)$ ones, but  the last two components are linearly dependent. The first two components yield
\ba
\label{tt-zeroth}
\frac{d M}{dR} &=& 4 \pi R^2 \rho\,, \\
\label{RR-zeroth}
\frac{d \nu}{dR} &=& 2\frac{4 \pi R^3 p + M}{R(R-2M)}\,.
\ea
Together with the EoS, we need one more equation to close the system of differential equations. Instead of using the $(\Theta,\Theta)$ or $(\phi,\phi)$ component of the Einstein equations, one can use the $R$ component of the conservation equations of the matter stress-energy tensor [Eq.~\eqref{mat-cons}] which is given by
\be
\frac{d\nu}{dR} = -\frac{2}{\rho + p} \frac{dp}{dR}\,.
\label{R-cons-zeroth}
\ee
By combining Eqs.~\eqref{RR-zeroth} and~\eqref{R-cons-zeroth}, we obtain the Tolman-Oppenheimer-Volkoff (TOV) equation
\be
\frac{dp}{dR} = -\frac{(4\pi R^3 p + M) (\rho + p)}{R(R-2M)}\,.
\label{TOV-zeroth}
\ee
%

%------------------------------------------------------------
\subsection{First Order in Spin}

\subsubsection{GR}

At $\mathcal{O}(\alpha'^0 \chi')$, the only non-vanishing component of the Einstein equations is the $(t,\phi)$ one. By using Eqs.~\eqref{tt-zeroth}--\eqref{TOV-zeroth}, the associated equation can be simplified to
\ba
&& \frac{\partial^2 \tilde{\omega}_{(1,0)}}{\partial R^2} +4 \frac{1 - \pi R^2 (\rho+p) e^{\lambda}}{R}  \frac{\partial \tilde{\omega}_{(1,0)}}{\partial R} \nn \\
& & + \frac{e^\lambda}{R^2} \left( \frac{\partial^2 \tilde{\omega}_{(1,0)}}{\partial \Theta^2} +3 \cot\Theta \frac{\partial \tilde{\omega}_{(1,0)}}{\partial \Theta} \right) \nn \\
& &  -16 \pi (\rho+p) e^\lambda \tilde{\omega}_{(1,0)}=0\,,
\label{1st-GR}
\ea
where
\begin{equation}
\label{eq:tildeomega-def}
\tilde{\omega}_{(1,0)} \equiv \Omega_\NS - \omega_{(1,0)}\,.
\end{equation}
We simplify the equation further by performing a Legendre decomposition:
\begin{equation}
\tilde{\omega}_{(1,0)} (R,\Theta) = \sum_\ell \tilde{\omega}_\ell (R) \left( - \frac{1}{\sin\Theta} \frac{\partial}{\partial \Theta} P_\ell (\cos\Theta) \right)\,,
\label{Legendre-omega}
\end{equation}
where $P_\ell$ is the $\ell$-th Legendre polynomial. Then, Eq.~\eqref{1st-GR} becomes 
\ba
&& \frac{d^2 \tilde{\omega}_{\ell}}{d R^2} +4 \frac{1 - \pi R^2 (\rho+p) e^\lambda }{R}  \frac{d \tilde{\omega}_{\ell}}{d R} \nn \\
& &- \left[ \frac{(\ell+2) (\ell -1)}{R^2} +16 \pi (\rho+p) \right] e^\lambda \tilde{\omega}_{\ell}=0\,.
\ea
By imposing asymptotic flatness and regularity at the center, one can show that $\tilde{\omega}_{\ell}$ must vanish for all $\ell$ except $\ell = 1$~\cite{hartle1967}. Thus, the above equation reduces to
\ba
\frac{d^2 \tilde{\omega}_1}{d R^2} + 4\frac{1- \pi R^2 (\rho +p) e^{\lambda}}{R}\frac{d \tilde{\omega}_1}{d R}  -16 \pi (\rho +p) e^{\lambda} \tilde{\omega}_1=0\,. \nn \\
\label{omega1RR}
\ea

\subsubsection{CS: Scalar Evolution Equation}

Next, we look at the scalar evolution equation at $\mathcal{O}(\alpha' \chi')$. By using Eqs.~\eqref{tt-zeroth}--\eqref{TOV-zeroth}, the equation can be written as
\ba
& &  \frac{\partial^2 \vartheta_{(1,1)}}{\partial R^2} + \frac{1 + e^\lambda \left[ 1 -4\pi R^2 (\rho-p)  \right]}{R} \frac{\partial \vartheta_{(1,1)}}{\partial R} \nn \\
& & + \frac{e^\lambda}{R^2} \left( \frac{\partial^2 \vartheta_{(1,1)}}{\partial \Theta^2} + \cot \Theta \frac{\partial \vartheta_{(1,1)}}{\partial \Theta} \right) \nn \\
&=&  8\pi \frac{\alpha}{\beta}  \delta \; \; e^{(\lambda - \nu)/2} \left( \sin\Theta \frac{\partial^2 \tilde{\omega}_{(1,0)}}{\partial R \partial \Theta} + 2 \cos\Theta \frac{\partial \tilde{\omega}_{(1,0)}}{\partial R} \right)\,, \nn \\
\label{scalar-evol-eq}
\ea
where 
\be
\delta \equiv \rho- \frac{M}{(4/3) \pi R^3}
\ee
denotes the density shift from the average. Since we are only interested in stationary and axisymmetric solutions, we consider a $\vartheta$ that only depends on the coordinates $R$ and $\theta$. As in Eq.~\eqref{Legendre-omega}, we perform a Legendre decomposition of $\vartheta$:
\be
\vartheta_{(1,1)} (R,\Theta) = \sum_\ell \vartheta_\ell (R) P_\ell (\cos \Theta)\,,
\ee
and find that the $\ell = 1$ term of Eq.~\eqref{scalar-evol-eq} is the only one with a source term.
The only homogeneous solution that satisfies both asymptotic flatness
and regularity at the center is the trivial $\vartheta_\ell = 0$ for $\ell \neq 1$.
For $\ell =1$, Eq.~\eqref{scalar-evol-eq} becomes
\ba
&& \frac{d^2 \vartheta_1}{d R^2} + \frac{1 + e^\lambda \left[ 1 -4\pi R^2 (\rho-p)  \right]}{R} \frac{d \vartheta_1}{d R}  - 2\frac{e^\lambda}{R^2} \vartheta_1 \nn \\
&= & 16\pi \frac{\alpha}{\beta}  \delta e^{(\lambda - \nu)/2} \frac{d \tilde{\omega}_1}{d R}\,. 
\label{vartheta1RR}
\ea

\subsubsection{CS: Field Equation}

Next, we consider the modified Einstein equation at $\mathcal{O}(\alpha'^2 \chi')$. As in GR, the only non-vanishing component of the field equations is the $(t,\phi)$ one. Notice that since $\vartheta$ is of $\mathcal{O}(\alpha' \chi')$, there is no $\mathcal{O}(\alpha'^2 \chi')$ contribution from $T_{\mu\nu}^\vartheta$. The $(t,\phi)$ component of the field equations is given by
\begin{equation}
\begin{aligned}
& \frac{\partial^2 \omega_{(1,2)}}{\partial R^2} + 4 \frac{1-\pi R^2 (\rho +p) e^\lambda}{R} \frac{\partial \omega_{(1,2)}}{\partial R}\\
& + \frac{e^\lambda}{R^2} \left( \frac{\partial^2 \omega_{(1,2)}}{\partial \Theta^2} + 3 \cot \Theta \frac{\partial \omega_{(1,2)}}{\partial \Theta} \right) -16 \pi (\rho +p) e^\lambda \omega_{(1,2)} \\
{}={}& \frac{128 \pi^2 \alpha e^{(\nu+\lambda)/2}}{R^3 \sin\Theta} \left[ \delta R \frac{\partial^2 \vartheta_{(1,1)}}{\partial R \partial \Theta} + \left( R \frac{d\rho}{dR} - \delta \right) \frac{\partial \vartheta_{(1,1)}}{\partial \Theta} \right]\,.
\end{aligned}
\label{1st-CS}
\end{equation}
As for $\tilde{\omega}_{(1,0)}$, we decompose $\omega_{(1,2)}$ via
\begin{equation}
\label{eq:w-def}
\omega_{(1,2)} (R,\Theta) = \sum_\ell w_\ell (R) \left( - \frac{1}{\sin\Theta} \frac{\partial}{\partial \Theta} P_\ell (\cos\Theta) \right)\,.
\end{equation}
Again, we find that Eq.~\eqref{1st-CS} with $\ell \neq 1$ becomes a homogeneous equation, leading to $w_\ell = 0$ for $\ell \neq 1$ after imposing asymptotic flatness and regularity at the center. For $\ell =1$, the equation reduces to
\ba
&& \frac{d^2 w_1}{d R^2} + 4 \frac{1-\pi R^2 (\rho +p) e^\lambda}{R} \frac{d w_1}{d R}  -16 \pi (\rho +p) e^\lambda w_1 \nn \\
&=& - \frac{128 \pi^2 \alpha e^{(\nu+\lambda)/2}}{R^3} \left[ \delta \; R \frac{d \vartheta_1}{d R} + \left( R \frac{d\rho}{dR} - \delta \right) \vartheta_1 \right]\,. \nn \\
\label{w1RR}
\ea
Notice that Eq.~\eqref{omega1RR} corresponds to the homogeneous version of Eq.~\eqref{w1RR}.

%------------------------------------------------------------
\subsection{Second Order in Spin}

In this subsection, we derive the stress-energy conservation and the field equations at quadratic order in spin. As mentioned previously, we do not consider the GR part since $\mathcal{O}(\alpha'^0 \chi'^2)$ terms do not appear in any equations at $\mathcal{O}(\alpha'^2 \chi'^2)$. Following the previous subsection, we perform the Legendre decompositions:
\begin{subequations}
\begin{align}
\label{eq:h-decomp}
h_{(2,2)} (R,\Theta) &= \textstyle\sum_\ell h_\ell (R) P_\ell (\cos \Theta)\,, \\
\label{eq:m-decomp}
m_{(2,2)} (R,\Theta) &= \textstyle\sum_\ell m_\ell (R) P_\ell (\cos \Theta)\,, \\
\label{eq:k-decomp}
k_{(2,2)} (R,\Theta) &= \textstyle\sum_\ell k_\ell (R) P_\ell (\cos \Theta)\,, \\
\label{eq:xi-decomp}
\xi_{(2,2)} (R,\Theta) &= \textstyle\sum_\ell \xi_\ell (R) P_\ell (\cos \Theta)\,.
\end{align}
\end{subequations}
We use the gauge freedom of the theory to set $k_0(R)=0$~\cite{hartle1967}.

\subsubsection{CS: Stress-Energy Conservation}
\label{sec:CS-SET-cons}

The non-vanishing components of the stress-energy conservation equations $\nabla^\mu T_{\mu\nu}^\MAT=0$ are $\nu = R$ and $\Theta$. First, let us look at the $\Theta$ component. By performing a Legendre decomposition, one finds that $\ell = 2$ is the only non-vanishing mode. By using the previously obtained results, one arrives at the algebraic condition
\be
\xi_2 = -\frac{R(R-2M)}{3 (4\pi R^3 p + M)} (3 h_2 - 2 e^{-\nu} R^2 \tilde{\omega}_1 w_1)
\label{xi2}
\ee
in the interior of the star.

Next, let us look at the $R$ component. The only non-vanishing modes are the $\ell =0$ and $\ell =2$ ones. For the latter, we obtain Eq.~\eqref{xi2} after we integrate the differential equation once with respect to $R$. For the former, again by integrating once, we obtain the algebraic relation
\be
h_0 = -\frac{ 4\pi R^3 p + M}{R(R-2 M)} \xi_0 - \frac{2}{3} e^{-\nu} R^2 \tilde{\omega}_1 w_1 + h_{0c}\,,
\label{xi0}
\ee
valid in the interior of the star, where $h_{0c}$ is an integration constant that corresponds to $h(R=0)$. 

\vspace{10mm}

\subsubsection{CS: Field Equations}

Let us now derive the modified field equations at $\mathcal{O}(\alpha'^2 \chi'^2)$. The only non-vanishing modes are the $\ell=0$ and $\ell=2$ ones. Let us first focus on the $\ell =0$ case, where our dependent variables are $h_0$, $m_0$ and $\xi_0$. Eq.~\eqref{xi0} already determines $h_0$, and thus, we are left with 2 degrees of freedom ($m_0$ and $\xi_0$). Following Ref.~\cite{hartle1967}, we use the $(t,t)$ and $(R,R)$ components of the field equations, and Eq.~\eqref{xi0}, to find
\bw
\allowdisplaybreaks[1]
\begin{align}
\label{m0R}
\frac{d m_0}{d R} &= -4\pi R^2 \frac{d\rho}{d R} \xi_0 + \frac{1}{6}R^2 e^{-(\nu + \lambda)} \left( 16 \pi \xi_\CS \delta\frac{d \tilde{\omega}_1}{d R} -R^2 \frac{d w_1}{d R} \right) \frac{d \tilde{\omega}_1}{d R} \nn \\*
& -\frac{16\pi}{3} e^{-\nu} R^4  (\rho +p) \tilde{\omega}_1 w_1 + \frac{2 \pi}{3}  \beta e^{-\lambda} R^2 \left( \frac{d \vartheta_1}{d R} \right)^2  + \frac{4\pi}{3} \beta \vartheta_1^2 \nn \\*
& -\frac{4\pi}{3} \alpha e^{-(\nu + \lambda)/2} \left\{ R\left( 9e^{-\lambda} +1+8 \pi R^2 p \right) \frac{d \vartheta_1}{d R} \frac{d \tilde{\omega}_1}{d R} -2\left[ 2e^{-\lambda} + 3 + 16 \pi R^2 (\rho + p) \right] \vartheta_1 \frac{d \tilde{\omega}_1}{d R} \right. \nn \\*
& -\left. 64\pi R^2 (\rho+p) \tilde{\omega}_1 \frac{d \vartheta_1}{d R}  -32 \pi R \left[ R \frac{d\rho}{dR} + (\rho + p) \left[ 2 - (1 + 8 \pi R^2 p) e^{\lambda} \right] \right] \vartheta_1 \tilde{\omega}_1  \right\} \,, \\
\label{h0R}
\frac{d \xi_0}{d R} &= \frac{1}{3[1-(1+8\pi R^2 p)e^\lambda ]} \left\{ \frac{6(1+8 \pi R^2 p)}{R} e^{2\lambda} m_0 + 3 \frac{e^{-\lambda} + 16\pi R^2 p - (1-8 \pi R^2 \rho) (1+8\pi R^2 p)e^{\lambda}}{R}e^{\lambda} \xi_0 \right. \nn \\*
& + \left. R^4 e^{-\nu} \frac{d \tilde{\omega}_1}{d R} \frac{d w_1}{d R} +4R^3 e^{-\nu} \left( \frac{d \tilde{\omega}_1}{dR} w_{1} + \tilde{\omega}_1 \frac{dw_1}{dR} \right)
+ 4\left( 3 e^{-\lambda} -1-8\pi R^2 p \right) R^2 e^{-(\nu -\lambda)} \tilde{\omega}_1 w_1 +4\pi \beta R^2 \left( \frac{d \vartheta_1}{d R} \right)^2 \right. \nn \\*
& - \left. 8\pi \beta e^{\lambda} \vartheta_1^2  + 8\pi \alpha e^{-(\nu-\lambda)/2} \left[ \left( 3 e^{-\lambda} -1-8 \pi R^2 p  \right) R \left( \frac{d \vartheta_1}{d R} \frac{d \tilde{\omega}_1}{d R} +16 \pi e^{\lambda} (\rho +p) \vartheta_1 \tilde{\omega}_1 \right) \right. \right. \nn \\*
& + \left. \left. 2 [8 \pi R^2 (\rho + p)-1] \vartheta_1 \frac{d \tilde{\omega}_1}{d R}   \right] \right\}\,. 
\end{align}
\allowdisplaybreaks[0]
%\ew
%
Using Eqs.~\eqref{xi0}, \eqref{m0R} and~\eqref{h0R}, one can also obtain
%\bw
\begin{align}
\label{h0R2}
\frac{d h_0}{d R} &= \frac{8 \pi R^2 p + 1}{R^2} e^{2\lambda} m_0 - \frac{4\pi R^2 (\rho +p)}{R-2M}h_0 + \frac{1}{6} R^3 e^{-\nu} \frac{d \tilde{\omega}_1}{d R} \frac{d w_1}{d R} - \frac{8\pi}{3} \frac{e^{-\nu} R^4 (\rho +p)}{R(R-2M)} \tilde{\omega}_1 w_1 -\frac{4\pi}{3} \beta \frac{e^{\lambda}}{R} \vartheta_1^2 + \frac{2\pi}{3} \beta R \left( \frac{d \vartheta_1}{d R} \right)^2 \nn \\
& +\frac{4\pi}{3} \alpha e^{(\lambda - \nu)/2} \left[ \left( 3 e^{-\lambda} -1-8 \pi R^2 p  \right) \left( \frac{d \vartheta_1}{d R} \frac{d \tilde{\omega}_1}{d R} +16 \pi e^{\lambda} (\rho +p) \vartheta_1 \tilde{\omega}_1 \right) +2 \frac{8 \pi R^2 (\rho + p)-1}{R} \vartheta_1 \frac{d \tilde{\omega}_1}{d R}  \right]\,.
\end{align}

We now focus on the $\ell = 2$ mode, where we have 4 unknown functions: $h_2$, $m_2$, $k_2$ and $\xi_2$. Eq.~\eqref{xi2} already gives us $\xi_2$, and thus, we need 3 more equations to close the system. We choose one of these to be the $(\Theta,\Theta)$ component of the field equations minus the $(\phi,\phi)$ component. This yields
\begin{align}
m_2 &= -e^{-\lambda} R h_2 -\frac{8\pi}{3} \beta e^{-\lambda} R \vartheta_1^2 - \frac{e^{-(\nu + \lambda)} R^5}{3} \left( e^{-\lambda} \frac{d w_1}{d R} \frac{d \tilde{\omega}_1}{d R} + 16 \pi (\rho + p) w_1 \tilde{\omega}_1 \right)  - \frac{16\pi}{3} \alpha e^{-(\nu + 3 \lambda)/2} R \left[ e^{-\lambda} R \frac{d \vartheta_1}{d R} \frac{d \tilde{\omega}_1}{d R} \right. \nn \\
& - \left. \left[ e^{-\lambda} + 4 \pi R^2 (\rho + p) \right] \vartheta_1 \frac{d \tilde{\omega}_1}{d R} - 8\pi R^2 (\rho + p) \tilde{\omega}_1 \frac{d \vartheta_1}{d R} - 8 \pi \left( R^2 \frac{d\rho}{dR} - e^{\lambda} \left( 4 \pi R^3 p + M \right) (\rho+p) \right) \vartheta_1 \tilde{\omega}_1 \right]\,.
\label{m2}
\end{align}
% \ew
%
For the remaining 2 equations, we use the $(R,\Theta)$ and $(R,R)$ components of the field equations. By using Eq.~\eqref{xi2}, they yield
\begin{align}
 \frac{d k_2}{d R} &= -\frac{d h_2}{d R} + \frac{\left( 3 e^{-\lambda}-1-8 \pi R^2 p\right) e^{\lambda}}{2R}h_2 + \frac{\left( e^{-\lambda} + 1 + 8\pi R^2 p \right) e^{2\lambda}}{2R^2}m_2 - \frac{8\pi}{3} \beta \vartheta_1 \frac{d \vartheta_1}{d R} \nn \\
& + \frac{16\pi}{3} \alpha e^{-(\nu + \lambda)/2} \left( \frac{d \vartheta_1}{d R} \frac{d \tilde{\omega}_1}{d R} - \frac{\vartheta_1}{R} \frac{d \tilde{\omega}_1}{d R} +8 \pi e^{\lambda} (\rho + p) \vartheta_1 \tilde{\omega}_1 \right)
\label{h2k2}
\intertext{and}
\frac{d h_2}{d R} &= \frac{[3-4 \pi R^2 (\rho +p)]e^{\lambda}}{R} h_2 - \frac{ 1 + (1 + 8 \pi R^2 p) e^{\lambda}}{2} \frac{d k_2}{dR} + 2 \frac{e^\lambda}{R} k_2 + \frac{e^{2 \lambda}(1+8\pi R^2 p)}{R^2} m_2 +\frac{4\pi}{3} \beta R\left( \frac{d \vartheta_1}{dR} \right)^2 \nn \\
& + \frac{4\pi}{3} \beta \frac{e^\lambda}{R} \vartheta_1^2  - \frac{e^{-\nu}}{6}  R^3 \frac{d \tilde{\omega}_1}{d R} \frac{d w_1}{d R}  + \frac{8\pi}{3} e^{\lambda-\nu} R^3 (\rho +p) \tilde{\omega}_1 w_1 + \frac{8 \pi}{3} \alpha  e^{(\lambda -\nu )/2 } \left( 3 e^{-\lambda} - 1 - 8 \pi R^2 p \right) \frac{d \vartheta_1}{dR} \frac{d \tilde{\omega}_1}{dR} \nn \\
& - \frac{16\pi}{3} \alpha  e^{(\lambda -\nu )/2} \frac{1+4\pi R^2 (\rho +p)}{R} \vartheta_1 \frac{d \tilde{\omega}_1}{dR} - \frac{64\pi^2}{3} \alpha e^{(3 \lambda - \nu )/2} \left( 3 e^{-\lambda} - 1 - 8 \pi R^2 p \right) (\rho + p) \vartheta_1 \tilde{\omega}_1\,. 
\label{RR2}
\end{align}
These equations are technically valid only inside the NS, because they
were developed from the matter conservation equations
[Eqs.~\eqref{xi2}-\eqref{xi0}]. However, we have checked that the
field equations outside the star are identical to those above after setting $\rho=0=p$.
\ew

%%%%%%%%%%%%%%%%%%%%%%%%%%%%%%%%%%%
\section{Isolated Neutron Star Solution: \\* Exterior Fields}
\label{sec:exterior}

In this section, we analytically solve the equations derived in the previous section outside the NS, where $\rho = 0 = p$. We use the superscript ``ext'' to refer to exterior quantities. We impose asymptotic flatness at spatial infinity as a boundary condition.

%------------------------------------------------------------
\subsection{Exterior Solutions}

\subsubsection{GR}

At $\mathcal{O}(\alpha'^0 \chi'^0)$, we solve Eqs.~\eqref{tt-zeroth} and~\eqref{RR-zeroth} to obtain
\ba
\left( e^\nu \right)^\ext = \left( e^{-\lambda} \right)^\ext &=& 1-\frac{2M_\NS}{R} \nn \\
&\equiv & f(R)\,, 
\label{nu-lambda-ext}
\ea
where $M_\NS = \mrm{const.}$ is the mass of the non-rotating NS, which is to be determined by matching $\nu^\ext$ or $\lambda^\ext$ with $\nu^\inter$ or $\lambda^\inter$ at the surface. At $\mathcal{O}(\alpha'^0 \chi'^1)$, we substitute the above equation in Eq.~\eqref{omega1RR} and obtain
\be
\tilde{\omega}_1^\ext = \Omega_\NS - \frac{2J}{R^3}\,,
\label{omega-ext}
\ee
where $J$ is an integration constant that corresponds to the spin angular momentum of the NS in GR. We define the mass and the spin angular momentum by expanding the metric about $R = \infty$ and extracting the appropriate coefficients~\cite{alihaimoud-chen}.

\subsubsection{CS: First Order in Spin}

Let us substitute Eqs.~\eqref{nu-lambda-ext} and~\eqref{omega-ext} in
the equation for $\vartheta_{1}$ [Eq.~\eqref{vartheta1RR}]. The solution to this equation is
\begin{align}
\vartheta_1^\ext & = \frac{5}{8} \frac{\alpha}{\beta} \frac{J}{M_\NS^2} \frac{1}{R^2} \left\{ 1+2\frac{M_\NS}{R} + \frac{18}{5} \frac{M_\NS^2}{R^2} \right. \nn \\
& + \left. C_\vartheta \frac{R^2}{M_\NS^2} \left[ 1 + \frac{R}{2M_\NS} \left( 1- \frac{M_\NS}{R} \right) \ln f(R) \right] \right\}\,,
\label{vartheta1-ext}
\end{align}
where $C_\vartheta$ is an integration constant that is to be
determined by matching this solution with the interior one at the NS
surface under continuous and differentiable boundary
conditions. Notice that an integration constant has been set to
0 by asymptotic flatness.

The modified field equation at $\mathcal{O}(\alpha'^2 \chi')$ can be solved by substituting $\vartheta_1^\ext$ in Eq.~\eqref{w1RR} to obtain
\begin{align}
w_1^\ext &= \frac{2 J_\CS}{R^3} - \frac{5}{8} \frac{\xi_\CS J}{M_\NS R^6} \left\{ 1+ \frac{12}{7} \frac{M_\NS}{R} + \frac{27}{10}\frac{M_\NS^2}{R^2} \right. \nn \\
& - \left. \frac{5}{32} C_\vartheta \frac{R^5}{M_\NS^5} \left[ 1+ \frac{M_\NS}{R} - \frac{54}{5} \frac{M_\NS^3}{R^3} \right. \right. \nn \\
& + \left. \left. \frac{R}{2M_\NS} \left( 1 - \frac{64}{5} \frac{M_\NS^3}{R^3} + \frac{48}{5} \frac{M_\NS^4}{R^4} \right) \ln f(R) \right] \right\}\,.
\label{w1-ext}
\end{align}
Here, $J_\CS$ is an integration constant that corresponds to a CS
correction to the spin angular momentum (there is also a correction to
the spin angular momentum from the log
term). Eqs.~\eqref{vartheta1-ext} and~\eqref{w1-ext} agree with those
found in~\cite{alihaimoud-chen}.

\subsubsection{CS: Second Order in Spin}

Next, we find the exterior solutions at $\mathcal{O}(\alpha'^2
\chi'^2)$. For the $\ell =0$ mode, we can solve the exterior version
of Eqs.~\eqref{m0R} and~\eqref{h0R2} to find
\bw
\ba
\label{h0_ext}
h_0^\ext &=&  \frac{5}{768} \frac{\xi_\CS J^2}{M_\NS^3 R^5 f(R)} \left\{ 1 + 100 \frac{M_\NS}{R} +66 \frac{M_\NS^2}{R^2} + \frac{684}{7} \frac{M_\NS^3}{R^3} - 648 \frac{M_\NS^4}{R^4}    -30 C_\vartheta \frac{R^4}{M_\NS^4} \left[ 1-2 \frac{M_\NS}{R} \right. \right. \nn\\ 
& & \left. \left. {} - \frac{4}{3} \frac{M_\NS^2}{R^2} - \frac{2}{3} \frac{M_\NS^3}{R^3} + \frac{24}{5} \frac{M_\NS^4}{R^4}  + \frac{R}{2 M_\NS} \left( 1-3 \frac{M_\NS}{R} + \frac{1}{3} \frac{M_\NS^2}{R^2} + \frac{1}{3} \frac{M_\NS^3}{R^3} + \frac{34}{5} \frac{M_\NS^4}{R^4} - \frac{24}{5} \frac{M_\NS^5}{R^5} \right) \ln f(R) \right] \right. \nn \\
& & \left. {}- \frac{15}{2} C_\vartheta^2 \frac{R^4}{M_\NS^4} \left[1 +\frac{R}{6M_\NS} \left( 1- \frac{M_\NS}{R} \right)\ln f(R) - \frac{R^2}{12 M_\NS^2} f(R) [\ln f(R)]^2  \right] \right\} + \frac{2J J_\CS}{R^4 f(R)} -  \frac{M_\CS}{R f(R)} \,, \\
\label{m0_ext}
m_0^\ext &=&  -\frac{25}{768} \frac{\xi_\CS J^2}{M_\NS^4 R^3} \left\{ 1 + 3\frac{M_\NS}{R} + \frac{322}{5} \frac{M_\NS^2}{R^2} + \frac{198}{5} \frac{M_\NS^3}{R^3} + \frac{6276}{175} \frac{M_\NS^4}{R^4} - \frac{17496}{25} \frac{M_\NS^5}{R^5} \right.  \nn \\
& & \left. {}+ 2 C_\vartheta \frac{R^2}{M_\NS^2} \left[ 1+\frac{M_\NS}{R} - \frac{4}{5} \frac{M_\NS^2}{R^2} - \frac{72}{5} \frac{M_\NS^3}{R^3} + \frac{R}{2M_\NS} \left( 1 + \frac{1}{5} \frac{M_\NS^2}{R^2} - \frac{78}{5} \frac{M_\NS^3}{R^3} +24 \frac{M_\NS^4}{R^4}  \right) \ln f(R) \right] \right. \nn \\
& & \left. {}-\frac{1}{2} C_\vartheta^2  \frac{R^4}{M_\NS^4} \left[1 + \frac{R}{M_\NS} \left( 1 - 2 \frac{M_\NS}{R} + \frac{1}{2} \frac{M_\NS^2}{R^2} \right) \ln f(R) + \frac{R^2}{4 M_\NS^2}  \left( 1-\frac{M_\NS}{R} \right) f(R) [\ln f(R)]^2  \right] \right\} \nn \\
& & {}-\frac{2J J_\CS}{R^3} +  M_\CS\,, 
\ea
\ew
where $M_\CS$ is an integration constant that corresponds to a part of the correction to the mass if one expands the above solutions about $R = \infty$. For the $\ell =2$ mode, we obtain similar solutions, shown in Appendix~\ref{l2}, but with another integration constant $C_Q$. If one is to deal with a BH, one needs to impose regularity at the horizon. By setting $J_\CS =0$, $C_\vartheta=0$, $M_\CS=0$ and $C_Q=(3015/14336) \xi_\CS (J^2/M_\NS^8)$, the exterior metric in $(t,r,\theta,\phi)$ coordinates reduces to the previously found BH solution~\cite{kent-CSBH}.

%----------------------------------------------------------
\subsection{Asymptotic Behavior at Spatial Infinity:\\* Scalar Dipole Charge and mass\\* Quadrupole Moment Deformation}

The asymptotic behaviors of $h_0^\ext$ and $w_1^\ext$ about spatial infinity is
\begin{equation}
\label{eq:asymptotic-deltaM-deltaJ}
h_0^\ext = -\frac{\delta M}{R} + \mathcal{O}\left( \frac{1}{R^{2}} \right), \quad w_1^\ext = \frac{2 \delta J}{R^3} + \mathcal{O}\left( \frac{1}{R^4} \right)
\end{equation}
with 
\ba
\delta M &=& M_\CS + \frac{25}{1536} \xi_\CS \frac{C_\vartheta^2 J^2}{M_\NS^7}\,,  \\
\delta J &=& J_\CS - \frac{25}{384} \xi_\CS \frac{C_\vartheta J}{M_\NS^4}\,.
\ea
Physically, $\delta M$ and $\delta J$ correspond to the CS corrections in the mass and the angular momentum respectively. 

Let us now define the observable mass $\tilde{M}$ and angular momentum $\tilde{J}$ as measured by an observer at spatial infinity via
\be
\tilde{M} \equiv M_\NS + \delta M, \quad \tilde{J} \equiv J + \delta J\,, 
\ee
such that 
\be
g_{tt}^{\ext} \sim -\left( 1- \frac{2 \tilde{M_\NS}}{R} \right)\,,
\qquad
g_{t\phi}^{\ext} \sim - \frac{2 \tilde{J}}{R} \sin^2 \theta\,,
\ee
near spatial infinity. The exterior metric and the scalar field obtained in the previous subsection can be rewritten in terms of $\tilde{M}$ and $\tilde{J}$ by replacing $M_\NS \to \tilde{M}- \delta M$ and $J \to \tilde{J}-\delta J$.

\if0%%%%%%%%%%%%%%%%%%%%%%%%%%%%%

With these replacements, the metric perturbations now have the asymptotic behaviors as
%
\ba
w_1^\ext &=& \frac{5}{24} \xi_\CS \frac{(C_\vartheta-3) \tilde{J}}{\tilde{M} R^6} + \mathcal{O}\left( \frac{1}{R^7} \right)\,,  \\
h_0^\ext &=& \frac{5}{6912} \xi_\CS \frac{(C_\vartheta-3)^2 \tilde{J}^2}{\tilde{M}^3 R^5} + \mathcal{O}\left( \frac{1}{R^6} \right)\,, \\
m_0^\ext &=& -\frac{25}{6912} \xi_\CS \frac{(C_\vartheta-3)^2 \tilde{J}^2}{\tilde{M}^4 R^3} + \mathcal{O}\left( \frac{1}{R^4} \right)\,, \\
h_2^\ext &=& \frac{5}{6912} \xi_\CS \frac{(C_\vartheta-3)^2 \tilde{J}^2}{\tilde{M}^3 R^5} + \mathcal{O}\left( \frac{1}{R^6} \right)\,, \\
k_2^\ext &=& \frac{1}{3840} \frac{25 \xi_\CS C_\vartheta (C_\vartheta-8) \tilde{J}^2 + 2048 C_Q \tilde{M}^8}{\tilde{M}^5 R^3} \nn \\
& & + \mathcal{O}\left( \frac{1}{R^4} \right)\,, \\
m_2^\ext &=& \frac{1}{3840} \frac{25 \xi_\CS C_\vartheta (C_\vartheta-8) \tilde{J}^2 + 2048 C_Q \tilde{M}^8}{\tilde{M}^5 R^2} \nn \\
& & + \mathcal{O}\left( \frac{1}{R^3} \right)\,. 
\ea
%
From Eq.~\eqref{metric-ansatz-rth}, 

\fi%%%%%%%%%%%%%%%%%%%%%%%%%%%%%%%%%5

The full exterior metric can then be written as $g_{\mu \nu}^{\ext} = g_{\mu \nu}^{\ext,\GR} + g_{\mu \nu}^{\ext,\CS}$, where $g_{\mu \nu}^{\ext,\GR}$ is the Hartle-Thorne metric with mass $\tilde{M}$ and angular momentum $\tilde{J}$, while $g_{\mu \nu}^{\ext,\CS}$ is a CS correction. The latter, in $(t,r,\theta,\phi)$ coordinates, has the following asymptotic behavior about spatial infinity
\ba
\label{eq:gtt-CS-asymp}
g_{tt}^{\ext, \CS} &=&  \frac{1}{3840} \frac{25 \xi_\CS C_\vartheta (C_\vartheta-8) \tilde{J}^2 + 2048 C_Q \tilde{M}^8}{\tilde{M}^5 r^3} \nn \\
& & {}\times (3 \cos^2 \theta -1) + \mathcal{O}\left( \frac{1}{r^4} \right) \\
g_{rr}^{\ext, \CS} &=&  \frac{1}{3840} \frac{25 \xi_\CS C_\vartheta (C_\vartheta-8) \tilde{J}^2 + 2048 C_Q \tilde{M}^8}{\tilde{M}^5 r^3} \nn \\
& & {}\times (3 \cos^2 \theta -1) + \mathcal{O}\left( \frac{1}{r^4} \right) \\
g_{\theta \theta}^{\ext, \CS} &=&  \frac{1}{3840} \frac{25 \xi_\CS C_\vartheta (C_\vartheta-8) \tilde{J}^2 + 2048 C_Q \tilde{M}^8}{\tilde{M}^5 r} \nn \\
& & {}\times (3 \cos^2 \theta -1) + \mathcal{O}\left( \frac{1}{r^2} \right) \\
g_{\phi \phi}^{\ext, \CS} &=&  \sin^2 \theta g_{\theta\theta}^{\ext, \CS} \\
g_{t \phi}^{\ext, \CS} &=&  \frac{5}{24} \xi_\CS \frac{(3- C_\vartheta) \tilde{J}}{\tilde{M} r^4} \sin^2 \theta + \mathcal{O}\left( \frac{1}{r^5} \right)\,. 
\ea
One can read off the correction to the gravitational potential per unit mass from Eq.~\eqref{eq:gtt-CS-asymp} as 
\be
\label{eq:UA_CS}
\delta U^{\CS} = -\frac{3}{r^{3}} Q \hat{S}^{i}\hat{S}^{j} n^{<ij>}
\ee
with 
\begin{equation}
Q \equiv \frac{1}{7680} \frac{25 \xi_\CS C_\vartheta (C_\vartheta-8) \tilde{J}^2 + 2048 C_Q \tilde{M}^8}{\tilde{M}^5} \,.
\label{eq:QA}
\end{equation}
Here, $\hat{S}^{i}$ is the unit spin angular momentum vector of the NS and $n^i$ is the unit vector to the field point. Notice that this definition of $\hat{S}^{i}$ \emph{differs} from that used in~\cite{quadratic}, because
in the latter $\hat{S}^{i}$ was not a proper unit vector. $Q$ corresponds to the correction to the quadrupole moment\footnote{This definition of quadrupole moment is different from the usual one given in e.g.~\cite{poisson-quadrupole,laarakkers} by a factor of 2. Although somewhat unconventional, we choose here to continue using the definitions of Ref.~\cite{kent-CSGW}.} and in the non-spinning BH case, Eq.~\eqref{eq:QA} reduces to $Q_\mrm{BH} = (201/3584) (\xi_\CS/\tilde{M}^4) \tilde{M}^3 \tilde{\chi}^2$, which agrees with~\cite{kent-CSGW}.

Finally, the asymptotic behavior of the scalar field is
\be
\vartheta^\ext =  \frac{5(3- C_\vartheta)}{24} \frac{\alpha}{\beta} \tilde{\chi} \frac{ \cos\theta}{ r^2} + \mathcal{O}\left( \frac{1}{r^3} \right)\,,
\ee
where $\tilde{\chi} \equiv \tilde{J}/\tilde{M}^2$. Comparing this to $\vartheta = \mu^i n_i/r^2 + \mathcal{O}(r^{-3})$ (see Eq.~(57) of Ref.~\cite{quadratic}) where $\mu^i$ is the scalar dipole charge, we can extract $\mu^i$:
\begin{equation}
\label{eq:mu-J-relation}
\mu^i = \frac{5(3-C_\vartheta)}{24} \frac{\alpha}{\beta} \tilde{\chi} \hat{S}^i\,.
\end{equation}
 For later convenience, we introduce the dimensionless scalar charge $\bar{\mu}$ in terms of $C_\vartheta$ and the compactness of the NS $C$:
\begin{equation}
\label{eq:mu-mubar-relation}
\bar{\mu} \equiv \frac{3-C_\vartheta}{3} \frac{1}{C^3}, \quad C \equiv \frac{\tilde{M}}{\RNS}\,.
\end{equation}
By using this $\bar{\mu}$, Eq.~\eqref{eq:mu-J-relation} can be rewritten as
\be
\label{eq:mu-S-relation}
\mu^{i} = \frac{5}{8}\frac{\alpha}{\beta}C^{3}\tilde{\chi} \hat{S}^{i}\bar{\mu}\,.
\ee
From Eq.~(59) of Ref.~\cite{quadratic}, we expect $\mu^i$ to be proportional to $C^3$, and hence, we have factored out $C^3$ from the definition of $\bar{\mu}$ above. In the non-spinning BH case, $\bar{\mu}$ reduces to $\bar{\mu}_\mrm{BH} = 8$, because then $C^{3} = 1/8$, so that $\bar{\mu}_{\mrm{BH}} C^{3} = 1$.  

Notice that the difference between $\tilde{M}$ and $M_\NS$ (and also between $\tilde{J}$ and $J$) in Eqs.~\eqref{eq:gtt-CS-asymp}--\eqref{eq:mu-S-relation} is irrelevant to the order of approximation considered here. Hence, we can freely set $\tilde{M} = M_\NS$ (and $\tilde{J} = J$) in these equations.
 
%-----------------------------------------------------------
\section{Isolated Neutron Star Solution:\\* Interior Fields}
\label{sec:interior}

In order to determine $\bar{\mu}$, $Q$, $\delta M$ and $\delta J$, we need to calculate the integration constants that appear in the definitions of these quantities. This can be achieved by matching the interior solutions to the exterior ones at the NS surface. In this section, we first explain the matching conditions at the surface. We then obtain the asymptotic behavior of the metric perturbations at the NS center to obtain the initial conditions for the numerical integration of the interior solution. We conclude with an explanation of the numerical procedure adopted in this paper. We use the superscript ``int'' to refer to interior quantities. 

\subsection{Boundary Conditions at the Surface}

In the previous subsection, we imposed asymptotic flatness at spatial infinity to find the exterior solutions but 4 integration constants remain, which are to be determined by the boundary conditions at the NS surface. At $\mathcal{O}(\alpha'^0 \chi'^0)$, we solve Eqs.~\eqref{tt-zeroth},~\eqref{RR-zeroth} and~\eqref{TOV-zeroth}, together with the EoS, for $\nu$, $\lambda$, $\rho$ and $p$. At the NS surface, we impose the continuity condition
\be
p(\RNS) =0, \quad e^{\nu (\RNS)} = e^{-\lambda (\RNS)} = 1-\frac{2 M_\NS}{\RNS}\,.
\label{BC-zeroth}
\ee
The first equation determines $\RNS$ while the second equation determines $M_\NS$.

At $\mathcal{O}(\alpha'^0 \chi')$, we solve Eq.~\eqref{omega1RR} for $\tilde\omega_{1}$. Since this is a second-order differential equation, we need to impose two boundary conditions at the surface. We choose one to be the continuity condition for $\tilde{\omega}_1$, i.e.
\be
\tilde{\omega}_1 (\RNS) = \Omega_\NS - \frac{2 J}{\RNS^3}\,.
\label{BC-omega1-1}
\ee
The other condition can be obtained by integrating Eq.~\eqref{omega1RR} from $\RNS-\epsilon$ to $\RNS + \epsilon$ and taking the limit as $\epsilon \rightarrow 0$:
\ba
\left[  \frac{d \tilde{\omega}_1}{dR} (\RNS) \right] &=& \lim_{\epsilon \rightarrow 0} \left\{ \frac{d \tilde{\omega}_1^\ext{}}{dR}  (\RNS + \epsilon) - \frac{ d \tilde{\omega}_1^\inter{}}{d R}  (\RNS - \epsilon) \right\}  \nn \\
&=& \lim_{\epsilon \rightarrow 0} \int^{\RNS + \epsilon}_{\RNS - \epsilon} \frac{d^2 \tilde{\omega}_1}{d R^2} dR\,, \nn \\
&=& \lim_{\epsilon \rightarrow 0} \int^{\RNS + \epsilon}_{\RNS - \epsilon} \left\{ - 4\frac{1- \pi R^2 (\rho +p) e^{\lambda}}{R}\frac{d \tilde{\omega}_1}{d R} \right. \nn \\ 
& & \left.  + 16 \pi (\rho +p) e^{\lambda} \tilde{\omega}_1\right\} dR\,,
\label{omega-junc}
\ea
where we have introduced the notation
\be
[ A(\RNS) ] \equiv \lim_{\epsilon \to 0} \left[ A^\ext (\RNS + \epsilon) - A^\inter (\RNS- \epsilon) \right]
\ee
for any quantity $A(R)$. Since the integrand is bounded, the right hand side of Eq.~\eqref{omega-junc} vanishes, leading to
\be
\left[ \frac{d \tilde{\omega}_1}{d R} (\RNS) \right] = 0\,.
\label{BC-omega1-2}
\ee

Let us introduce the moment of inertia $I$, defined by~\cite{hartle1967}
\be
\label{eq:I-definition}
I \equiv \frac{J}{\Omega_\NS}\,.
\ee
Equation~\eqref{BC-omega1-1} can then be rewritten as
\be
\tilde{\omega}_1 (\RNS) = \Omega_\NS \left( 1 - \frac{2 I}{\RNS^3} \right)\,.
\label{BC-omega1-3}
\ee
By Eqs.~\eqref{tt-zeroth},~\eqref{RR-zeroth},~\eqref{omega1RR} and~\eqref{BC-omega1-2}, $I$ can be expressed in integral form as~\cite{hartle1967,kalogera-psaltis}
\be
I = \frac{8\pi}{3} \frac{1}{\Omega_\NS} \int_0^{\RNS} \frac{e^{-(\nu + \lambda )/2} R^5 (\rho + p) \tilde{\omega}_1}{R-2M} dR\,.
\label{I}
\ee 
This expression is valid both in GR, as well as in dynamical CS gravity~\cite{alihaimoud-chen}.

As in the $\mathcal{O}(\alpha'^0 \chi')$ case, the scalar evolution equation [Eq.~\eqref{vartheta1RR}] is a second-order differential equation. Since all terms in this equation are bounded at the surface, we impose the continuity and differentiability conditions
\be
[ \vartheta_1 (\RNS) ] = 0 = \left[ \frac{d \vartheta_1}{d R} (\RNS) \right]\,.
\label{BC-vartheta}
\ee
At $\mathcal{O}(\alpha'^2 \chi')$, the non-vanishing component of the field equations is given by Eq.~\eqref{w1RR}, which is also a second-order differential equation, but here the situation is slightly different. First, we impose the continuity condition at the surface 
\be
[ w_1 (\RNS) ] = 0\,.
\label{BC-w-1}
\ee
Next, we integrate Eq.~\eqref{w1RR} from $\RNS-\epsilon$ to $\RNS + \epsilon$ and take the limit $\epsilon \to 0$ as
\ba
& & \left[  \frac{d w_1}{dR} (\RNS) \right] \nn \\
&=& \lim_{\epsilon \rightarrow 0} \int^{\RNS + \epsilon}_{\RNS - \epsilon} \frac{d^2 w_1}{d R^2} dR\,, \nn \\
&=& \lim_{\epsilon \rightarrow 0} \int^{\RNS + \epsilon}_{\RNS - \epsilon} \left\{ - 4 \frac{1-\pi R^2 (\rho +p) e^\lambda}{R} \frac{d w_1}{d R} \right. \nn \\
& & \left. +16 \pi (\rho +p) e^\lambda w_1 \right. \nn \\ 
& & \left.  - \frac{128 \pi^2 \alpha e^{(\nu+\lambda)/2}}{R^3} \left[ \delta R \frac{d \vartheta_1}{d R} + \left( R \frac{d\rho}{dR} - \delta \right) \vartheta_1 \right] \right\} dR\,, \nn \\
&=& - 128 \pi^2 \alpha \lim_{\epsilon \rightarrow 0} \int^{\RNS + \epsilon}_{\RNS - \epsilon} \frac{e^{(\nu+\lambda)/2} \vartheta_1}{R^2} \frac{d\rho}{dR} dR\,.
\label{w-junc}
\ea
where in the last equality, we only kept the term in the integrand that is not bounded\footnote{The factor $d \rho /dR$ is not bounded since $\rho$ is discontinuous at the surface.}. This equation can be further simplified by integrating by parts:
\ba
\left[  \frac{d w_1}{dR} (\RNS) \right] &=& - 128 \pi^2 \alpha \lim_{\epsilon \rightarrow 0} \int^{\RNS + \epsilon}_{\RNS - \epsilon}  \left\{ \frac{d}{dR} \left( \frac{e^{(\nu+\lambda)/2} \vartheta_1}{R^2} \rho \right) \right. \nn \\
& & \left. - \rho \frac{d}{dR} \left( \frac{e^{(\nu+\lambda)/2} \vartheta_1}{R^2} \right)  \right\} dR\,.
\ea
Since the integrand in the second term is bounded, this term vanishes, so that one finally finds the jump condition
\ba
\left[  \frac{d w_1}{dR} (\RNS) \right] &=& 128 \pi^2 \alpha  \frac{\vartheta_1 (\RNS)}{\RNS^{2}} \rho (\RNS)\,.
\ea
This agrees with the condition found in~\cite{alihaimoud-chen} in the limit $\epsilon \to 0$.
Since the density of the NS at the surface is typically about 7 orders of magnitude lower than the mean density, the above jump condition shows that $d w_1/dR$ is almost continuous at the surface~\cite{alihaimoud-chen}. Therefore, in the numerical calculation below, we adopt the differentiability condition~\cite{alihaimoud-chen}
\be
\left[  \frac{d w_1}{dR} (\RNS) \right] = 0\,.
\label{BC-w-2}
\ee
At $\mathcal{O}(\alpha'^2 \chi'^2)$, the evolution equations are first-order. We thus impose the continuity conditions
\begin{equation}
\begin{aligned}
\left[h_0 (\RNS) \right] &= 0 = [m_0 (\RNS) ]\,,\\
[h_2 (\RNS) ] &= 0 = [k_2 (\RNS) ]\,.
\end{aligned}
\label{BC-2nd-order}
\end{equation}
% 

%--------------------------------------
\subsection{Asymptotic Behavior at the Stellar Center}
\label{sec:asymptotia-center}

Before we numerically integrate the interior equations, we must understand the asymptotic behavior of the pressure, density, metric perturbations, and the scalar field at the NS center. First, let us focus on the pure GR case. In the non-spinning sector, we can use Eqs.~\eqref{tt-zeroth},~\eqref{RR-zeroth} and~\eqref{TOV-zeroth} to  asymptotically expand $\rho$, $p$, $M$ and $\nu$ about $R \ll M$ via
\begin{align}
\label{rho0}
\rho (R) &= \rho_c + \rho_2 R^2 + \mathcal{O}(R^3)\,, \\
\label{p-center}
p (R) &= p_c - \frac{2\pi}{3} (\rho_c + p_c) (\rho_c + 3 p_c) R^2 + \mathcal{O}(R^3)\,, \\
\label{center-M}
M(R) &= \frac{4\pi}{3} \rho_c R^3 + \frac{4\pi}{5} \rho_2 R^5 + \mathcal{O}(R^6)\,, \\ 
\label{nu0}
\nu (R) &= \nu_c + \frac{4\pi}{3} (\rho_c + 3 p_c) R^2 + \mathcal{O}(R^3) \quad (R \to 0^{+})\,,
\end{align}
where $\rho_c$, $p_c=p(\rho_c)$ and $\nu_c$ are the density, pressure and $\nu$ at the NS center, respectively. $\rho_2$ can be expressed in terms of $\rho_c$ and $p_c$ through Eq.~\eqref{p-center}, the TOV equation and the EoS. $\nu_c$ is determined by matching the interior solution with the exterior one at the surface. To first-order in spin, the solution to Eq.~\eqref{omega1RR} that is regular at the center is 
\be
\tilde{\omega}_1 (R) = \tilde{\omega}_c + \frac{8\pi}{5} (\rho_c + p_c) \tilde{\omega}_c R^2 + \mathcal{O}(R^3) \quad (R \to 0^{+})\,,
\label{omega0}
\ee
where $\tilde{\omega}_c$ is a constant that is to be determined by the matching condition at the surface.

Let us now focus on the scalar field and the CS corrections to the metric perturbation. For the former, the solution to Eq.~\eqref{vartheta1RR} that is regular at the center has the asymptotic form
\be
\vartheta_1 (R) = \vartheta'_c R + \frac{2\pi}{15} (5\rho_c - 3p_c) \vartheta'_c R^3 + \mathcal{O}(R^4) \quad (R \to 0^{+})\,.
\label{vartheta0}
\ee
$\vartheta'_c$ is a constant that corresponds to $d\vartheta/dR$ at the center. From Eq.~\eqref{w1RR}, the metric perturbation at first order in spin has the asymptotic behavior
\ba
w_1 (R) &=& w_c + \frac{8\pi}{5} \left[ (\rho_c + p_c) w_c -16\pi \alpha e^{\nu_c/2} \rho_2 \vartheta'_c \right] R^2  \nn \\
& &{}+ \mathcal{O}(R^3) \quad (R \to 0^{+})\,.
\label{w0}
\ea
Again, $\vartheta'_c$ and $w_c$ are to be determined by matching at the surface. 
For the $\ell =0$ mode corrections at quadratic order in spin, Eqs.~\eqref{xi0}--\eqref{h0R} yield 
\ba
\label{center-h0}
h_0 (R) &=& h_{0c} + \frac{128\pi^2}{3} \alpha (\rho_c + p_c) (\rho_c + 3 p_c) e^{-\nu_c/2} \tilde{\omega}_c \vartheta'_c R^2 \nn \\
& & {}+ \mathcal{O}(R^3), \\
\label{center-m0}
m_0 (R) &=& \frac{2\pi}{3} \vartheta'_c \left[ 64 \pi \alpha (\rho_c + p_c) e^{-\nu_c/2} \tilde{\omega}_c + \beta \vartheta'_c \right] R^3 \nn \\
& & {}+ \mathcal{O}(R^4), \\ 
\label{center-xi0}
\xi_0 (R) &=& -\frac{\tilde{\omega}_c e^{-\nu_c/2} }{3\pi(\rho_c + 3 p_c)}  \left[ 64 \pi^2 \alpha (\rho_c + p_c)  \vartheta'_c + e^{-\nu_c/2} w_c \right]  R \nn \\
& & {}+ \mathcal{O}(R^2)  \quad (R \to 0^{+})\,.
\label{center0}
\ea
Similarly, Eqs.~\eqref{xi2} and~\eqref{m2}--\eqref{RR2} yield the asymptotic behaviours for the $\ell =2$ mode:
\ba
\label{center-h2}
h_2 (R) &=& h_{2c}'' R^2 + \mathcal{O}(R^3)\,, \\
\label{center-k2}
k_2 (R) &=& -\frac{1}{3} \left[ 3 h_{2c}'' -128 \pi^2 \alpha (\rho_c + p_c) e^{-\nu_c/2} \vartheta'_c \tilde{\omega}_c \right. \nn \\
& & \left. + 8 \pi \beta \vartheta'_c{}^2 \right] R^2 + \mathcal{O}(R^3)\,, \\
\label{center-m2}
m_2 (R) &=& -\frac{1}{3} \left[ 3 h_{2c}'' -128 \pi^2 \alpha (\rho_c + p_c) e^{-\nu_c/2} \vartheta'_c \tilde{\omega}_c \right. \nn \\
& & \left.  + 8 \pi \beta \vartheta'_c{}^2 \right] R^3 + \mathcal{O}(R^4)\,, \\
\label{center-xi2}
\xi_2 (R) &=& - \frac{3 h_{2c}'' - 2 e^{-\nu_c} \tilde{\omega}_c w_c}{4\pi (\rho_c + 3 p_c)} R + \mathcal{O}(R^2)  \quad (R \to 0^{+})\,. \nn \\
\ea
Here, $h_{2c}''$ is a constant that needs to be determined by matching at the NS surface.

%--------------------------------------
\subsection{Numerical Method}

Let us now explain how we solve the interior equations numerically to obtain the corresponding interior solutions.

\subsubsection{Solution to $\mathcal{O}(\chi'{}^0)$}

First, we must obtain the GR solutions at zeroth order in spin. There are 4 unknown functions that need to be determined, $\nu$, $\lambda$ [or equivalently $M$ through Eq.~\eqref{M-lambda}], $\rho$ and $p$. The 4 equations that we need to solve are Eqs.~\eqref{tt-zeroth},~\eqref{RR-zeroth},~\eqref{TOV-zeroth}, given the EoS. Notice that Eqs.~\eqref{tt-zeroth},~\eqref{TOV-zeroth} and the EoS form a closed system for $M$, $\rho$ and $p$. We solve these equations as an initial value problem using an adaptive step-size, fourth-order Runge-Kutta method (from the GSL library~\cite{gsl}) with an accuracy of $10^{-3}$ on each step, starting at $R=R_\epsilon$ toward $R=\RNS$, with the initial conditions given in Eqs.~\eqref{rho0} and~\eqref{center-M}, where $R_\epsilon$ is the core radius $R_\epsilon/\RNS \ll 1$. The radius of the NS can be obtained by finding the radius $\RNS$ where $p(\RNS)=0$ and the mass of the NS is given by $M_\NS = M(\RNS)$. We repeat the calculation for various $\rho_c$ to obtain a mass-radius relation. 

The remaining equation, Eq.~\eqref{RR-zeroth}, can be solved for $\nu$, again using an adaptive step-size, fourth-order Runge-Kutta method with an accuracy of $10^{-3}$ per step, from $R=R_\epsilon$ to $\RNS$ and with the initial condition given in Eq.~\eqref{nu0}. However, one must be careful that $\nu(\RNS)$ satisfies the boundary condition given in Eq.~\eqref{BC-zeroth}. Suppose we obtained the trial solution $\nu_\mrm{tr}(R)$ by using the initial condition $\nu(R_\epsilon) =(4\pi/3) (\rho_c + 3 p_c) R_\epsilon^2$ (i.e. $\nu_c = 0$). Since Eq.~\eqref{RR-zeroth} is shift-invariant, $\nu_\mrm{tr}(R)$ plus a constant $C_\nu$ is also a solution. The new solution $\nu_\mrm{tr}(R)+C_\nu$ will satisfy the correct boundary condition provided
\be
e^{\nu_\mrm{tr}(\RNS)+C_\nu} = f(\RNS)\,,
\ee
which then yields~\cite{yunespsaltis}
\be
C_\nu = \ln f(\RNS) - \nu_\mrm{tr} (\RNS)\,.
\ee

\subsubsection{Solution to $\mathcal{O}(\chi')$}
\label{GR-1st}

Next, we solve Eq.~\eqref{omega1RR}, a second-order differential equation, for $\tilde{\omega}_1$.  We take advantage of the fact that Eq.~\eqref{omega1RR} is linear and homogeneous~\cite{kalogera-psaltis} and solve this equation with the initial condition given in Eq.~\eqref{omega0}. We choose a specific value for $\tilde{\omega}_c$ and solve the equation from $R=R_\epsilon$ to $\RNS$ using a fourth-order Runge-Kutta method to obtain the trial solution $\tilde{\omega}_\mrm{tr}(R)$ and the trial moment of inertia $I_\mrm{tr}$ calculated from Eq.~\eqref{I}. The solution we seek is one that satisfies the boundary condition of Eq.~\eqref{BC-omega1-3}. It is improbable that $\tilde{\omega}_\mrm{tr}$ will satisfy this boundary condition. However, due to the homogeneity of the differential equation, the product of $\tilde{\omega}_\mrm{tr}$ and a constant is also a solution. Hence, we can construct a new solution via $C_{\tilde{\omega}} \tilde{\omega}_\mrm{tr}$, where $C_{\tilde{\omega}}$ is a constant, which will satisfy the boundary condition, provided that 
\be
C_{\tilde{\omega}} \tilde{\omega}_\mrm{tr} (\RNS) = \Omega_\NS \left( 1 - \frac{2 C_{\tilde{\omega}} I_\mrm{tr}}{\RNS^3} \right)\,, 
\ee
which then yields
\be
C_{\tilde{\omega}} = \frac{\Omega_\NS \RNS^3}{\tilde{\omega}_\mrm{tr}(\RNS) \RNS^3 + 2 I_\mrm{tr} \Omega_\NS}\,.
\ee
Clearly, $\Omega_{\NS}$ (like $\rho_{c}$ or $p_{c}$) is a quantity that must be specified {\emph{a priori}} and controls 
how rapidly the NS is rotating.
%
%
%\subsubsection{Scalar Evolution Equation}
%\label{scale-field-meta-code}

The scalar field satisfies Eq.~\eqref{vartheta1RR}, which is not homogeneous or shift-invariant. Since the differential equation is linear, we take the following approach~\cite{hartle1967}. First, we solve Eq.~\eqref{vartheta1RR} with arbitrary values for $\vartheta_1 (R_\epsilon)$ and $\vartheta_1' (R_\epsilon)$ such that the solution satisfies regularity at the NS center [i.e. with an arbitrary value of $\vartheta_1'$ in Eq.~\eqref{vartheta0}] to obtain a particular solution $\vartheta_p^\inter (R)$ in the interior region. Next, we look for a homogeneous solution $\vartheta_h^\inter (R)$ in the interior region by solving Eq.~\eqref{vartheta1RR} with a vanishing source term on the right-hand-side with arbitrary $\vartheta_1 (R_\epsilon)$ and $\vartheta_1' (R_\epsilon)$ such that $\vartheta_h^\inter (R)$ satisfies regularity at the NS center. The asymptotic behavior of $\vartheta_h^\inter (R)$ that is regular at the NS center is also given by Eq.~\eqref{vartheta0}. With $\vartheta_p^\inter (R)$ and $\vartheta_h^\inter (R)$ at hand, one can construct a generic solution to Eq.~\eqref{vartheta1RR} in the interior region that satisfies regularity at the NS center as 
\be
\vartheta_1^\inter (R) = \vartheta_p^\inter (R) + C_\vartheta^h \vartheta_h^\inter (R)\,,
\label{theta-sol-generic}
\ee
where $C_\vartheta^h$ is a constant. The constants $C_\vartheta$ in Eq.~\eqref{vartheta1-ext} and $C_\vartheta^h$ in Eq.~\eqref{theta-sol-generic} are determined by the matching condition at the NS surface given in Eq.~\eqref{BC-vartheta}. One can solve Eq.~\eqref{BC-vartheta} for  $C_\vartheta$ and $C_\vartheta^h$ algebraically in terms of $\vartheta_p^\inter (\RNS)$ and $\vartheta_h^\inter (\RNS)$.

The CS correction to the gravitomagnetic sector of the metric is controlled by Eq.~\eqref{w1RR}, which is a second order differential equation for $w_1$. Again, this equation is not shift invariant or scale invariant, and thus, we must solve it as we did the scalar field above. The matching conditions at the NS surface are given in Eqs.~\eqref{BC-w-1} and~\eqref{BC-w-2}, while the asymptotic behavior of the solution that is regular at the NS center is given in Eq.~\eqref{w0}. The asymptotic behavior of the homogeneous solution is given by setting $\alpha = 0$ in Eq.~\eqref{w0}, which is the same as Eq.~\eqref{omega0} with $\tilde{\omega}_c$ being replaced by $w_c$. 

We have that the solutions for $\vartheta_1$ and $w_1$ obtained in this way are identical to solving Eqs.~\eqref{vartheta1RR} and~\eqref{w1RR} via a purely numerical shooting method.

\subsubsection{Solution to $\mathcal{O}(\chi'{}^2)$}

For the $\ell=0$ mode at $\mathcal{O}(\alpha'^2 \chi'^2)$, we need to solve Eqs.~\eqref{m0R} and~\eqref{h0R} to obtain $h_{0c}$ in Eq.~\eqref{xi0}. Since these are two coupled first-order differential equations, we can solve them as an initial value problem with the initial conditions given in Eqs.~\eqref{center-m0} and~\eqref{center-xi0}. Then, we impose the continuity condition at the surface to obtain $M_\CS$ in Eq.~\eqref{m0_ext}. The interior solution for $h_0$ can be obtained through Eq.~\eqref{xi0}. By imposing the continuity of $h_0$ at the surface, one can determine $h_{0c}$.

For the $\ell = 2$ mode, we need to solve Eqs.~\eqref{h2k2} and~\eqref{RR2} for $h_2$ and $k_2$. We take the same steps explained above for solving Eqs.~\eqref{vartheta1RR} and~\eqref{w1RR}. We first obtain the particular solutions $h_p^\inter$ and $k_p^\inter$ with an arbitrary choice of $h_c''$ in Eqs.~\eqref{center-h2} and~\eqref{center-k2}. Next, we look for homogeneous solutions $h_p^\inter$ and $k_p^\inter$. The initial conditions at the NS center is given by setting $\alpha=0$ and $\beta=0$ in Eqs.~\eqref{center-h2} and~\eqref{center-k2}. As in Eq.~\eqref{theta-sol-generic}, one can construct generic solutions for $h_2$ and $k_2$ that are regular at the NS center via
\ba
h_2^\inter (R) &=& h_p^\inter (R) + C_Q^h h_h^\inter (R)\,, \\
k_2^\inter (R) &=& k_p^\inter (R) + C_Q^h k_h^\inter (R)\,, 
\ea
where $C_Q^h$ is a constant that is to be determined, together with another constant $C_Q$ in Eqs.~\eqref{h2-ext} and~\eqref{k2-ext}, with the matching condition at the NS surface given in Eq.~\eqref{BC-2nd-order}. One can solve Eq.~\eqref{BC-2nd-order} for $C_Q$ and $C_Q^h$ algebraically in terms of $h_2(\RNS)$ and $k_2(\RNS)$. We have checked that the solutions obtained in this fashion are identical to numerically solving Eqs.~\eqref{h2k2} and~\eqref{RR2} via a Riccati method~\cite{dieci1,dieci2,takata}\footnote{We had difficulty in stably carrying out the shooting method due to the nearly scale-invariant structure of Eqs.~\eqref{h2k2} and~\eqref{RR2} for most of the region in parameter space.}.

%------------------------------------------------------------
\section{Numerical Solutions}
\label{sec:solutions}

In this section, we show the numerically obtained NS solutions using 4 different EoSs: APR~\cite{APR}, SLy~\cite{SLy,shibata-fitting}, LS220~\cite{LS, ott-EOS} and Shen~\cite{Shen1,Shen2,ott-EOS}. 

\begin{figure*}[ht]
\begin{center}
\begin{tabular}{l}
\includegraphics[width=8.5cm,clip=true]{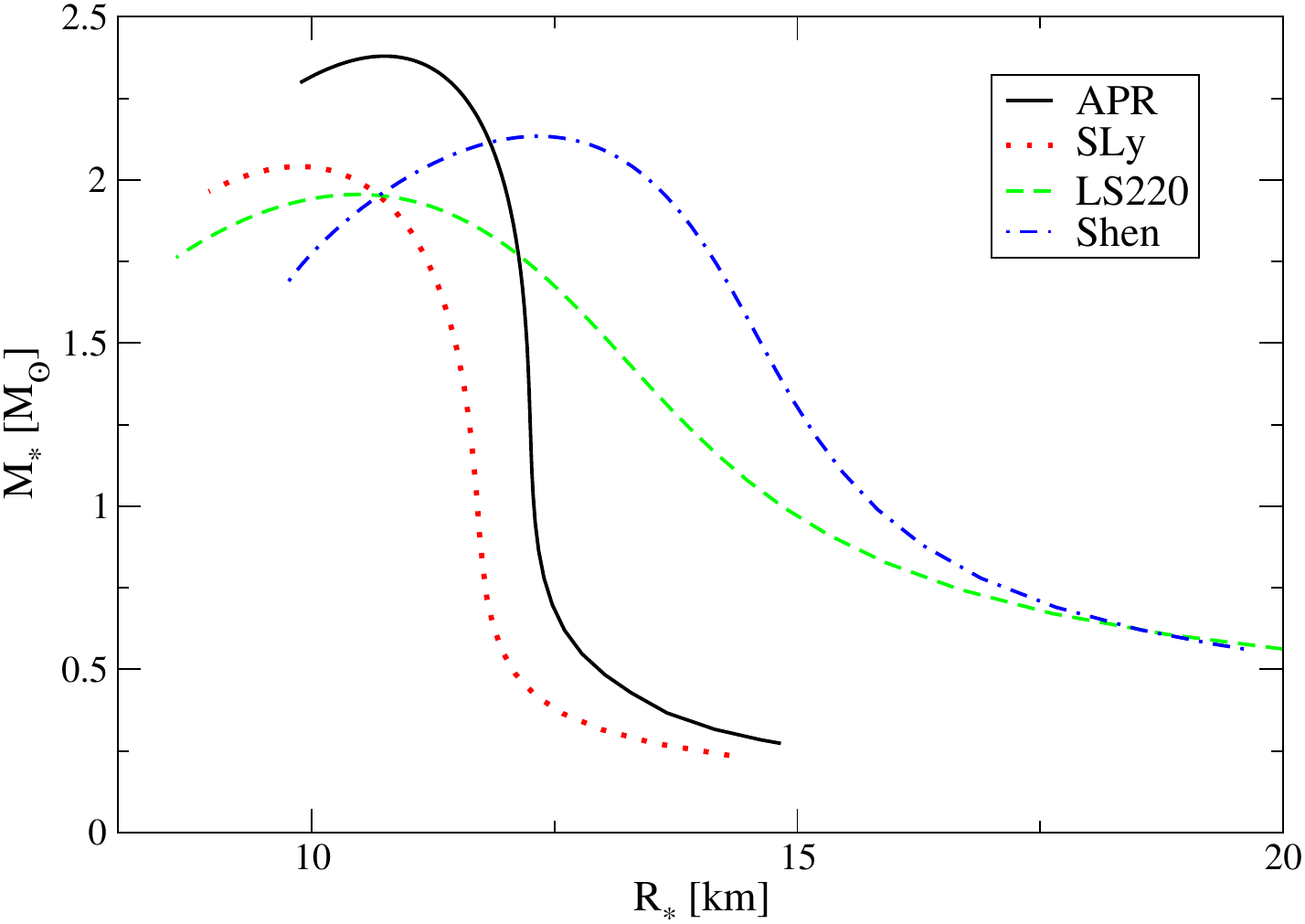}  \qquad
\includegraphics[width=8.5cm,clip=true]{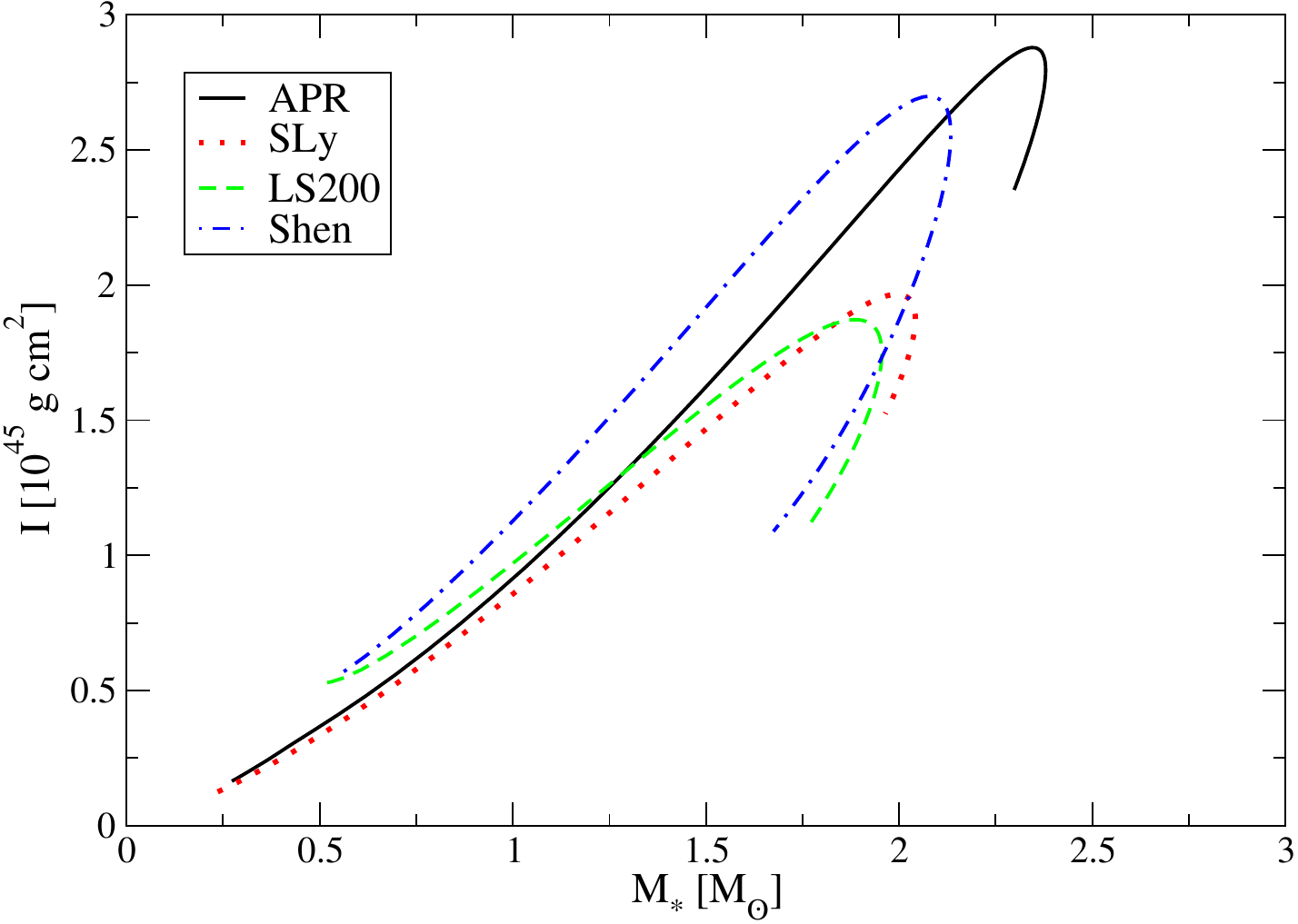} 
\end{tabular}
\caption{\label{fig:MR} 
(Color online)
The mass-radius relation (left) and the moment-of-inertia
[Eq.~\eqref{eq:I-definition}] vs. mass relation (right) in GR for NSs
with various EoSs: APR (solid), SLy (dotted), LS220 (dashed), and Shen
(dot-dashed).}
\end{center}
\end{figure*}

\begin{figure*}[tp]
\includegraphics[width=9.0cm,clip=true]{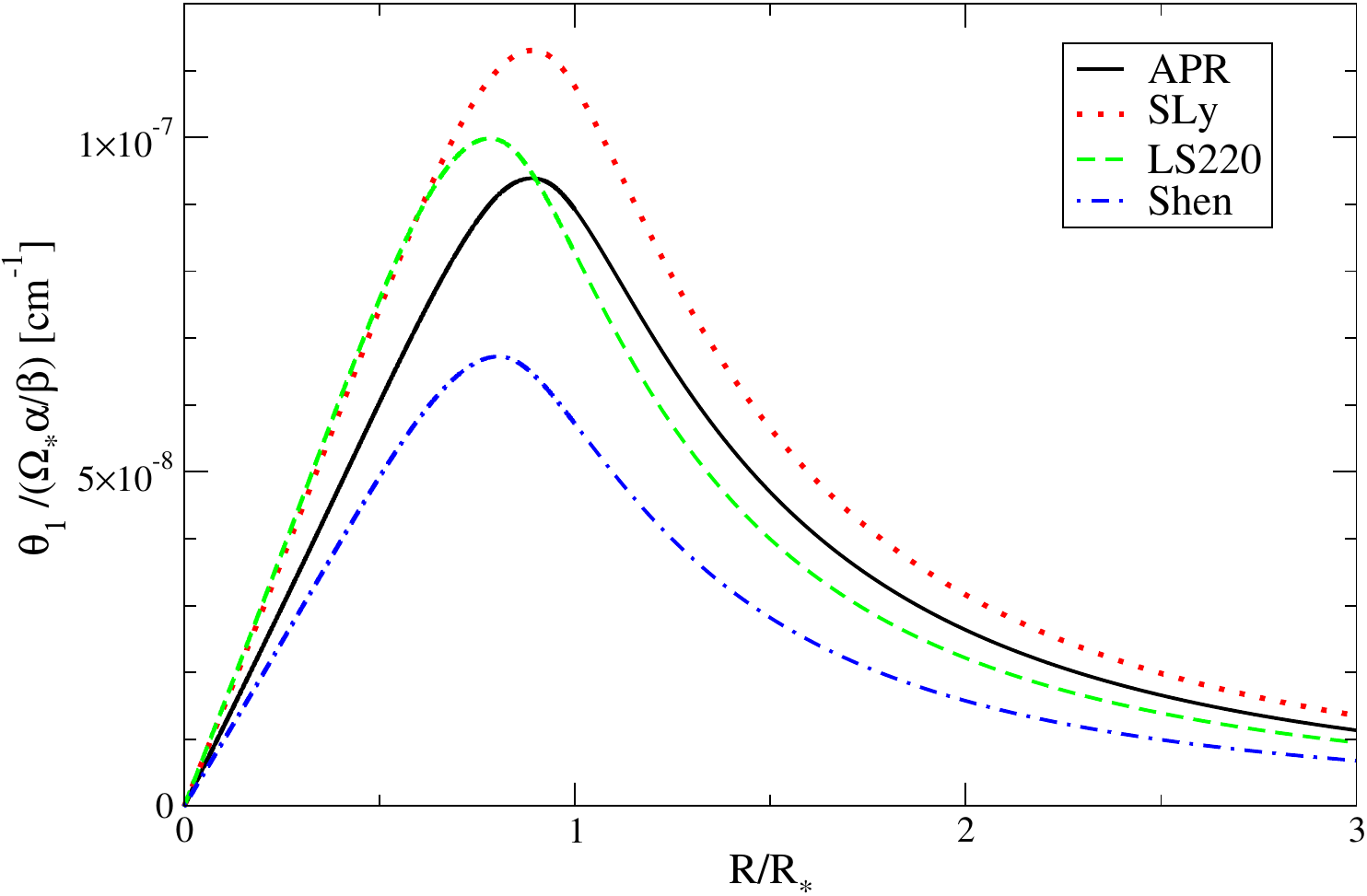}  
\includegraphics[width=8.0cm,clip=true]{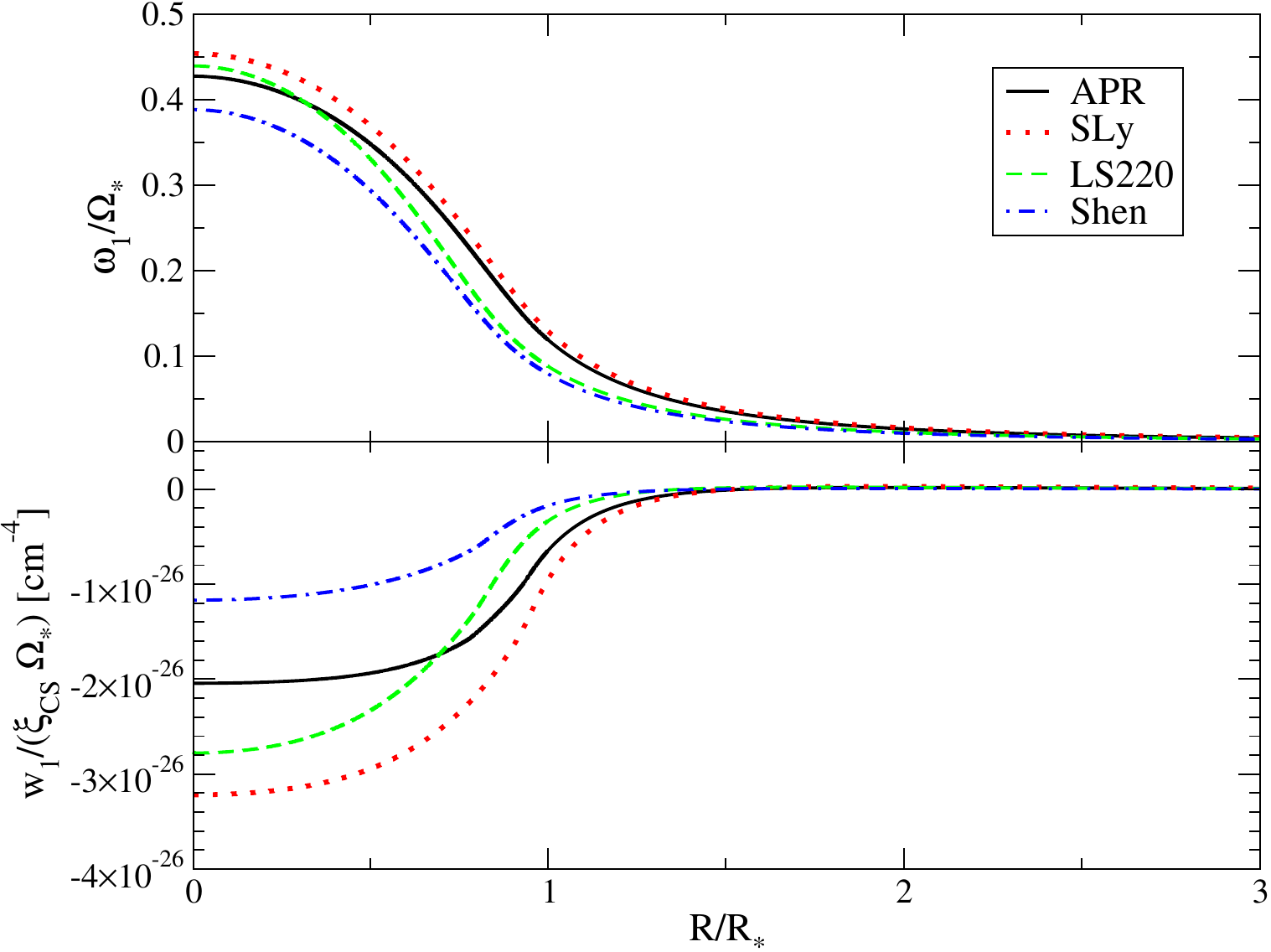} 
\caption{\label{fig:theta}
(Color online)
Left: Scalar field profile as a function of NS radius with $M_\NS =
1.4M_\odot$ and for various EoSs. The linear behaviour of $\vartheta$ near
$R\to 0$ is smooth and differentiable, after multiplying by
$P_{1}(\cos\Theta)$. 
Right: Correction to the metric rotation $\omega$ as a function of radius, at
linear order in spin, for various EoSs with $M_\NS = 1.4M_\odot$. The top
plot shows the $\mathcal{O}(\alpha'^{0})$ (GR) contribution $\omega_1/\Omega_\NS$ 
[Eq.~\eqref{eq:omega-1-def}], while the bottom plot shows the $\mathcal{O}(\alpha'^{2})$ (CS) 
contribution $w_1/(\xi_\CS\Omega_\NS)$ [Eq.~\eqref{eq:w-def}].
In all cases, only the $\ell=1$ multipole moment
survives at linear order in spin in the small coupling
limit. See Sec.~\ref{sec:asymptotia-center} for details on asymptotia.}
\end{figure*}

We begin by showing that the results obtained here reproduce previously obtained results, but for a wider range of EoSs. The left panel of Fig.~\ref{fig:MR} shows results at zeroth order in spin, i.e.~the NS mass-radius relation. Observe that the SLy curve agrees with~\cite{shibata-fitting}. Notice also that the 4 EoSs we adopt in this paper are consistent with the existence of the recently found $(1.97\pm0.04 ) M_\odot$ millisecond pulsar J1614-2230~\cite{1.97NS}. The right panel of Fig.~\ref{fig:MR} presents results at linear order in spin, i.e.~the moment of inertia as a function of mass. The APR curve agrees exactly with the results of~\cite{pani-NS-EDGB}). The left panel of Fig.~\ref{fig:theta} shows $\vartheta_1 /(\Omega_{\NS} \alpha/\beta)$ as a function of $R/\mathcal{R}_\NS$ for a NS of mass $M_\NS=1.4M_\odot$. Observe that $\vartheta_1$ has its maximum near the NS surface. The behavior of these curves is consistent with Fig.~5 in~\cite{alihaimoud-chen}.  The top right panel of Fig.~\ref{fig:theta} shows $\omega_1 /\Omega_{\NS}$ versus $R/\mathcal{R}_\NS$, with
\begin{equation}
\label{eq:omega-1-def}
\omega_1 \equiv \Omega_\NS - \tilde{\omega}_1\,.
\end{equation}
One can see that frame-dragging is strongest near the NS center.  These curves are consistent with Fig.~6 of~\cite{alihaimoud-chen}. In the bottom right panel of the same figure, we show $w_1/(\xi_\CS \Omega_\NS)$, which is also peaked in the inner region. This behavior is also consistent with Fig.~6 of~\cite{alihaimoud-chen}.

%% Begin zeroth order figs
%% End zeroth order figs

%% Begin first order figs
%% End first order figs

{\renewcommand{\arraystretch}{1.2}
\begin{table*}[ptb]
%\capstart
\begin{centering}
\begin{tabular}{cr@{\hspace{1em}}r@{.}l r@{.}l r@{.}l r@{.}l r@{.}l r@{.}l}
\hline
\hline
\noalign{\smallskip}
& EoS &  \multicolumn{2}{c}{$a$} &  \multicolumn{2}{c}{$b$}
&  \multicolumn{2}{c}{$c$} &  \multicolumn{2}{c}{$d$} &  \multicolumn{2}{c}{$e$} &  \multicolumn{2}{c}{Error (\%)} \\
\hline
\multirow{4}{*}{$\bar{\mu}$} &APR & 2&90 (0.022)  & -2&30 (0.067) & 1&54 (0.074) & -0&546 (0.036) & 0&0748 (0.0062) & 0&11\\
& SLy & 2&82 (0.027)  & -2&31 (0.087) & 1&62 (0.10) & -0&581 (0.052) & 0&0777 (0.0095) & 0&12\\
& LS220 & 5&87 (0.071) & -8&07 (0.25) & 6&60 (0.31) & -2&63 (0.17) & 0&400 (0.033) & 0&49\\
& Shen & 5&35 (0.049) & -6&86 (0.16) & 5&39 (0.20) & -2&07 (0.099) & 0&307 (0.018) & 0&36\\
\hline
\multirow{4}{*}{$[Q/(J^2/M_\NS)] (0.1/\zeta)$}& APR & -0&802 (0.034)  & -0&675 (0.10) & 0&578 (0.11) & -0&192 (0.053) & 0&0150 (0.0091) & 0&16\\
& SLy & -1&79 (0.055)  & 1&51 (0.18) & -1&37 (0.20) & 0&675 (0.10) & -0&138 (0.018) & 0&23\\
& LS220 & 6&72 (0.21) & -16&8 (0.72) & 15&6 (0.91) & -6&66 (0.49) & 1&06 (0.097) & 1&2\\
& Shen & 5&01 (0.13) & -13&0 (0.42) & 11&8 (0.49) & -4&94 (0.25) & 0&778 (0.045) & 0&67\\
\hline
\multirow{4}{*}{$-(\delta M/M_\odot) (\Omega_\mrm{1ms}/\Omega_\NS)^2 (0.1/\zeta)$ }& APR & -2&42 (0.046) & -0&503 (0.14) & 0&752 (0.15) & -0&266 (0.069) & 0&0234 (0.012) & 0&18\\
& SLy & -3&48 (0.15)  & 2&18 (0.48) & -2&29 (0.55) & 1&33 (0.27) & -0&297 (0.0487) & 0&84\\
& LS220 & 4&53 (0.24) & -14&0 (0.85) & 12&3 (1.1) & -4&81 (0.58) & 0&673 (0.11) & 2&0\\
& Shen & 3&00 (0.16) & -10&1 (0.51) & 8&60 (0.61) & -3&24 (0.30) & 0&445 (0.055) & 1&5\\
\hline
\multirow{4}{*}{$(\delta J/J) (0.1/\zeta)$ }& APR & -1&43 (0.030)  & -1&73 (0.088) & 1&92 (0.095) & -0&815 (0.044) & 0&141 (0.0074) & 0&13\\
& SLy & -0&242 (0.26)  & -6&30 (0.80) & 7&80 (0.91) & -3&97 (0.44) & 0&766 (0.078) & 1&0\\
& LS220 & 5&22 (0.11) & -16&1 (0.37) & 15&9 (0.46) & -7&27 (0.24) & 1&30 (0.047) & 0&56\\
& Shen & 3&94 (0.097) & -12&9 (0.32) & 12&1 (0.37) & -5&18 (0.18) & 0&869 (0.032) & 0&98\\
\noalign{\smallskip}
\hline
\hline
\end{tabular}
\end{centering}
\caption{
\label{table:C1}
Numerical coefficients for the fitting formula of $\bar{\mu}$, $Q$,
$\delta M$, and $\delta J$ as functions of mass using the functional
form in Eq.~\eqref{C1-C2-fit}. Standard errors on each coefficient are
in parentheses. For definitions of dimensionless scalar dipole susceptibility $\bar{\mu}$,
quadrupole correction $Q$, mass shift $\delta M$, and angular momentum
shift $\delta J$, see Eqs.~\eqref{eq:mu-mubar-relation},
\eqref{eq:QA}, and \eqref{eq:asymptotic-deltaM-deltaJ}. The last column shows the maximum fractional error between values that are obtained numerically and from the fitting formula.}
\end{table*}
}

%% Begin second order figs
\begin{figure*}[tp]
\begin{center}
\includegraphics[width=8.5cm,clip=true]{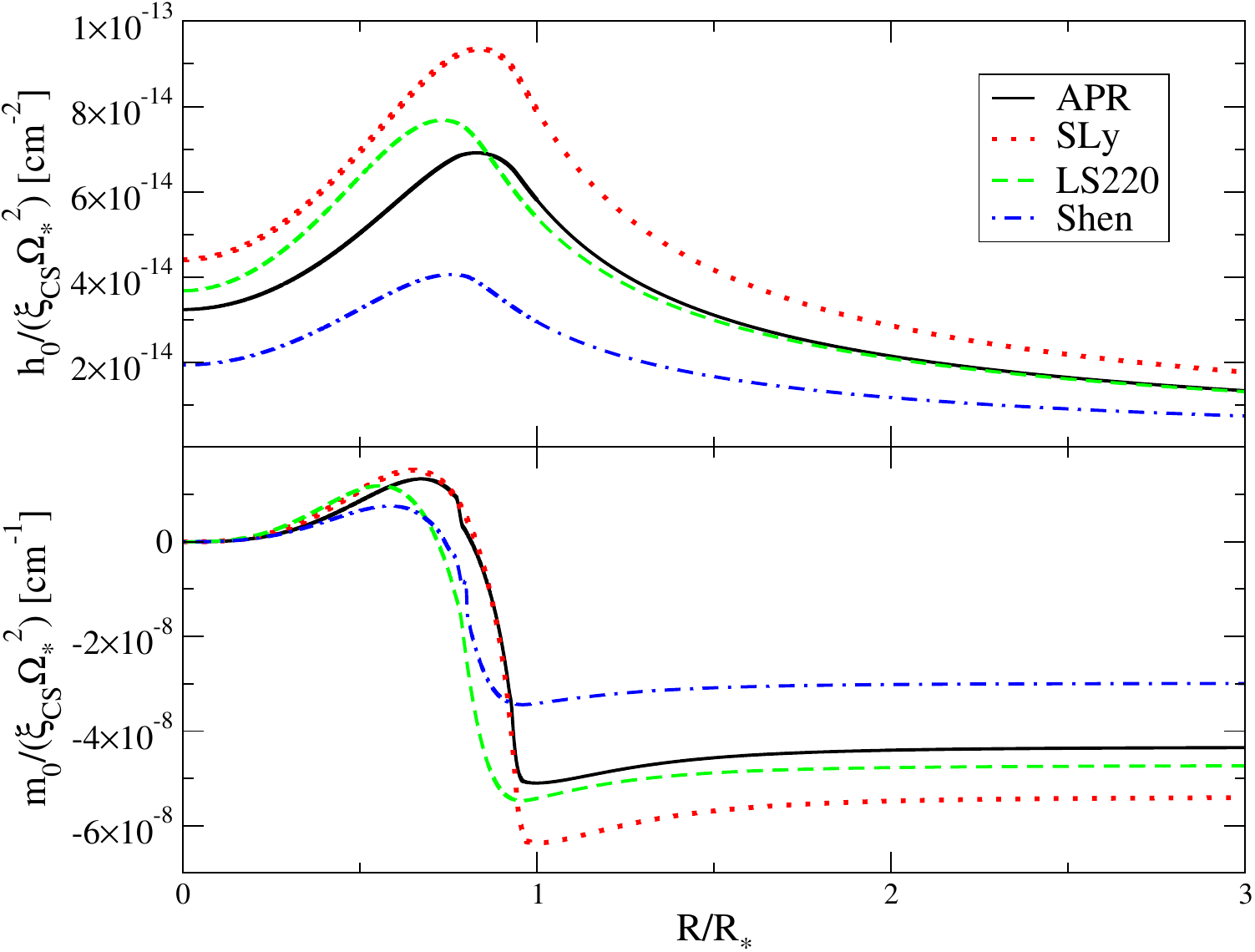} \qquad
\includegraphics[width=8.5cm,clip=true]{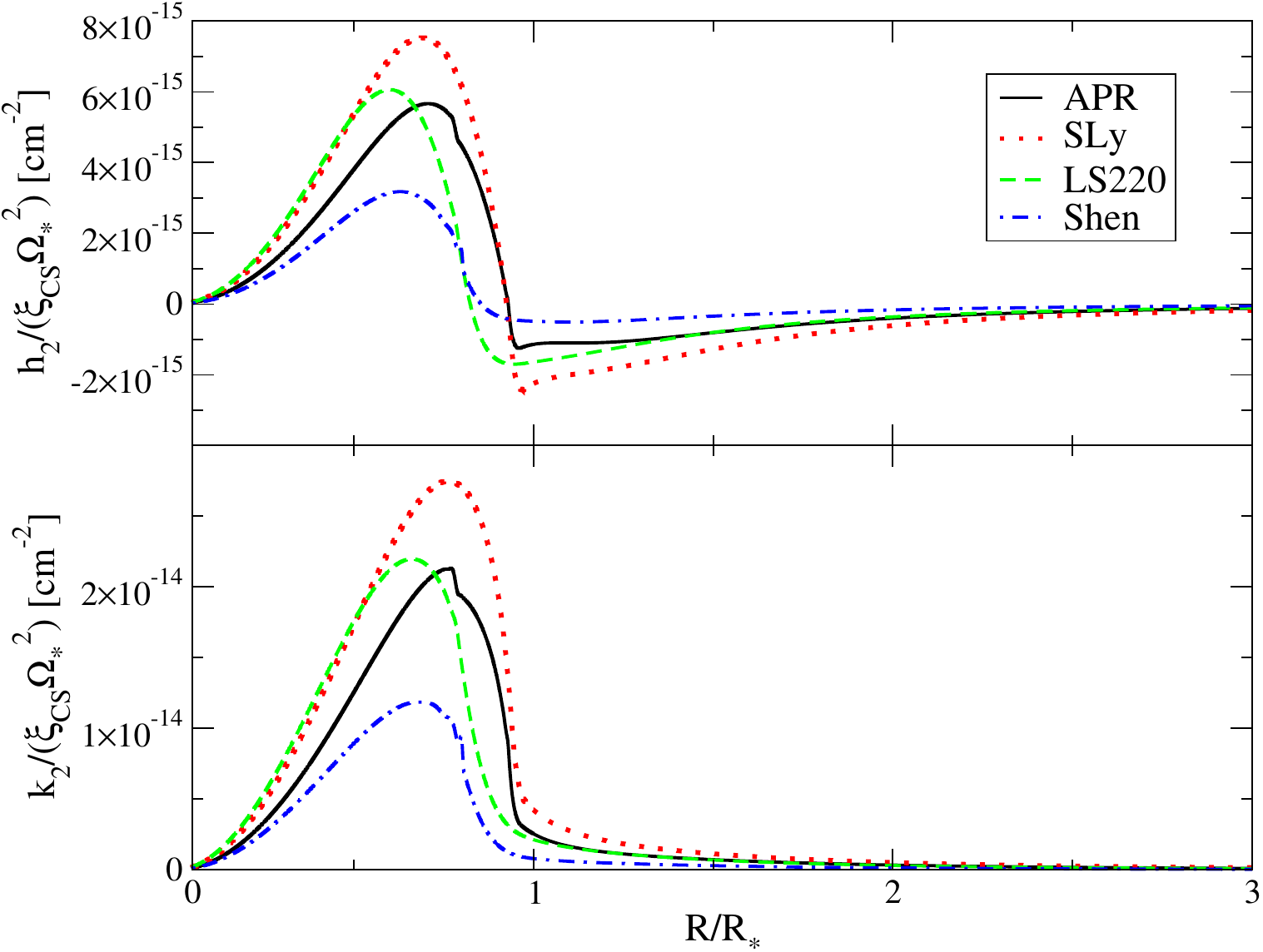}  
\caption{\label{fig:CS-h0} 
(Color online)
$\ell=0$ (left) and $\ell=2$ (right) metric perturbations at second order in spin as functions of
radius, for various EoSs with $M_\NS = 1.4M_\odot$:
$h_0 /(\xi_\CS \Omega_\NS^2)$ (top left) and $m_0 /(\xi_\CS\Omega_\NS^2)$
(bottom left) as defined in Eqs.~\eqref{eq:h-decomp} and
\eqref{eq:m-decomp}; $h_2 /(\xi_\CS \Omega_\NS^2)$ (top right) and 
$k_2 /(\xi_\CS \Omega_\NS^2)$ (bottom right) 
as defined in Eqs.~\eqref{eq:h-decomp} and \eqref{eq:k-decomp}.}
\end{center}
\end{figure*}

\begin{figure*}[htp]
\begin{center}
\begin{tabular}{l}
\includegraphics[width=8.5cm,clip=true]{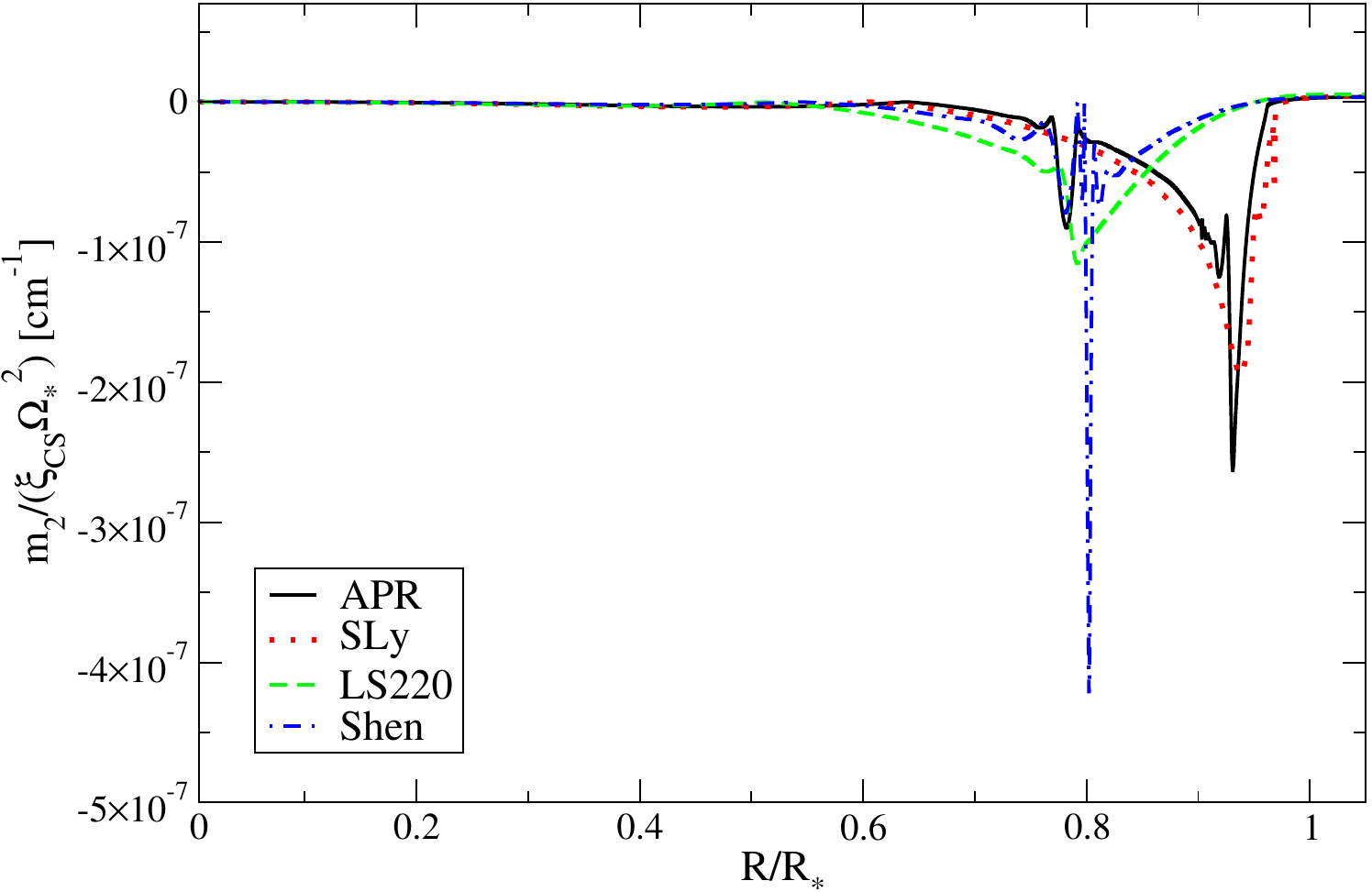}  
\includegraphics[width=8.5cm,clip=true]{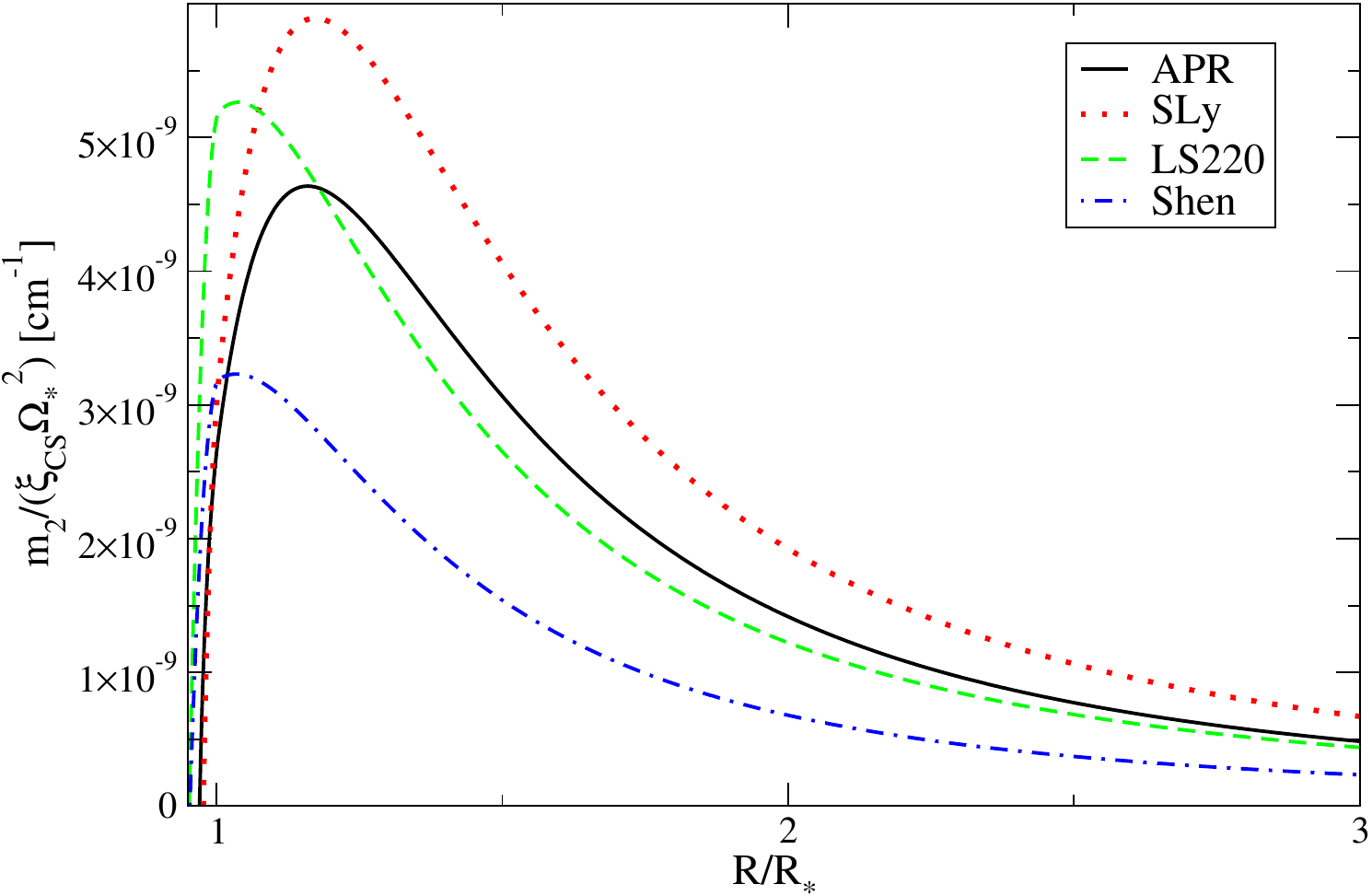}  
\end{tabular}
\caption{\label{fig:CS-m2}
$\ell=2$ metric perturbation at second order in spin as a function of
radius, for various EoSs with $M_\NS = 1.4M_\odot$:
$m_2 /(\xi_\CS \Omega_\NS^2)$ [Eq.~\eqref{eq:m-decomp}] for the
interior (left) and the exterior (right). Notice the difference in
scales. The features in the interior correspond to the locations of
nuclear phase transitions (see Fig.~\ref{fig:rho-r}).}
\end{center}
\end{figure*}

\begin{figure}[htp]
\begin{center}
\includegraphics[width=8.5cm,clip=true]{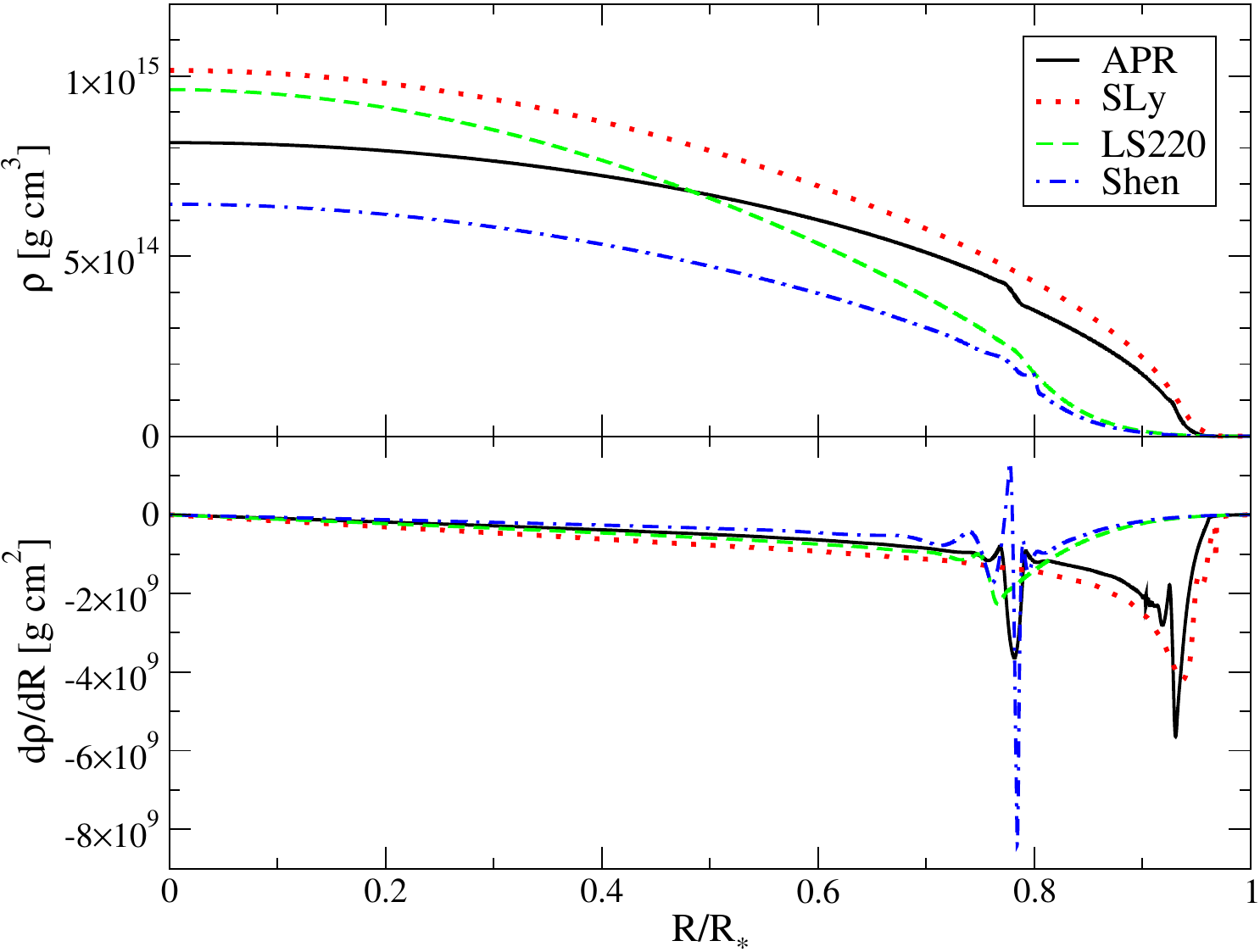}  
\caption{\label{fig:rho-r}
Mass density $\rho$ (top) and radial derivative $d\rho/dR$ (bottom) in
the NS interior, for various EoSs with $M_\NS = 1.4M_\odot$. Recall
that $R$ is defined by surfaces of constant $\rho$, so there is no
angular dependence to $\rho$. Nuclear phase transitions are visible as
sudden changes in slope. These give rise to the features in the metric
functions seen in Fig.~\ref{fig:CS-m2}.}
\end{center}
\end{figure}
%% End second order figs

Let us now present new results, valid to quadratic order in spin. Figures~\ref{fig:CS-h0} and~\ref{fig:CS-m2} show the metric perturbations at quadratic order in spin for the $\ell = 0$ and $\ell=2$ modes respectively. The interior behavior of $h_0$, $h_{2}$ and $k_{2}$ is similar to that of $\vartheta_1$. The strange behavior of $m_2$ in the interior reflects the strange behavior of $d\rho(R)/dR$, plotted in Fig.~\ref{fig:rho-r}. The oscillatory structure in these quantities disappears for a polytropic EoS. The oscillations in the EoS, which then propagate to $m_2$, are due to nuclear phase transitions~\cite{APR,SLy,LS,Shen1,Shen2}. We recall that we used an adaptive step-size, fourth-order Runge-Kutta method with an accuracy of $10^{-3}$ per step to solve the differential equations; therefore, the strange behavior alluded to above is not a numerical artifact and it is well-resolved.

Figure~\ref{fig:mu-M} already presented $\bar{\mu}$, $[Q/(J^2/M_\NS)] (0.1/\zeta)$, $\delta M (\Omega_\mrm{1ms}/\Omega_\NS)^2 (0.1/\zeta)$ and $(\delta J/J) (0.1/\zeta)$ as a function of $M_\NS/M_\odot$ for various EoSs. This behavior can be nicely fit by
\ba
\label{C1-C2-fit}
A_i &=& \exp\left( a_i + b_i x  +c_i x^2  + d_i x^3 + e_i x^4 \right)\,, \\
A_i &\equiv & \left( \bar{\mu}, \ \frac{Q}{J^2/M_\NS} \frac{0.1}{\zeta}, \ - \frac{\delta M}{M_\odot} \frac{\Omega_\mrm{1ms}^2}{\Omega_\NS^2} \frac{0.1}{\zeta}, \ \frac{\delta J}{J} \frac{0.1}{\zeta} \right)\,, \nn \\
 \ea
with $x \equiv M_\NS/M_\odot$. The estimated numerical coefficients are shown in Table~\ref{table:C1}, together with standard errors. Figure~\ref{fig:mu-M} also shows these fits as a function of $M_\NS/M_\odot$. Notice that the CS correction increases the spin angular momentum which is consistent with the result reported in~\cite{alihaimoud-chen}. Moreover, the CS correction increases the quadrupole moment while it decreases the observed mass.

%%%%%%%%%%%%%%%%%%%%%%%%%%%%%%%%%%%
\section{Neutron Star Binary Evolution}
\label{sec:binary-evolution}

Let us now study the evolution of a NS binary with component masses $m_A$, radii $\mathcal{R}_{\NS A}$, spin angular momenta $\chi_A m_A^2 \hat{S}_A^i$, dimensionless scalar charges $\bar{\mu}_A$ and quadrupole moment shifts $Q_A$. We use the subscript $A =(1,2)$ to denote the $A^\mathrm{th}$ binary component. We also use the subscript ``12'' to denote relative differences. The binary orbit can be described by 5 intrinsic and extrinsic parameters: semimajor axis $a$, eccentricity $e$, the angle of periastron $\omega$, inclination $\iota$, and the angle of the ascending node $\Omega$ (not to be confused with the gravitomagnetic metric perturbation $\omega$ or the NS angular velocity $\Omega_\NS$).

A binary's orbital evolution is affected by dissipative effects and conservative effects. The former can be obtained by calculating the energy flux and the angular momentum flux radiated via gravitational and scalar radiation, which we explain in detail in Appendix~\ref{sec:E-flux}. In this section, we focus on the latter. Here, we are only interested in CS corrections on the conservative effect relative to the leading GR contribution, and hence, we consider CS corrections to Newtonian dynamics. We follow Gauss' perturbation
method~\cite{smart,robertson-noonan,TEGP} with the conventions and notation
of~\cite{TEGP}. One must then calculate the perturbing accelerations
$\delta a_{A}^{i}$ and from them the relative perturbing acceleration
$\delta a_{12}^{i}$ (in any convenient coordinate system). This vector is then
decomposed by projecting onto a \emph{time-varying}
orthonormal triad with $e_{1}^{i}=n_{12}^{i}$ and
$e_{2}^{i}=\hat{L}^{i}$ (and $e_{3}=e_{2}\times e_{1}$ so as to complete the
orthonormal triad). In this triad, the components of
$\delta a_{12}^{i}$ are defined as~\cite{TEGP}
\begin{subequations}
\label{eq:accel-decomp}
\begin{align}
\accR &\equiv \delta a_{12}^{i} e_{1,i}\,, \\
\accW &\equiv \delta  a_{12}^{i} e_{2,i}\,, \\
\accS &\equiv \delta a_{12}^{i} e_{3,i}\,,
\end{align}
\end{subequations}
where inner products are taken with a flat Euclidean metric.
With this decomposition, the osculating orbital elements evolve
secularly as
\begin{subequations}
\label{eq:oscul-evol}
\begin{align}
\dot\omega &= - \frac{p \accR}{he}\cos\phi +
\frac{(p+r) \accS }{h e}\sin\phi -\dot\Omega \cos\iota\,, \\
\dot{e} &= \frac{1-e^{2}}{h}\left[ a \accR\sin\phi +
  \frac{\accS}{e} \left( \frac{ap}{r}-r \right) \right]\,, \\
\dot{a} &= \frac{2a^{2}}{h} \left( \frac{\accS p}{r}+\accR e\sin\phi \right)\,, \\
{\textstyle\frac{d}{dt}\iota} &= \frac{\accW r}{h}\cos(\omega+\phi)\,, \\
\dot\Omega &= \frac{\accW r}{h}\sin(\omega+\phi) \csc\iota\,.
\end{align}
\end{subequations}
Here, 
\be
p \equiv a(1-e^2)
\ee
is the semi-latus rectum, $r$ and $\phi$ are the quantities related to the instantaneous orbital elements, given by
\ba
r &\equiv & \frac{p}{1+e \cos\phi}\,, \\
r^2 \frac{d \phi}{dt} &\equiv & h \equiv \sqrt{m p}\,,
\ea
with $m \equiv m_1 + m_2$ and $h$ is the orbital angular momentum per unit mass.
In all of Eqs.~\eqref{eq:oscul-evol}, the right-hand sides are to be
orbit averaged as defined in Eq.~\eqref{eq:orbit-ave-def} (this is
appropriate when the time derivatives of the osculating elements are
much smaller than the orbital timescale and there are no
resonances). Of course, all of the $\phi$ dependence in $\accR,
\accW, \accS$ and $r$ must be included in the orbit
averaging.

In Sec.~\ref{sec:bin-evol-metric}, we calculate the secular
evolution from the near-zone metric deformation due to the CS
correction to the quadrupole. In order to calculate the secular
evolution due to the scalar field, we derive the effective
dipole interaction in Sec.~\ref{sec:dipole-interaction}. With the
acceleration due to the scalar interaction, we calculate the
scalar's correction to the binary's secular evolution.

%--------------------------------------
\subsection{Metric Quadrupole Correction}
\label{sec:bin-evol-metric}

Since both the mass monopole and current dipole of each body
is reabsorbed into physically measured quantities, the leading
order CS correction is the quadrupole deformation to the metric given by Eq.~\eqref{eq:QA}. From Eq.~\eqref{eq:UA_CS}, we can
find the acceleration on body 1 due to body 2, 
\begin{equation}
\label{eq:aU_1i}
\delta a^{U}_{1i} = -\pd_{i}\delta U_{2}^{\CS} = -15 Q_{2}
\frac{n_{12}^{<ijk>}}{r_{12}^{4}} \hat{S}_{2}^{j} \hat{S}_{2}^{k}\,,
\end{equation}
where $r_{12}$ is the separation of the binary components and $n_{12}^i$ is the unit vector from body 2 to body 1.
The acceleration on body 2 due to body 1 is simply the above
expression with $1\leftrightarrow 2$. This gives the relative
acceleration
\begin{equation}
  \delta a^{U}_{12i} = -15 \frac{n_{12}^{<ijk>}}{r_{12}^{4}}
\left(Q_{1}\hat{S}_{1}^{j}\hat{S}_{1}^{k} +
  Q_{2}\hat{S}_{2}^{j}\hat{S}_{2}^{k} \right)\,.
\end{equation}

The orbit averages are performed on eccentric orbits lying in the
$x$---$y$ plane with the major axis along the $\hat{x}$-axis, parametrized
as~\cite{PetersMathews}
\begin{align}
x_{1}^{i} &= d_{1}( \cos \phi, \sin \phi, 0)\,, \\
x_{2}^{i} &= -d_{2}( \cos \phi, \sin \phi, 0)\,,
\end{align}
\begin{align}
d_{1} &= \frac{m_{2}}{m} d\,, &
d_{2} &= \frac{m_{1}}{m} d\,, &
d &= \frac{a(1-e^{2})}{1+e\cos\phi}\,,
\end{align}
\begin{equation}
\dot\phi = \frac{\left[ma(1-e^{2})\right]^{1/2}}{d^{2}}\,,
\end{equation}
from which $v$ and all other derived quantities can be calculated,
e.g.~$n_{12}^{i}= (x_{1}^{i} - x_{2}^{i})/d = (\cos\phi,\sin\phi,0)$.

Taking the above expression, decomposing as in
Eqs.~\eqref{eq:accel-decomp}, and performing orbit averaging with the period
$P=2\pi a^{3/2}/M^{1/2}$, we find
\begin{widetext}
\begin{subequations}
\begin{align}
\ave{ \dot\omega}_h &=
\frac{3}{a^{7/2}\sqrt{m}(1-e^{2})^2}
Q_{1} \left[ -1 + \frac{3}{2}\left(
    \hat{S}_{1,x}^{2}+\hat{S}_{1,y}^{2} \right)
  -\hat{S}_{1,z}\cot\iota \left( \hat{S}_{1,x}\sin\omega +
    \hat{S}_{1,y}\cos\omega \right)
 \right] \plusonetotwo\,,
\\
\ave{ {\textstyle \frac{d}{dt}}\iota }_h &=
\frac{3}{a^{7/2}\sqrt{m}(1-e^{2})^2}
Q_{1} \hat{S}_{1,z} \left( \hat{S}_{1,x}\cos\omega -
    \hat{S}_{1,y}\sin\omega  \right) \plusonetotwo\,,
\\
\label{eq:ave-dotOmega}
\ave{ \dot\Omega }_h &=
\frac{3}{a^{7/2}\sqrt{m}(1-e^{2})^2}
Q_{1}
\hat{S}_{1,z}\csc\iota \left( \hat{S}_{1,x}\sin\omega +
    \hat{S}_{1,y}\cos\omega \right) \plusonetotwo\,,
\end{align}
\end{subequations}
and $\langle\dot{e}\rangle_h = 0 = \langle \dot{a}\rangle_h$, and where
the $i=x,y,z$ components of $\hat{S}_{A,i}$ are taken in the
(non-inertial) coordinate system where the binary's orbit is in the
$x$---$y$ plane, with pericenter along the $+\hat{x}$
direction.
\end{widetext}

%-----------------------------
\subsection{Dipole Interaction and Scalar Force Correction}
\label{sec:dipole-interaction}

To derive the effective dipole interaction, we start by finding an
effective interaction Lagrangian between one of the body's scalar dipole moments
and the scalar field. This comes from the cross-interaction part of the kinetic term of the scalar field in the action, i.e., by decomposing the kinetic term of the scalar field in the Lagrangian density as
\be
\mathcal{L}_\mrm{kin} = -\frac{\beta}{2} (\partial_\mu \vartheta)(\partial^\mu \vartheta) = \mathcal{L}_\mrm{kin,1} + \mathcal{L}_\mrm{kin,2} + \mathcal{L}_\mrm{int}\,,
\ee
where
\ba
\mathcal{L}_\mrm{kin,A} & \equiv & -\frac{\beta}{2} (\partial_\mu \vartheta_A)(\partial^\mu \vartheta_A)\,, \\
\label{eq:L_int}
\mathcal{L}_\mrm{kin,int} & \equiv & -\beta (\partial_\mu \vartheta_1)(\partial^\mu \vartheta_2)\,, 
\ea
and $\mathcal{L}_\mrm{kin,int}$ corresponds to the dipole interaction Lagrangian density. By substituting $\vartheta_A = \mu_A^i n_{A,i}/r_A^2 = - \mu_A^i \partial_i (1/r_A)$ into Eq.~\eqref{eq:L_int}, we obtain
\ba
\mathcal{L}_\mrm{kin,int} &=& -\beta \mu_1^j \mu_2^k \partial_{ij} \left( \frac{1}{r_1} \right) \partial_{ik} \left( \frac{1}{r_2} \right) \nn \\
&=& -\beta \mu_1^j \mu_2^k \partial^{(1)}_{ij} \left( \frac{1}{r_1} \right) \partial^{(2)}_{ik} \left( \frac{1}{r_2} \right)\,,
\ea
where $\partial^{(A)}_{i}$ denotes the partial derivative with respect to $x_A^i$.

The dipole interaction Lagrangian can be obtained by performing the volume integral of $\mathcal{L}_\mrm{kin,int}$ as
\begin{align}
L_\mrm{int} &= -\beta \mu_1^j \mu_2^k \int \partial^{(1)}_{ij} \left( \frac{1}{r_1} \right) \partial^{(2)}_{ik} \left( \frac{1}{r_2} \right) d^3x \nn \\
&= -\beta \mu_1^j \mu_2^k \partial^{(1)}_{ij} \partial^{(2)}_{ik} \int \frac{1}{r_1} \frac{1}{r_2} d^3x\,.
\end{align}
By applying Hadamard regularization, as explained e.g.~in Appendix B of~\cite{quadratic}, the above integration can be performed to yield
\begin{align}
L_\mrm{int} &= 2 \pi \beta \mu_1^j \mu_2^k \partial^{(1)}_{ij} \partial^{(2)}_{ik} r_{12} = 12 \pi \beta \mu_1^j \mu_2^k \frac{n_{12\left\langle jk \right\rangle}}{r_{12}^3} \nn \\
&= 4\pi\beta \frac{1}{r_{12}^{3}} \left[ 3(\mu_{1}\cdot n_{12})
  (\mu_{2}\cdot n_{12}) - (\mu_{1}\cdot\mu_{2})  \right] \,.
\end{align}
As there are no derivatives on particle locations, this gives an
effective pairwise interaction potential 
$$U_{\inter} = -L_{\inter}.$$
This expression agrees with Eq.~(5) of~\cite{kent-CSGW}. We can find the force with (minus) the particle derivative of the effective pairwise interaction potential as
\begin{equation}
F^{(\vartheta)}_{A,i} = -\pd^{(A)}_{i} U_{\inter} \,.
\end{equation}

From the above we can compute the relative dipole-dipole force
(with $\mu$ the reduced mass)
\begin{align}
a^{(\vartheta)}_{12,i} &= \frac{75}{128} \frac{1}{\mu}\frac{\xi_\CS}{m^4} \chi_{1} \chi_{2}
C_1^3 C_2^3 \bar{\mu}_1 \bar{\mu}_2 \left( \frac{m}{r_{12}} \right)^{4} 
 \left\{\hat{S}_{1}^{i}(n_{12}\cdot\hat{S}_{2})
 \right. 
 \nn \\
 &+ \left. \frac{1}{2} 
n_{12}^{i} \left[   (\hat{S}_{1}\cdot \hat{S}_{2}) - 5 (n_{12}\cdot
  \hat{S}_{1})(n_{12}\cdot \hat{S}_{2}) \right]
\right\} - (1 \leftrightarrow 2)\,. 
\end{align}
Decomposing $a_{12}^{(\vartheta)}$, inserting this into the Gauss
equations, and averaging, we find
\begin{widetext}
\begin{subequations}
\begin{align}
\ave{\dot{\omega}}_\vartheta &= \frac{75}{256} \frac{1}{\mu}\frac{\xi_\CS}{m^4} \frac{\chi_{1}\chi_{2}}{(1-e^{2})^{2}} C_1^3 C_2^3 \bar{\mu}_1 \bar{\mu}_2 \left( \frac{m}{a} \right)^{7/2}
\left\{
\frac{1}{2}\left(\hat{S}_{1,x}\hat{S}_{2,x}
 +\hat{S}_{1,y}\hat{S}_{2,y}\right)
-\hat{S}_{1,z}\hat{S}_{2,z} \right.\nonumber\\
&\qquad\left.{}-\cot\iota \left[
\hat{S}_{1,z}\left(
\hat{S}_{2,x}\sin\omega + \hat{S}_{2,y}\cos\omega
\right) 
\right]
\right\}\plusonetotwo\,, \\
\ave{{\textstyle \frac{d}{dt}}\iota}_\vartheta &= \frac{75}{256} \frac{1}{\mu}\frac{\xi_\CS}{m^4} \frac{\chi_{1}\chi_{2}}{(1-e^{2})^{2}} C_1^3 C_2^3 \bar{\mu}_1 \bar{\mu}_2 \left( \frac{m}{a} \right)^{7/2}
\left[ \hat{S}_{1,z} \left( \hat{S}_{2,x}\cos\omega -
    \hat{S}_{2,y}\sin\omega \right) \right] \plusonetotwo\,, \\
\ave{\dot{\Omega}}_\vartheta &= \frac{75}{256} \frac{1}{\mu}\frac{\xi_\CS}{m^4} \frac{\chi_{1}\chi_{2}}{(1-e^{2})^{2}} C_1^3 C_2^3 \bar{\mu}_1 \bar{\mu}_2 \left( \frac{m}{a} \right)^{7/2}
\csc\iota
\left[ \hat{S}_{1,z} \left( \hat{S}_{2,x}\sin\omega +
    \hat{S}_{2,y}\cos\omega \right) \right] \plusonetotwo \,,
\end{align}
\end{subequations}
and $\langle \dot{e} \rangle_\vartheta = 0 = \langle \dot{a} \rangle_\vartheta$.
\end{widetext}

{\renewcommand{\arraystretch}{1.2}
\begin{table*}
%\capstart
\begin{centering}
\begin{tabular}{rcc@{\hspace{1.5em}}c}
\hline
\hline
\noalign{\smallskip}
Timing parameter &&   Pulsar A & Pulsar B  \\ 
\hline
Spin frequency &$f_\nu$ (Hz) & 44.054069392744(2) & 0.36056035506(1) \\
Orbital period &$P_{b}$ (day) & 0.10225156248(5) & --- \\
Eccentricity &$e$ & 0.0877775(9) & --- \\
Projected semimajor axis &$(a/c) \sin \iota$ (s) & 1.415032(1) &1.5161(16) \\
Longitude of periastron &$\omega$ ($^\circ$) & 87.0331(8) & $87.0331+180.0$ \\
\hline
Advance rate of periastron & $\langle \dot{\omega} \rangle$ ($^\circ$ /yr) & 16.89947(68) & [16.96(5)]\footnote{An independent parameter fit of $\langle \dot{\omega} \rangle$ for pulsar B is consistent with the more precise result of pulsar A~\cite{kramer-double-pulsar}.} \\
Orbital decay rate &$\dot{P}_{b}$ & $-1.252(17) \times 10^{-12}$ & --- \\
Gravitational redshift &$\gamma$ (ms) & 0.3856(26) & --- \\
Shapiro delay range &$r$ ($\mu$s) & 6.21(33) & --- \\
Shapiro delay shape &$s$ & 0.99974(-39,+16) & --- \\
\hline
Inclination &$\iota$ ($^\circ$) &\multicolumn{2}{c}{88.69(-76,+50)} \\ 
Mass function &($M_\odot$) & 0.29096571(87) & 0.3579(11) \\
Mass ratio && \multicolumn{2}{c}{1.0714(11)} \\
Mass ($M_\odot$) && 1.3381(7) & 1.2489(7) \\
\noalign{\smallskip}
\hline
\hline
\end{tabular}
\end{centering}
\caption{Timing parameters of the double binary pulsar PSR J0737-3039~\cite{kramer-double-pulsar}. Measurement uncertainties on last digits are shown in parentheses.}
\label{table:double-pulsar}
\end{table*}
}

%{\renewcommand{\arraystretch}{1.2}
\begin{table}
%\capstart
\begin{centering}
\begin{tabular}{cc}
\hline
\hline
\noalign{\smallskip}
EoS & $\langle \dot{\omega} \rangle_h/\langle \dot{\omega} \rangle_\GR$  \\ 
\noalign{\smallskip}
\hline
\noalign{\smallskip}
APR & -$3.7 \times10^{-9} $ \\
SLy & -$5.1 \times10^{-8} $ \\
LS & -$3.2 \times10^{-9} $ \\
Shen & -$3.0 \times10^{-8} $ \\
\noalign{\smallskip}
\hline
\hline
\end{tabular}
\end{centering}
\caption{\label{table:ratio}
$\langle \dot{\omega} \rangle_h/\langle \dot{\omega} \rangle_\GR$ for PSR J0737-3039 with various NS EoSs, where we set $\zeta = 1$ (this ratio is linearly proportional to $\zeta$).}
\end{table}
%}

\section{Applications to NS Observations}
\label{sec:applications}

In this section, we apply the results derived in the previous sections to observed NS systems. We consider the recently found massive millisecond pulsar J1614-2230~\cite{1.97NS} and the double binary pulsar system PSR J0737-3039~\cite{burgay,lyne,kramer-double-pulsar}. In this section, we consider the maximum CS corrections allowed within the weak coupling approximation, i.e.~$\zeta =1$. Clearly, such a value for $\zeta$ violates the small coupling approximation. But as we shall see, even with such a large $\zeta$ value, which leads to a strong GR deviation, NS observations will still not be accurate enough to allow for a bound on the theory.

%-----------------------------------------------
\subsection{Massive Millisecond Pulsar J1614-2230}

From measurements of the Shapiro time delay, the mass of the pulsar J1614-2230 has been determined to be $(1.97 \pm 0.04)M_\odot$~\cite{1.97NS}. As mentioned previously, the effect of a CS modification is to \emph{reduce} the magnitude of the observed mass relative to the GR expectation. In Fig.~\ref{fig:MR-CS}, we plotted the mass-radius relation in both GR and dynamical CS gravity. We have set the spin period of the NS to be the one observed for J1614-2230, $P_\mrm{spin} = 3.1508076534271(6)$ms. When making these plots, we did not include $\mathcal{O}(\alpha'^0 \chi'^2)$ contribution, but we have checked that such contributions only affect the results to 0.2$\%$ at most. As expected, the maximum mass decreases as we increase the CS dimensionless coupling constant $\zeta$. Therefore, we can try to place bounds on the theory by requiring that the maximum observed mass be greater than 1.93$M_\odot$. Of course, such a method to test modified gravity theories is not new; it has been considered in e.g.~\cite{eling-AE-NS} for Einstein-Aether theory and~\cite{pani-EDGB-NS} for Einstein-Dilaton-Gauss-Bonnet theory.

The horizontal dashed line in Fig.~\ref{fig:MR-CS} is the lower bound on the mass of PSR J1614-2230. For the LS220 EoS, the CS maximum NS mass with $\zeta=1$ becomes less than 1.93$M_\odot$. This means that we can place a meaningful constraint on the theory \emph{assuming} that this is the correct EoS. However, the CS maximum NS mass with $\zeta=1$ with the APR, SLy and Shen EoSs are all above 1.93$M_\odot$. Therefore, whether a meaningful constraint of dynamical CS gravity can be placed depends strongly on the EoS. Since the EoS has not been observationally measured, this degeneracy prevents us from constraining dynamical CS gravity with mass-radius relations.

%----------------------------------------------
\subsection{Double Pulsar Binary PSR J0737-3039}

The observed timing parameters of PSR J0737-3039~\cite{burgay,lyne,kramer-double-pulsar} are summarized in Table~\ref{table:double-pulsar}. One can test GR from measurements of the post-Keplerian (PK) parameters~\cite{damour-taylor}: the (orbital averaged) advance rate of the periastron $\langle \dot{\omega} \rangle$, the orbital decay rate $\dot{P}$, gravitational redshift parameter $\gamma$, and the range $r$ and the shape $s$ of the Shapiro time delay.

Let us first look at the CS effect on $\langle \dot{\omega} \rangle$. Since the spin of the secondary pulsar is 100 times smaller than the primary one, we only consider CS corrections that depend on $\chi_1$, meaning we only keep $\langle \dot{\omega} \rangle_h$ and neglect $\langle \dot{\omega} \rangle_\vartheta$. By taking the ratio of the former to the GR expression for $\langle \dot{\omega} \rangle_\GR = 3 (2 \pi/P)^{5/3} m^{2/3} (1-e^2)^{-1}$~\cite{stairs,TEGP}, we obtain
\begin{align}
\label{eq:dotomega-ratio}
\frac{\langle \dot{\omega} \rangle_h}{\langle \dot{\omega} \rangle_\GR} &= \frac{Q_1}{m^3} \frac{1}{1-e^2} \frac{m}{a} \left[ -1 + \frac{3}{2}\left(
    \hat{S}_{1,x}^{2}+\hat{S}_{1,y}^{2} \right)\right. \nn \\
& \quad\left.  -\hat{S}_{1,z}\cot\iota \left( \hat{S}_{1,x}\sin\omega +
    \hat{S}_{1,y}\cos\omega \right)
 \right]\,.
\end{align}
Since this is proportional to $m/a$, the CS correction to $\langle \dot{\omega} \rangle$ is of 1PN order relative to GR.

$\dot{P}$ is proportional to $\dot{E}$ which is given in Eqs.~\eqref{eq:dotEtheta} and~\eqref{eq:dotEh} and both are proportional to $(m/a)^7$. By comparing this with the GR expectation, which is proportional to $(m/a)^5$, the CS correction to $\dot{P}$ is of 2PN relative order. Similarly, one can show that the CS correction to $\gamma$, $r$, and $s$ are also of relative 2PN order. Therefore, the dominant CS correction to the binary evolution is in $\langle \dot{\omega} \rangle$ and we can neglect CS corrections to all other PK parameters. We can also neglect the CS corrections to the spin-precession equations, which appear at 2PN relative order (see Appendix~\ref{sec:spin-precession}). 

In order to place constraints on dynamical CS gravity, one needs to know the orientation of the spin of the primary pulsar since Eq.~\eqref{eq:dotomega-ratio} depends on $\hat{S}_1^i$. Unfortunately, this quantity is currently unconstrained. In 2004, Ref.~\cite{jenet-ransom} attempted to measure this quantity by modeling the intensity variation of pulsar B caused by the emission from pulsar~A~\cite{jenet-ransom}. Doing so, one obtained $\theta_{S_1}=167^\circ \pm 10^\circ$ and  $\phi_{S_1} = 246^\circ \pm 5^\circ$, where $\theta_{S_1}$ is the angle between $\hat{S}_1^i$ and the unit orbital angular momentum vector $\hat{L}^i$, and $\phi_{S_1}$ is the angle between $\hat{S}_1^i$ projected onto the orbital plane and the direction of the ascending node. However, since then, Ref.~\cite{manchester} found no observable major profile changes in the emission of pulsar A, and hence the model of Ref.~\cite{jenet-ransom} is ruled out.  Nevertheless, following Ref.~\cite{lattimer-schutz}, we adopt the above values for $\hat{S}_1^i$ to give an order of magnitude estimate of the possible magnitude of the constraint, if this quantity were measured in the future.

By using Kepler's Law $a^3 = m (P/2 \pi)^2$ and the CS quadrupole moment deformation found in Sec.~\ref{sec:solutions}, we can evaluate Eq.~\eqref{eq:dotomega-ratio} for NSs with various EoSs. The results for $\zeta =1$ are summarized in Table~\ref{table:ratio}. One sees that in the small coupling approximation, the ratio of $|\langle \dot{\omega} \rangle_h / \langle \dot{\omega} \rangle_\GR|$ is of $\mathcal{O}(10^{-8})$ at most. For other choice of the unit vector $\hat{S}_1^i$, $|\langle \dot{\omega} \rangle_h / \langle \dot{\omega} \rangle_\GR|$ increases by at most 0.075\%. This means that in order to place meaningful constraints on dynamical CS gravity from binary pulsar observations, we need to observe $\langle \dot{\omega} \rangle$ \emph{and} at least 2 other PK parameters (or the mass functions or the mass ratio) within $10^{-8}$ accuracy (the latter are needed to determine the masses of the pulsars). With the current observational data, we can determine the masses of the pulsars using the mass ratio and $s$, and test dynamical CS gravity with $\langle \dot{\omega} \rangle$. However, since the former only have accuracies of $\mathcal{O}(10^{-3})$, we cannot place any constraint on this theory from the double binary pulsar system. This conclusion also holds for other binary pulsar systems.

Notice that $|\langle \dot{\omega} \rangle_h/\langle \dot{\omega} \rangle_\GR|$ as shown in Table~\ref{table:ratio} is smaller than $|\langle \dot{\omega} \rangle_\mrm{SO}/\langle \dot{\omega} \rangle_\GR| \sim 10^{-5}$ where $\langle \dot{\omega} \rangle_\mrm{SO}$ is the advance rate of the periastron due to GR spin-orbit coupling and it is given by~\cite{damour-schaefer,lattimer-schutz}
\ba
\frac{\langle \dot{\omega} \rangle_\mrm{SO}}{\langle \dot{\omega} \rangle_\GR} &=& -\frac{1}{6 \sqrt{1-e^2}} \frac{4 m_1 + 3 m_2}{m_1} \frac{I_1 f_{\nu_1}}{m^2} \left( \frac{m}{a} \right)^{1/2} \nn \\
& & \times (2 \cos \theta_{S_1} + \cot \iota \sin \theta_{S_1} \sin \phi_{S_1})\,,
\ea
where $f_{\nu_1}$ is the spin frequency of pulsar A. Since this ratio depends on the moment of inertia $I$, which encodes information of the internal structure of the NS, it would be difficult to test dynamical CS gravity even if the measurement accuracies of the PK parameters improved, unless $I$ were determined to high accuracy. 

The results obtained here suggests that pulsar binaries cannot currently be used to constrain dynamical CS gravity. If so, GW observations are the only way to test dynamical CS gravity in the dynamical, strong-field regime, as demonstrated in~\cite{kent-CSGW}.

%%%%%%%%%%%%%%%%%%%%%%%%%%%%%%%%%%%
\section{Conclusions and Discussions}
\label{sec:conclusions}

In this paper, we extended the previous work in~\cite{yunespsaltis,alihaimoud-chen} to construct slowly-rotating NSs in dynamical CS gravity in the small-coupling and slowly-rotating approximations to quadratic order in spin. At this order, we found a negative CS mass correction and a positive CS quadrupole moment deformation. We applied the former correction to test the theory by requiring that the observed maximum mass be greater than the lower bound on the mass of the massive millisecond pulsar J1614-2230. Unfortunately, we could not obtain any meaningful constraint due to degeneracies with the unknown EoS of nuclear matter. Next, we used the quadrupole deformation to derive the correction to the evolution of the NS binary. Among all PK parameters, we found that the dominant CS correction appears in the advance rate of the periastron $\langle \dot{\omega} \rangle$. We applied our results to the double binary pulsar PSR J0737-3039~\cite{burgay,lyne,kramer-double-pulsar} but found that the CS correction is too small to be constrained, i.e.~it is of relative 1PN order. This correction is even smaller than the one arising from GR spin-orbit coupling, which in turn depends on the moment of inertia $I$. Hence, we would not be able to constrain the theory unless we knew $I$ to high precision. The results obtained here indicate that GW observations are the only way to test the theory in the strong and dynamical field regime~\cite{kent-CSGW}.   
 
One possible way to extend our work is to construct gravitational waveforms for NS binaries. This can be done by following the analysis of Ref.~\cite{kent-CSGW}. As in the BH binary case discussed in the reference, the corrections to GR waveforms will enter at 2PN order. However, one needs to be careful about constructing gravitational waveforms for spinning NS binaries in GR, since the spin of the NS induces a quadrupole deformation. This deformation contains information about the internal structure of the NS, which also enters at 2PN order~\cite{poisson-quadrupole}. In order to obtain 2PN waveforms, one must thus first construct slowly-rotating NS solutions in GR to quadratic-order in spin. This spin-induced 2PN effect will be strongly degenerate with the 2PN CS correction. However, this degeneracy might be broken as follows. The \emph{tidally} induced quadrupole moment enters at 5PN order~\cite{flanagan-hinderer}, but is enhanced by a factor of $\mathcal{O}(\RNS/M_\NS)^5$. The tidal effect may be detected using future ground-based GW interferometers~\cite{flanagan-hinderer,read-markakis-shibata,hinderer-lackey-lang-read,lackey-kyutoku,damour-nagar-villain}. Since the internal structure can be determined to some extent from tides, the degeneracy between the spin-induced term and the CS correction might be broken. There is also a GR spin-spin interaction term at 2PN order in the phase of the gravitational waveform. Again, the degeneracies between spins and the CS correction term might be broken if the binaries are precessing. However, we expect that BH binaries are more useful in constraining the theory than NS binaries because (i) the degeneracy with the spin-induced term would not be broken completely, (ii) the CS deformation to the quadrupole moment for a NS is smaller than that of a stellar-mass BH due to cancellation between the first and the second terms in Eq.~\eqref{eq:QA}, and (iii) the spin of the NS binaries just before coalescence is expected to be rather small~\cite{bildsten-cutler,hinderer-lackey-lang-read,damour-nagar-villain}, making the CS correction even smaller. Also, the radius of a NS is larger than that of a stellar-mass BH, and hence, the latter allows tests at smaller length scales, which means that BH binaries have a stronger potential to constrain the theory. 

Another possible avenue of future work is to study oscillations of NSs in dynamical CS gravity. By considering perturbations of the gravitational, scalar and the matter fields, one can investigate e.g. f-, p-, r- and w-modes~\cite{kokkotas-living}. The first two modes can be studied under the Cowling approximation, where one neglects the perturbations of the gravitational and the scalar fields (f- and p-modes in alternative theories of gravity have been studied in e.g.~\cite{sotani-fp-ST} for scalar-tensor theory and~\cite{sotani-fp-TeVeS} for tensor-vector-scalar (TeVeS) theory). This amounts to solving the perturbed matter equation $\nabla^{\mu} \delta T_{\mu\nu}^\mrm{mat} =0$, where $\delta T_{\mu\nu}^\mrm{mat}$ is the perturbed matter stress-energy tensor. At zeroth order in spin, since the background is exactly the same as GR, we expect that f- and p-modes to be identical to those in GR. To study w-modes, one needs to include gravitational and scalar field perturbations. W-modes in alternative theories of gravity have been investigated in e.g.~\cite{sotani-w-ST} for scalar-tensor theory and~\cite{sotani-w-TeVeS} for TeVeS. In these theories, one finds that axial metric perturbations decouple from scalar and polar metric ones. Hence, w-modes can be studied by solving the axial perturbation equation as an eigenvalue problem. In dynamical CS gravity, we expect that polar perturbations will decouple from the rest, as in the case for BHs~\cite{cardoso-gualtieri,garfinkle,molina}. However, again, since the background is the same as in GR, the polar perturbation equations should be identical to those in GR. At linear order in spin, one can look at r-modes by studying the toroidal oscillation of the NS in the Cowling approximation. As in GR~\cite{kojima,beyer} and in TeVeS~\cite{sotani-toroidal}, the spectrum for toroidal oscillations of slowly-rotating stars should also be continuous in dynamical CS gravity. However, this might be an artifact of the slow-rotation limit, since in GR, the spectrum becomes discrete for fast-rotating NSs~\cite{ruoff,gaertig}. In principle, if the maximum and/or the minimum of the toroidal oscillation frequencies and the moment of inertia are independently determined, one can break the degeneracy between the gravitational theory and the EoS, and hence test GR~\cite{sotani-toroidal}. This looks challenging from an observational point of view.  At quadratic order in spin, there can be corrections to f-, p-, and w-modes, but obviously, they are suppressed by $\mathcal{O}(\zeta \chi^2)$.

%%%%%%%%%%%%%%%%%%%%%%%%%%%%%%%%%%%%
\acknowledgments 
We would like to thank Will Farr for pointing out the Riccati method.
NY acknowledges support from NSF grant PHY-1114374, as well as support provided by the National Aeronautics and Space Administration from grant NNX11AI49G, under sub-award 00001944. TT is supported by the Grant-in-Aid for Scientific Research (Nos. 21244033, 21111006, 24103006 and 24103001). LCS acknowledges that support for this work was provided by the National Aeronautics and Space Administration through Einstein Postdoctoral Fellowship Award Number PF2-130101 issued by the Chandra X-ray Observatory Center, which is operated by the Smithsonian Astrophysical Observatory for and on behalf of the National Aeronautics Space Administration under contract NAS8-03060. Some calculations used the computer algebra-systems \textsc{Maple}, in combination with the \textsc{GRTensorII} package~\cite{grtensor}. Other calculations were carried out with the \textsc{xTensor} package for \textsc{Mathematica}~\cite{2008CoPhC.179..586M,2009GReGr..41.2415B}.

%%%%%%%%%%%%%%%%%%%%%%%%%%%%%%%%%%%%%%%%%%%%%%
\appendix

\begin{widetext}

\section{The $\ell=2$ Mode Exterior Solutions at Second Order in Spin}
\label{l2}

\allowdisplaybreaks[1]
We solve Eqs.~\eqref{m2}--~\eqref{RR2} and obtain the $\ell=2$ mode exterior solution at $\mathcal{O}(\alpha'^2 \chi'^2)$:
\ba
\label{h2-ext}
h_2^\ext &=& -\frac{3015}{14336} \frac{\xi_\CS J^2 R}{M_\NS^9 f(R)} \left\{  1 - 3 \frac{M_\NS}{R} + \frac{4}{3} \frac{M_\NS^2}{R^2} + \frac{2}{3} \frac{M_\NS^3}{R^3} + \frac{8}{15} \frac{M_\NS^4}{R^4} + \frac{8}{15} \frac{M_\NS^5}{R^5} + \frac{35792}{63315} \frac{M_\NS^6}{R^6} - \frac{3296}{1407} \frac{M_\NS^7}{R^7} \right. \nn \\*
& & \left. - \frac{28576}{12663} \frac{M_\NS^8}{R^8} - \frac{7520}{603} \frac{M_\NS^9}{R^9} + \frac{41792}{3015} \frac{M_\NS^{10}}{R^{10}} - \frac{14336}{335} \frac{M_\NS^{11}}{R^{11}}  + \frac{R f(R)^2}{2 M_\NS} \ln f(R) \right. \nn \\*
& & \left. -\frac{70}{603} C_\vartheta  \left[ 1-3 \frac{M_\NS}{R} + \frac{4}{3} \frac{M_\NS^2}{R^2} + \frac{2}{3} \frac{M_\NS^3}{R^3} - \frac{608}{15} \frac{M_\NS^5}{R^5} + \frac{1024}{15} \frac{M_\NS^6}{R^6} + \frac{512}{5} \frac{M_\NS^7}{R^7} \right. \right. \nn \\*
& & \left. \left. +\frac{R f(R)}{2 M_\NS} \left( 1 - 2 \frac{M_\NS}{R} - \frac{8}{3} \frac{M_\NS^4}{R^4} - \frac{848}{15} \frac{M_\NS^5}{R^5} - \frac{48}{5} \frac{M_\NS^6}{R^6} + \frac{256}{5} \frac{M_\NS^7}{R^7} \right) \ln f(R) \right] \right. \nn \\*
& & \left. +\frac{140}{603} C_\vartheta^2 f(R) \left[ 1 - \frac{M_\NS}{R} + \frac{R}{2M_\NS} \left( 1-2 \frac{M_\NS}{R} + \frac{4}{3} \frac{M_\NS^2}{R^2} \right) \ln f(R) + \frac{1}{6} \left( 1-\frac{M_\NS}{R} \right) [\ln f(R)]^2  \right] \right\} \nn \\*
&& +\frac{15}{4} \frac{J_\CS J R}{M_\NS^5 f(R)} \left[ 1 - 3 \frac{M_\NS}{R} + \frac{4}{3} \frac{M_\NS^2}{R^2} + \frac{2}{3} \frac{M_\NS^3}{R^3} + \frac{8}{15} \frac{M_\NS^4}{R^4} - \frac{8}{15} \frac{M_\NS^5}{R^5} - \frac{16}{15} \frac{M_\NS^6}{R^6} + \frac{R}{2 M_\NS} f(R)^2 \ln f(R) \right] \nn \\*
& & +  C_Q \frac{R}{M_\NS f(R)} \left[ 1 - 3 \frac{M_\NS}{R} + \frac{4}{3} \frac{M_\NS^2}{R^2} + \frac{2}{3}\frac{M_\NS^3}{R^3} + \frac{R}{2 M_\NS} f(R)^2 \ln f(R) \right]\,, \\
\label{m2-ext}
m_2^\ext &=& \frac{3015}{14336} \frac{\xi_\CS J^2 R^2}{M_\NS^9} \left\{ 1 - 3\frac{M_\NS}{R} + \frac{4}{3} \frac{M_\NS^2}{R^2} + \frac{2}{3} \frac{M_\NS^3}{R^3} + \frac{8}{15} \frac{M_\NS^4}{R^4} + \frac{2024}{9045}\frac{M_\NS^5}{R^5} - \frac{1136}{21105} \frac{M_\NS^6}{R^6} + \frac{183584}{12663} \frac{M_\NS^7}{R^7} \right.  \nn \\*
& & \left.  - \frac{249952}{4221} \frac{M_\NS^8}{R^8} - \frac{19552}{1005} \frac{M_\NS^9}{R^9} - \frac{607808}{3015} \frac{M_\NS^{10}}{R^{10}} + \frac{43008}{67} \frac{M_\NS^{11}}{R^{11}} + \frac{R}{2 M_\NS} f(R)^2 \ln f(R)  \right. \nn \\*
& & \left. -\frac{70}{603} C_\vartheta  \left[ 1 - 3 \frac{M_\NS}{R} + \frac{4}{3} \frac{M_\NS^2}{R^2} + 6 \frac{M_\NS^3}{R^3} - \frac{32}{3} \frac{M_\NS^5}{R^5} + \frac{4864}{15} \frac{M_\NS^6}{R^6} - \frac{3072}{5} \frac{M_\NS^7}{R^7} \right. \right. \nn \\*
&& \left. \left.  + \frac{R}{2 M_\NS} \left( 1 -2 \frac{M_\NS}{R} + \frac{16}{3} \frac{M_\NS^3}{R^3} + \frac{8}{3} \frac{M_\NS^4}{R^4} -48 \frac{M_\NS^5}{R^5} + \frac{1648}{5} \frac{M_\NS^6}{R^6} - \frac{1792}{5} \frac{M_\NS^7}{R^7} \right) f(R) \ln f(R) \right] \right. \nn \\*
& & \left. + \frac{140}{603}C_\vartheta^2 f(R) \left[ 1-\frac{7}{3} \frac{M_\NS}{R} + \frac{1}{2} \frac{R}{M_\NS} \left( 1- \frac{14}{3} \frac{M_\NS}{R} + 4 \frac{M_\NS^2}{R^2} \right)  \ln f(R) \right. \right. \nn \\*
& & \left. \left. -\frac{R}{3 M_\NS} \left( 1-\frac{M_\NS}{R} \right) \left( 1-\frac{3}{2} \frac{M_\NS}{R} \right) [\ln f(R)]^2  \right] \right\} \nn \\*
& & -\frac{15}{4} \frac{J_\CS J R^2}{M_\NS^5} \left[ 1 - 3 \frac{M_\NS}{R} + \frac{4}{3} \frac{M_\NS^2}{R^2} + \frac{2}{3} \frac{M_\NS^3}{R^3} + \frac{8}{15} \frac{M_\NS^4}{R^4} - \frac{56}{15} \frac{M_\NS^5}{R^5} + \frac{16}{3} \frac{M_\NS^6}{R^6} + \frac{R}{2 M_\NS} f(R)^2 \ln f(R) \right] \nn \\*
& & -  C_Q \frac{R^2}{M_\NS} \left[ 1-3\frac{M_\NS}{R} + \frac{4}{3} \frac{M_\NS^2}{R^2} + \frac{2}{3} \frac{M_\NS^3}{R^3} + \frac{R}{2M_\NS} f(R)^2 \ln f(R) \right]\,, \\
%\ea
%
%\ba
\label{k2-ext}
k_2^\ext &=& \frac{3015}{14336} \frac{\xi_\CS J^2 R}{M_\NS^9} \left\{ 1 + \frac{M_\NS}{R} - \frac{2}{3} \frac{M_\NS^2}{R^2} + \frac{8}{15} \frac{M_\NS^4}{R^4} + \frac{2272}{1809} \frac{M_\NS^5}{R^5} + \frac{151264}{63315} \frac{M_\NS^6}{R^6} + \frac{2368}{1809} \frac{M_\NS^7}{R^7} + \frac{35936}{4221} \frac{M_\NS^8}{R^8} \right. \nn \\*
& & \left. + \frac{30784}{3015} \frac{M_\NS^9}{R^9} + \frac{14336}{335} \frac{M_\NS^{10}}{R^{10}} + \frac{R}{2M_\NS} \left( 1-2\frac{M_\NS^2}{R^2} \right) \ln f(R) \right. \nn \\*
& & \left. - \frac{70}{603} C_\vartheta  \left[ 1 + \frac{M_\NS}{R} - \frac{2}{3} \frac{M_\NS^2}{R^2} - \frac{8}{3} \frac{M_\NS^3}{R^3} - \frac{16}{3} \frac{M_\NS^4}{R^4} - \frac{176}{3} \frac{M_\NS^5}{R^5} - \frac{512}{5} \frac{M_\NS^6}{R^6} \right. \right. \nn \\*
&& \left. \left. + \frac{R}{2 M_\NS} \left( 1 - 2 \frac{M_\NS^2}{R^2} - \frac{8}{3} \frac{M_\NS^3}{R^3} - \frac{16}{3} \frac{M_\NS^4}{R^4} - \frac{176}{3} \frac{M_\NS^5}{R^5} - \frac{352}{5} \frac{M_\NS^6}{R^6} + \frac{512}{5} \frac{M_\NS^7}{R^7} \right) \ln f(R) \right] \right. \nn \\* 
& & \left. + \frac{140}{603} C_\vartheta^2  \left[ 1-\frac{1}{3} \frac{M_\NS}{R} + \frac{R}{2M_\NS} \left( 1 - \frac{8}{3} \frac{M_\NS}{R} + 2 \frac{M_\NS^2}{R^2} \right) \ln f(R) - \frac{R}{3 M_\NS} \left( 1-\frac{1}{2} \frac{M_\NS}{R} \right) f(R) [\ln f(R)]^2  \right] \right\} \nn \\*
& & - \frac{15}{4} \frac{J_\CS J R}{M_\NS^5} \left[ 1 + \frac{M_\NS}{R} - \frac{2}{3} \frac{M_\NS^2}{R^2} + \frac{8}{15} \frac{M_\NS^4}{R^4} + \frac{16}{15} \frac{M_\NS^5}{R^5}+ \frac{R}{2 M_\NS} \left( 1-2\frac{M_\NS^2}{R^2} \right) \ln f(R)  \right] \nn \\*
& & - C_Q \frac{R}{M_\NS} \left[ 1+\frac{M_\NS}{R} - \frac{2}{3} \frac{M_\NS^2}{R^2} + \frac{R}{2 M_\NS} \left( 1 - 2 \frac{M_\NS^2}{R^2} \right) \ln f(R)  \right]\,, 
%\ea
%
%\ba
\allowdisplaybreaks[0]%
\ea%
%\end{widetext}
%
\end{widetext}%
where $C_Q$ is another integration constant. One might think that the above metric components diverge at infinity. However, one can show that $h_2^\ext = \mathcal{O}(R^{-3})$, $m_2^\ext = \mathcal{O}(R^{-2})$ and $k_2^\ext = \mathcal{O}(R^{-3})$ as $R \rightarrow \infty$.

%--------------------------------------------------
\section{Energy Flux \\* and Angular Momentum Flux}
\label{sec:E-flux}

In this appendix, we look at the dissipative effects of the binary evolution, i.e.~the energy and angular momentum fluxes radiated via gravitational and scalar radiation.

%--------------------------------------------------
\subsection{Energy Flux}

The rate of change of the orbital binding energy $\dot{E}_{b}$ must be balanced by the energy flux $\mathcal{F}$
carried away from the system by all propagating degrees of freedom. In dynamical CS
gravity, there are two such quantities, the metric perturbation $(h)$ and the scalar field $(\vartheta)$, and thus
$\dot{E}_{b} = - {\cal{F}}_{(h)} - {\cal{F}}_{(\vartheta)} = \dot{E}^{(h)} + \dot{E}^{(\vartheta)}$. 
Each of these contributions can be split into a GR term plus a CS term, noting of course that 
$\dot{E}_{\GR}^{(\vartheta)} = 0$, we then have $\dot{E}_{b} =  \dot{E}^{(h)} + \delta \dot{E}^{(\hDef)} 
+ \delta \dot{E}^{(\vartheta)}$, where the $\delta$'s are to remind us that these are CS corrections. 

In either case, for any field $\varphi$ with stress-energy tensor
$T^{(\varphi)}$, the energy flux is (see Sec.~VI of~\cite{quadratic}) 
\begin{equation}
\label{eq:Edotdefinition}
\dot{E}^{(\varphi)} = \lim_{r\to\infty} \int_{S^2_r} 
\left<T^{(\varphi)}_{ti} n^i \right>_{\omega}  r^2 d\Omega\,,
\end{equation}
where the orbit average of any quantity $Q$ is defined as
\begin{equation}
\label{eq:orbit-ave-def}
\left<Q\right>_{\omega} = \frac{1}{T}\oint Q dt = \frac{1}{T} \int_0^{2\pi}
\frac{Q(\phi) \mathrm{d}\phi}{\dot{\phi}}\,,
\end{equation}
where $T$ is the orbital period, $\phi$ is the orbital phase, and
the Jacobian $\dot\phi$ must of course be included.

%--------
\subsubsection{Scalar Field}
Let us consider the energy flux associated with the CS scalar field $\vartheta$. 
The scalar stress-energy tensor was given in Eq.~\eqref{eq:Tab-theta}. 
The radiative far-zone solution for the scalar field $\vartheta^{\FZ}$ is~\cite{quadratic},
\begin{equation}
\vartheta^{\FZ} = \frac{1}{r} \ddot{\mu}_{ij}  n^{ij} \,,
\label{eq:thetaFZquadrupole}
\end{equation}
where $\mu_{ij}$ is the magnetic-type quadrupole tensor of the source,
\begin{equation}
\mu^{ij} \equiv x_1^{(i} \mu_1^{j)} + x_2^{(i} \mu_2^{j)} \,.
\label{eq:quadrupoledef}
\end{equation}
There are other contributions to $\vartheta^{\FZ}$, but they are
suppressed by the ratio of the orbital and radiation-reaction
or precession timescales, so we neglect them here. 

Inserting this far-zone solution into $T^{(\vartheta)}$ and this into the energy flux
formula, we obtain
\begin{equation}
\delta  \dot{E}^{(\vartheta)} =\, -\frac{4\pi}{15}\beta
\left\langle \left[ 2 \dddot{\mu}_{ij}\dddot{\mu}^{ij} +
    \left(\dddot{\mu}^i{}_i\right)^2 \right]
\right\rangle_{\omega}\,,
\end{equation}
which upon expansion returns
\begin{align}
\label{eq:delta_Edot_theta}
\delta\dot{E}^{(\vartheta)} &= -\frac{5}{768} \frac{\xi_\CS}{m^4} \left<
  \frac{m^{6}}{r_{12}^{6}} \left(
\Delta^{2} v_{12}^{2}+ 2 (\Delta \cdot v_{12})^{2}
\right.\right.
\nonumber \\
&+ \left. \left.
3 \Delta^{2}(n_{12}\cdot v_{12})^{2}  - 12 (n_{12}\cdot v_{12}) (\Delta\cdot n_{12})(\Delta\cdot v_{12})
\right.\right.
\nonumber \\
&+ \left. \left.
18 (\Delta \cdot n_{12})^{2} (n_{12}\cdot v_{12})^{2}
\right)  \right>_{\omega}\,.
\end{align}
We have here defined
\begin{equation}
\Delta^{i} =  \frac{m_{2}}{m} C_{1}^{3} \chi_{1} \hat{S}_{1}^{i} \bar{\mu}_{1}
- \frac{m_{1}}{m} C_{2}^{3} \chi_{2} \hat{S}_{2}^{i} \bar{\mu}_{2}\,,
\label{Deltai-in-mubars}
\end{equation}
so as to agree with the definition of $\Delta^{i}$ in~\cite{quadratic}
when the two bodies are BHs. 
In the circular orbit limit, $(n_{12}\cdot v_{12}) \to 0$ and our Eq.~\eqref{eq:delta_Edot_theta} agrees with Eq.~(137) of~\cite{quadratic}.

By performing the orbit average, we find that the correction to the rate of change to the 
orbital binding energy due to the scalar field is 
\begin{equation}
\delta\dot{E}^{(\vartheta)} = -\frac{5}{768}\zeta\left(\frac{m}{a}
\right)^{7} \frac{ 2 \Delta_{1}^{2}f_{1}(e) + 2\Delta_{2}^{2}
  f_{2}(e)+\Delta_{3}^{2}f_{3(e)}}{(1-e^{2})^{11/2}}\,,
  \label{eq:dotEtheta}
\end{equation}
where
\begin{align}
f_{1}(e) &= 1+\frac{77}{8}e^{2} + \frac{37}{4}e^{4}+\frac{81}{128}e^{6}\,, \\
f_{2}(e) &= 1+\frac{75}{8}e^{2} + 8 e^{4}+\frac{63}{128}e^{6}\,, \\
f_{3}(e) &= 1+\frac{19}{2}e^{2} + \frac{69}{8}e^{4}+\frac{9}{16}e^{6}\,.
\end{align}
We could rewrite these expressions in terms of the structure constants of Eq.~\eqref{Deltai-in-mubars},
but the resulting expression is rather long and unilluminating. 
It may seem surprising at first that the different components of
$\Delta^{i}$ play unequal roles, but the binary orbit sets up a preferred
coordinate system that treats components of $\Delta^{i}$ differently.

%------
\subsubsection{Metric Perturbation}
The leading-order correction to the effective GW stress-energy tensor
is given by (see~\cite{Stein:2010pn} and Sec.~VI of~\cite{quadratic})
\begin{equation}
T_{\mu \nu}^{{(\hDef)}} = \frac{1}{16 \pi}  \left< \hGR^{\TT}_{ij ,(\mu}
\hDef_{\TT}^{ij}{}_{,\nu)}\right>_{\lambda}\,,
\label{SET}
\end{equation}
where $\langle \rangle_{\lambda}$ is a short-wavelength average,
$h_{ij}$ is the metric perturbation in GR,
$\hDef_{ij}$ is the correction to the metric perturbation due
to CS gravity, and TT stands for the transverse-traceless projection,
\begin{equation}
H^{\TT}_{ij} = \Lambda_{ij,kl} H_{kl}\,, \qquad
\Lambda_{ij,kl}=P_{ik} P_{jl}-\frac{1}{2} P_{ij} P_{kl}\,,
\end{equation}
with $P_{ij}=\delta_{ij}-n_{ij}$ the projector onto the plane
perpendicular to the line from the source to a FZ field point.
The leading-order expressions for $h_{ij}$ and $\hDef_{ij}$ in the far-zone are (see
Eq.~(118) of~\cite{quadratic})
\begin{align}
\label{hGRFZ}
h_{ij} &= \frac{2\ddot{I}_{ij}}{r}\,, \\
\label{hCSFZ}
\hDef_{ij} &= \frac{8 \pi \beta}{r r_{12}^{3}} \left\{2 \mu_{1}^{(i} \mu_{2}^{j)} -12 n_{12}^{(i} \mu_{1}^{j)} \left(n_{12}^{k} \mu_{2k}\right) 
\right. \\ 
& + \left. 3 n_{12}^{ij} \left[5 \left(n_{12}^{k} \mu_{1k} \right)
    \left(n_{12}^{l} \mu_{2l} \right) - \mu_{1k} \mu_{2}^{k} \right]
\right\} \plusonetotwo\,,\nn
\end{align}
where $I_{ij}=m_{1}x^{1}_{i}x^{1}_{j} \plusonetotwo$ is the mass
quadrupole tensor.

The effective  stress-energy tensor of Eq.~\eqref{SET} must be inserted into
Eq.~\eqref{eq:Edotdefinition} to calculate the correction to the rate of change
of the orbital binding energy due to the CS correction to the metric perturbation.  
Inserting the previous expressions and performing the integral over $d\Omega$
gives
\begin{widetext}
\begin{align}
\label{eq:delta_dotE_h}
\delta\dot{E}^{(\hDef)} = \frac{128\pi}{5} \beta m_{1} m_{2} \Big<
\frac{1}{r_{12}^{6}} &\Big[
-6 (v_{12}\cdot\mu_{1})(v_{12}\cdot\mu_{2})
+9 v_{12}^{2}(n_{12}\cdot\mu_{1})(n_{12}\cdot\mu_{2})
-3 v_{12}^{2}(\mu_{1}\cdot\mu_{2}) \nn \\ 
&+14 (n_{12}\cdot v_{12})(n_{12}\cdot\mu_{1})(v_{12}\cdot\mu_{2})
+14 (n_{12}\cdot v_{12})(n_{12}\cdot\mu_{2})(v_{12}\cdot\mu_{1}) \nn \\ 
&+5 (n_{12}\cdot v_{12})^{2} (\mu_{1}\cdot \mu_{2})
-34 (n_{12}\cdot v_{12})^{2} (n_{12}\cdot\mu_{1}) (n_{12}\cdot\mu_{2})
\Big]\Big>_{\omega}\,.
\end{align}
\end{widetext}
Similar to the scalar field case, in the circular orbit, Eq.~\eqref{eq:delta_dotE_h} divided by the GR energy flux $\dot{E}_\GR$ agrees with Eq.~(142) of~\cite{quadratic}. 
Performing the orbit-averaging yields
\begin{align}
\delta\dot{E}^{(\hDef)} &=
-\frac{192\pi}{5} \frac{\beta m_{1}m_{2}m}{a^{7}}
\nonumber \\
\times &
\frac{g_{1}(e)\mu_{1}^{x}\mu_{2}^{x} + g_{2}(e)\mu_{1}^{y}\mu_{2}^{y}
  + 2 g_{3}(e)\mu_{1}^{z}\mu_{2}^{z} }{(1-e^{2})^{11/2}}\,, 
\end{align}
where
\begin{align}
g_{1}(e) &= 1 -5e^{2} -\frac{161}{24}e^{4}-\frac{43}{96}e^{6}\,, \\
g_{2}(e) &= 1 +\frac{59}{3}e^{2} +\frac{419}{24}e^{4}+\frac{33}{32}e^{6}\,, \\
g_{3}(e) &= 1 +\frac{43}{6}e^{2} +\frac{41}{8}e^{4}+\frac{13}{48}e^{6}\,.
\end{align}
Rewriting the energy flux in terms of the dimensionless structure constants, 
we find
\begin{align}
  \label{eq:dotEh}
\delta\dot{E}^{(\hDef)} &=
-\frac{15}{16} \frac{\xi_\CS}{m^4}
\eta\left( \frac{m}{a} \right)^{7}
\chi_{1}\chi_{2}\bar{\mu}_{1}\bar{\mu}_{2}C_{1}^{3}C_{2}^{3} 
\nonumber \\
&\times \frac{g_{1}(e)\hat{S}_{1}^{x}\hat{S}_{2}^{x} + g_{2}(e)\hat{S}_{1}^{y}\hat{S}_{2}^{y}
  + 2 g_{3}(e)\hat{S}_{1}^{z}\hat{S}_{2}^{z} }{(1-e^{2})^{11/2}}\,.
\end{align}

%%%%%%%%%%%%%%%%%%%%%%%%%%%%%%%%%%%%%%%%%%%%%%%%%%
\subsection{Angular momentum flux}
In this appendix, we present the rate of change of the orbital angular momentum, as induced by the propagating scalar field and metric perturbation. 

\subsubsection{Scalar Field}
The angular momentum flux for the scalar field $\vartheta$, which has
stress-energy tensor $T^{(\vartheta)}$ is~\cite{PetersMathews}
\begin{align}
\label{eq:Ldot}
\delta \dot{L}^{(\vartheta)}_{i} &= -\epsilon_{ijk} \lim_{r\to\infty} \int_{S^{2}_{r}}
\left<  T^{(\vartheta)}_{kl} x^{j} n^{l}  \right>_{\omega} r^{2} d\Omega \\
&= -\epsilon_{ijk} \lim_{r\to\infty} \int_{S^{2}_{r}}
\left<  T^{(\vartheta)}_{kl} n^{jl} \right>_{\omega} r^{3} d\Omega\,.
\end{align}
Only the parts of $T^{(\vartheta)}_{kl}$ which decay as $r^{-3}$ contribute a
finite part. One can verify that the parts that decay as $r^{-2}$ actually vanish identically
prior to taking the limit to spatial infinity. Inserting the same far-zone solution as before
[Eq.~\eqref{eq:thetaFZquadrupole}] into the stress energy tensor [Eq.~\eqref{eq:Tab-theta}], 
and inserting this into Eq.~\eqref{eq:Ldot}, we have
\begin{align}
\delta \dot{L}^{(\vartheta)}_{i} = \frac{16\pi}{15}\beta \epsilon_{ijk}
\left< \dddot{\mu}_{jp} \ddot{\mu}_{kp}
\right>_{\omega}\,.
\end{align}
Expanding in terms of time derivatives of the quadrupole tensor gives
\begin{align}
\delta \dot{L}^{(\vartheta)}_{i} &= \frac{5}{768} \frac{\xi_\CS}{m^4} m \left<
\left(\frac{m}{r_{12}}\right)^{5}
\Big[
(\Delta \cdot v_{12}) (\Delta \times n_{12})_{i} 
\right. 
\nonumber \\
&+ \left.
(\Delta \cdot n_{12}) (v_{12}\times \Delta)_{i} +
\Delta^{2} (v_{12}\times n_{12})_{i}
\Big] \right>_{\omega}\,.
\end{align}
Finally, performing the orbit-averaging as before, with $\hat{z}$ lying
perpendicular to the orbital plane, gives
\begin{align}
\delta \dot{L}^{(\vartheta)}_{i} &= \frac{5}{768} \frac{\xi_\CS}{m^4}
\left( \frac{m}{a} \right)^{6} \sqrt{am}
\frac{1+3 e^{2}+\frac{3}{8}e^{4}}{(1-e^{2})^{4}}
\nonumber \\
&\times
\left[ (\Delta \cdot \hat{L}) \Delta_{i} - 2 \Delta^{2} \hat{L}_{i} \right]\,.
\end{align}

\subsubsection{Gravitational Field}
The angular momentum flux associated with the metric perturbation at null infinity can be written as 
(see Eq.~(4.22') in~\cite{Thorne:1980rm}) 
\begin{align}
\label{eq:LdotGW}
\delta \dot{L}^{(h)}_{i} &= - \frac{\epsilon_{ipq}}{16 \pi}   \lim_{r\to\infty} \int_{S^{2}_{r}}
\left<  h_{pa}^{\TT} \dot{h}_{aq}^{\TT} r^{2} - \frac{1}{2} n_{p} h_{ab,q}^{\TT} \dot{h}_{ab}^{\TT} r^{3}\right>_{\lambda,\omega} d\Omega\,.
\end{align}
This corrects a well-known mistake in the paper of Peters~\cite{Peters:1964zz}. If the metric perturbation
consists of a GR term plus a CS correction, we can then calculate the CS modification to the angular momentum flux via
\begin{align}
\label{eq:LdotGWCS}
\delta \dot{L}^{(\hDef)}_{i} &= \frac{1}{16 \pi}  \epsilon_{ipq} \lim_{r\to\infty} \int_{S^{2}_{r}}
\left<  \left(\hDef_{pa}^{\TT} \dot{h}_{aq}^{\TT} + h_{pa}^{\TT} \dot{\hDef}_{aq}^{\TT} \right) r^{2} 
\right.
\nonumber \\
&- \left.
 \frac{1}{2} n_{p} \left( \hDef_{ab,q}^{\TT} \dot{h}_{ab}^{\TT} + h_{ab,q}^{\TT} \dot{\hDef}_{ab}^{\TT} \right) r^{3}\right>_{\lambda,\omega} d\Omega\,.
\end{align}

As before, only the parts of the integrand that decay as $r^{-2}$ and $r^{-3}$ in the first and second
terms respectively contribute a finite part. One can verify that any seemingly divergent terms actually
vanish prior to taking the limit to spatial infinity. Inserting the same far-zone solution as before
[Eqs.~\eqref{hGRFZ}-\eqref{hCSFZ}] into the above expressions, inserting the expressions for
the time dependence of $\mu_{ij}$ and orbit averaging, we finally obtain
\begin{equation}
\delta \dot{L}_{i}^{\hDef} = \frac{15}{16} \frac{\xi_\CS}{m^4} \eta \left(\frac{m}{a}\right)^{6} \sqrt{a m} \chi_{1} \chi_{2} \bar{\mu}_{1} \bar{\mu}_{2} C_{1}^{3} C_{2}^{3} \left(1 - e^{2} \right)^{-4} \ell_{i}\,,
\end{equation}
where we have defined
\begin{align}
\ell_{x} &= \left(1 + \frac{15}{4} e^{2} + \frac{1}{2} e^{4} \right) \left(\hat{S}_{1}^{z} \hat{S}_{2}^{x} + \hat{S}_{1}^{x} \hat{S}_{2}^{z} \right)\,,
\\
\ell_{y} &= \left(1 + \frac{9}{4} e^{2} + \frac{1}{4} e^{4}\right)
\left(\hat{S}_{1}^{z} \hat{S}_{2}^{y} + \hat{S}_{1}^{y} \hat{S}_{2}^{z} \right)\,,
\\
\ell_{z} &= - \left[
\left(1 - \frac{1}{8} e^{4} \right)  \hat{S}_{1}^{x} \hat{S}_{2}^{x}
+ \left(1 + 6 e^{2} + \frac{7}{8} e^{4} \right) \hat{S}_{1}^{y} \hat{S}_{2}^{y}
\right. 
\nonumber \\
&+ \left.
2 \left( 1 + 3 e^{2} + \frac{3}{8} e^{4}\right) \hat{S}_{1}^{z} \hat{S}_{2}^{z} \right]\,.
\end{align}
One can, of course, check that in the zero eccentricity limit $\delta \dot{E}^{\hDef} = \Omega  \delta \dot{L}_{z}^{\hDef}$, where $\Omega = 2 \pi/T$ is the orbital frequency.

%------------------------------------------------
\section{Spin Precession due to \\* Scalar Interaction}
\label{sec:spin-precession}

Using the relationship in Eq.~\eqref{eq:mu-S-relation}, we can rewrite
the scalar dipole-dipole interaction as an additional spin-spin
interaction,
\begin{multline}
\delta\mathcal{L}_{SS} = \frac{25}{256} \frac{\xi_\CS}{m^4}
\frac{\bar{\mu}_{1}\bar{\mu}_{2}}{\eta^2}
C_{1}^{3}C_{2}^{3} \\
\times \frac{1}{r_{12}^{3}} \left[
 3(S_{1}\cdot n_{12})(S_{2}\cdot n_{12}) - (S_{1}\cdot S_{2})
\right]\,,
\end{multline}
where $C_{A}$ is the compactness of body $A$.
Compare this with the leading spin-spin interaction present in GR,
e.g.~Eq.~(5b) of~\cite{Kidder:1993do}. It is clear that this
contribution is of the same form with only a different prefactor. This
interaction will lead to additional spin precession (to be added to
that already present from GR) of the form
\begin{multline}
\label{eq:Sdot}
\delta\dot{S}^{i}_{1,\mrm{SS}} = -\frac{25}{256}\frac{\xi_\CS}{m^4}
\frac{\bar{\mu}_{1}\bar{\mu}_{2}}{\eta^2}
C_{1}^{3}C_{2}^{3} \\
\times \frac{1}{r_{12}^{3}} \epsilon_{ijk} S_{1}^{k}
\left\{ \frac{1}{2} S_{2}^{j} +3(n_{12}\cdot S_{2})n_{12}^{j} \right\} \,.
\end{multline}
and similarly for body 2 with $1\leftrightarrow 2$.

The dipole-dipole interaction does not modify the spin-orbit coupling
at all, so the orbital angular momentum $L^{i}$ does not appear in the
above expression as it does in GR. We expect that the
$\mathcal{O}(\alpha'{}^2)$ conservative correction to $L^{i}$ appears at 2PN
order or higher, so the correction to the
orbit-induced precession will only appear at higher than 2 PN order.

This effect is interesting in that it comes in at the \emph{same} PN
order as in GR. However, currently, there
are no binary systems with sufficiently well-modeled spin precession
that could be used to measure or place constraints on
Chern-Simons gravity. Indeed, the spin precession periods for
currently known pulsar binary systems are of the order of hundreds of
years~\cite{Clifton:2008gr}. Therefore, it seems unlikely that this
phenomenon will soon be used for constraining modified theories.

Precession is also caused by a monopole-quadrupole interaction~\cite{goldstein,lai-shapiro}, which enters at the same PN order as the spin-spin interaction~\cite{poisson-quadrupole}. Since there is a quadrupole moment deformation in dynamical CS gravity, there would be a CS correction to the monopole-quadrupole interaction of the form
\be
\langle \dot{S_1} \rangle_Q = 3 \frac{\mu}{m^2} \frac{Q_1^\CS}{m_1^3 \chi_1} \left( \frac{m}{r_{12}} \right)^3 \left( \hat{L} \cdot \hat{S}_1 \right) \epsilon_{ijk} \hat{L}^j S_1^k\,,
\ee
and similarly for body 2.
Notice that this is of the same PN order as the correction to the spin-spin interaction shown in Eq.~\eqref{eq:Sdot}. 

%%%%%%%%%%%%%%%%%%%%%%%%%%%%%%%%%%%%%%%%%%%%
%%%%%%%%%%%%%%%%%%%%%%%%%%%%%%%%%%%%%%%%%%%%
\bibliography{master}

%merlin.mbs 2010-03-15 4.21a (PWD, AO, DPC)
\begin{thebibliography}{3}%
\makeatletter
\providecommand \@ifxundefined [1]{%
 \@ifx{#1\undefined}
}%
\providecommand \@ifnum [1]{%
 \ifnum #1\expandafter \@firstoftwo
 \else \expandafter \@secondoftwo
 \fi
}%
\providecommand \@ifx [1]{%
 \ifx #1\expandafter \@firstoftwo
 \else \expandafter \@secondoftwo
 \fi
}%
\providecommand \natexlab [1]{#1}%
\providecommand \enquote  [1]{``#1''}%
\providecommand \bibnamefont  [1]{#1}%
\providecommand \bibfnamefont [1]{#1}%
\providecommand \citenamefont [1]{#1}%
\providecommand \href@noop [0]{\@secondoftwo}%
\providecommand \href [0]{\begingroup \@sanitize@url \@href}%
\providecommand \@href[1]{\@@startlink{#1}\@@href}%
\providecommand \@@href[1]{\endgroup#1\@@endlink}%
\providecommand \@sanitize@url [0]{\catcode `\\12\catcode `\$12\catcode
  `\&12\catcode `\#12\catcode `\^12\catcode `\_12\catcode `\%12\relax}%
\providecommand \@@startlink[1]{}%
\providecommand \@@endlink[0]{}%
\providecommand \url  [0]{\begingroup\@sanitize@url \@url }%
\providecommand \@url [1]{\endgroup\@href {#1}{\urlprefix }}%
\providecommand \urlprefix  [0]{URL }%
\providecommand \Eprint [0]{\href }%
\@ifxundefined \urlstyle {%
  \providecommand \doi  [0]{\begingroup \@sanitize@url \@doi}%
  \providecommand \@doi [1]{\endgroup \@@startlink {\doibase
  #1}doi:\discretionary {}{}{}#1\@@endlink }%
}{%
  \providecommand \doi  [0]{doi:\discretionary{}{}{}\begingroup
  \urlstyle{rm}\Url }%
}%
\providecommand \doibase [0]{http://dx.doi.org/}%
\providecommand \Doi [0]{\begingroup \@sanitize@url \@Doi }%
\providecommand \@Doi  [1]{\endgroup\@@startlink{\doibase#1}\@@Doi}%
\providecommand \@@Doi [1]{#1\@@endlink}%
\providecommand \selectlanguage [0]{\@gobble}%
\providecommand \bibinfo  [0]{\@secondoftwo}%
\providecommand \bibfield  [0]{\@secondoftwo}%
\providecommand \translation [1]{[#1]}%
\providecommand \BibitemOpen [0]{}%
\providecommand \bibitemStop [0]{}%
\providecommand \bibitemNoStop [0]{.\EOS\space}%
\providecommand \EOS [0]{\spacefactor3000\relax}%
\providecommand \BibitemShut  [1]{\csname bibitem#1\endcsname}%
%</preamble>
\bibitem [{\citenamefont {{Yagi}}\ \emph {et~al.}(2013)\citenamefont {{Yagi}},
  \citenamefont {{Stein}}, \citenamefont {{Yunes}},\ and\ \citenamefont
  {{Tanaka}}}]{Yagi:2013mbt}%
  \BibitemOpen
  \bibfield  {author} {\bibinfo {author} {\bibfnamefont {K.}~\bibnamefont
  {{Yagi}}}, \bibinfo {author} {\bibfnamefont {L.~C.}\ \bibnamefont {{Stein}}},
  \bibinfo {author} {\bibfnamefont {N.}~\bibnamefont {{Yunes}}}, \ and\
  \bibinfo {author} {\bibfnamefont {T.}~\bibnamefont {{Tanaka}}},\ }\Doi
  {10.1103/PhysRevD.87.084058} {\bibfield  {journal} {\bibinfo  {journal}
  {Phys. Rev.},\ }\textbf {\bibinfo {volume} {D87}},\ \bibinfo {pages} {084058}
  (\bibinfo {year} {2013})},\ \Eprint {http://arxiv.org/abs/1302.1918}
  {arXiv:1302.1918 [gr-qc]} \BibitemShut {NoStop}%
%%CITATION = ARXIV:1302.1918;%%
\bibitem [{\citenamefont {{Yagi}}\ \emph {et~al.}(2012)\citenamefont {{Yagi}},
  \citenamefont {{Stein}}, \citenamefont {{Yunes}},\ and\ \citenamefont
  {{Tanaka}}}]{quadratic}%
  \BibitemOpen
  \bibfield  {author} {\bibinfo {author} {\bibfnamefont {K.}~\bibnamefont
  {{Yagi}}}, \bibinfo {author} {\bibfnamefont {L.~C.}\ \bibnamefont {{Stein}}},
  \bibinfo {author} {\bibfnamefont {N.}~\bibnamefont {{Yunes}}}, \ and\
  \bibinfo {author} {\bibfnamefont {T.}~\bibnamefont {{Tanaka}}},\ }\Doi
  {10.1103/PhysRevD.85.064022} {\bibfield  {journal} {\bibinfo  {journal}
  {Phys. Rev.},\ }\textbf {\bibinfo {volume} {D85}},\ \bibinfo {eid} {064022}
  (\bibinfo {year} {2012})},\ \Eprint {http://arxiv.org/abs/1110.5950}
  {arXiv:1110.5950 [gr-qc]} \BibitemShut {NoStop}%
\bibitem [{\citenamefont {{Yagi}}\ \emph {et~al.}(2016)\citenamefont {{Yagi}},
  \citenamefont {{Stein}}, \citenamefont {{Yunes}},\ and\ \citenamefont
  {{Tanaka}}}]{quadratic-erratum}%
  \BibitemOpen
  \bibfield  {author} {\bibinfo {author} {\bibfnamefont {K.}~\bibnamefont
  {{Yagi}}}, \bibinfo {author} {\bibfnamefont {L.~C.}\ \bibnamefont {{Stein}}},
  \bibinfo {author} {\bibfnamefont {N.}~\bibnamefont {{Yunes}}}, \ and\
  \bibinfo {author} {\bibfnamefont {T.}~\bibnamefont {{Tanaka}}},\ }\Doi
  {10.1103/PhysRevD.93.029902} {\bibfield  {journal} {\bibinfo  {journal}
  {Phys. Rev.},\ }\textbf {\bibinfo {volume} {D93}},\ \bibinfo {pages} {029902}
  (\bibinfo {year} {2016})},\ \Eprint {http://arxiv.org/abs/1110.5950}
  {arXiv:1110.5950 [gr-qc]} \BibitemShut {NoStop}%
\end{thebibliography}%


\begin{thebibliography}{102}%
\makeatletter
\providecommand \@ifxundefined [1]{%
 \@ifx{#1\undefined}
}%
\providecommand \@ifnum [1]{%
 \ifnum #1\expandafter \@firstoftwo
 \else \expandafter \@secondoftwo
 \fi
}%
\providecommand \@ifx [1]{%
 \ifx #1\expandafter \@firstoftwo
 \else \expandafter \@secondoftwo
 \fi
}%
\providecommand \natexlab [1]{#1}%
\providecommand \enquote  [1]{``#1''}%
\providecommand \bibnamefont  [1]{#1}%
\providecommand \bibfnamefont [1]{#1}%
\providecommand \citenamefont [1]{#1}%
\providecommand \href@noop [0]{\@secondoftwo}%
\providecommand \href [0]{\begingroup \@sanitize@url \@href}%
\providecommand \@href[1]{\@@startlink{#1}\@@href}%
\providecommand \@@href[1]{\endgroup#1\@@endlink}%
\providecommand \@sanitize@url [0]{\catcode `\\12\catcode `\$12\catcode
  `\&12\catcode `\#12\catcode `\^12\catcode `\_12\catcode `\%12\relax}%
\providecommand \@@startlink[1]{}%
\providecommand \@@endlink[0]{}%
\providecommand \url  [0]{\begingroup\@sanitize@url \@url }%
\providecommand \@url [1]{\endgroup\@href {#1}{\urlprefix }}%
\providecommand \urlprefix  [0]{URL }%
\providecommand \Eprint [0]{\href }%
\providecommand \doibase [0]{http://dx.doi.org/}%
\providecommand \selectlanguage [0]{\@gobble}%
\providecommand \bibinfo  [0]{\@secondoftwo}%
\providecommand \bibfield  [0]{\@secondoftwo}%
\providecommand \translation [1]{[#1]}%
\providecommand \BibitemOpen [0]{}%
\providecommand \bibitemStop [0]{}%
\providecommand \bibitemNoStop [0]{.\EOS\space}%
\providecommand \EOS [0]{\spacefactor3000\relax}%
\providecommand \BibitemShut  [1]{\csname bibitem#1\endcsname}%
\let\auto@bib@innerbib\@empty
%</preamble>
\bibitem [{\citenamefont {De~Felice}\ and\ \citenamefont
  {Tsujikawa}(2010)}]{fR}%
  \BibitemOpen
  \bibfield  {author} {\bibinfo {author} {\bibfnamefont {A.}~\bibnamefont
  {De~Felice}}\ and\ \bibinfo {author} {\bibfnamefont {S.}~\bibnamefont
  {Tsujikawa}},\ }\href@noop {} {\bibfield  {journal} {\bibinfo  {journal}
  {Living Rev. Rel.}\ }\textbf {\bibinfo {volume} {13}},\ \bibinfo {pages} {3}
  (\bibinfo {year} {2010})},\ \Eprint {http://arxiv.org/abs/1002.4928}
  {arXiv:1002.4928 [gr-qc]} \BibitemShut {NoStop}%
%%CITATION = 1002.4928;%%
\bibitem [{\citenamefont {{Will}}(1993)}]{TEGP}%
  \BibitemOpen
  \bibfield  {author} {\bibinfo {author} {\bibfnamefont {C.~M.}\ \bibnamefont
  {{Will}}},\ }\href@noop {} {\emph {\bibinfo {title} {Theory and Experiment in
  Gravitational Physics, by Clifford M.~Will, pp.~396.~ISBN
  0521439736.~Cambridge, UK: Cambridge University Press, March 1993.}}},\
  edited by\ \bibinfo {editor} {\bibnamefont {{Will, C.~M.}}}\ (\bibinfo {year}
  {1993})\BibitemShut {NoStop}%
\bibitem [{\citenamefont {Will}(2006)}]{will-living}%
  \BibitemOpen
  \bibfield  {author} {\bibinfo {author} {\bibfnamefont {C.~M.}\ \bibnamefont
  {Will}},\ }\href {http://www.livingreviews.org/lrr-2006-3} {\bibfield
  {journal} {\bibinfo  {journal} {Living Reviews in Relativity}\ }\textbf
  {\bibinfo {volume} {9}} (\bibinfo {year} {2006})},\ \Eprint
  {http://arxiv.org/abs/gr-qc/0510072} {arXiv:gr-qc/0510072} \BibitemShut
  {NoStop}%
%%CITATION = GR-QC/0510072;%%
\bibitem [{\citenamefont {Stairs}(2003)}]{stairs}%
  \BibitemOpen
  \bibfield  {author} {\bibinfo {author} {\bibfnamefont {I.~H.}\ \bibnamefont
  {Stairs}},\ }\href@noop {} {\bibfield  {journal} {\bibinfo  {journal} {Living
  Rev.Rel.}\ }\textbf {\bibinfo {volume} {6}},\ \bibinfo {pages} {5} (\bibinfo
  {year} {2003})},\ \Eprint {http://arxiv.org/abs/astro-ph/0307536}
  {arXiv:astro-ph/0307536 [astro-ph]} \BibitemShut {NoStop}%
%%CITATION = ASTRO-PH/0307536;%%
\bibitem [{\citenamefont {{Burgay}}\ \emph {et~al.}(2003)\citenamefont
  {{Burgay}}, \citenamefont {{D'Amico}}, \citenamefont {{Possenti}},
  \citenamefont {{Manchester}}, \citenamefont {{Lyne}}, \citenamefont
  {{Joshi}}, \citenamefont {{McLaughlin}}, \citenamefont {{Kramer}},
  \citenamefont {{Sarkissian}}, \citenamefont {{Camilo}}, \citenamefont
  {{Kalogera}}, \citenamefont {{Kim}},\ and\ \citenamefont
  {{Lorimer}}}]{burgay}%
  \BibitemOpen
  \bibfield  {author} {\bibinfo {author} {\bibfnamefont {M.}~\bibnamefont
  {{Burgay}}}, \bibinfo {author} {\bibfnamefont {N.}~\bibnamefont {{D'Amico}}},
  \bibinfo {author} {\bibfnamefont {A.}~\bibnamefont {{Possenti}}}, \bibinfo
  {author} {\bibfnamefont {R.~N.}\ \bibnamefont {{Manchester}}}, \bibinfo
  {author} {\bibfnamefont {A.~G.}\ \bibnamefont {{Lyne}}}, \bibinfo {author}
  {\bibfnamefont {B.~C.}\ \bibnamefont {{Joshi}}}, \bibinfo {author}
  {\bibfnamefont {M.~A.}\ \bibnamefont {{McLaughlin}}}, \bibinfo {author}
  {\bibfnamefont {M.}~\bibnamefont {{Kramer}}}, \bibinfo {author}
  {\bibfnamefont {J.~M.}\ \bibnamefont {{Sarkissian}}}, \bibinfo {author}
  {\bibfnamefont {F.}~\bibnamefont {{Camilo}}}, \bibinfo {author}
  {\bibfnamefont {V.}~\bibnamefont {{Kalogera}}}, \bibinfo {author}
  {\bibfnamefont {C.}~\bibnamefont {{Kim}}}, \ and\ \bibinfo {author}
  {\bibfnamefont {D.~R.}\ \bibnamefont {{Lorimer}}},\ }\href {\doibase
  10.1038/nature02124} {\bibfield  {journal} {\bibinfo  {journal} {Nature}\
  }\textbf {\bibinfo {volume} {426}},\ \bibinfo {pages} {531} (\bibinfo {year}
  {2003})},\ \Eprint {http://arxiv.org/abs/arXiv:astro-ph/0312071}
  {arXiv:astro-ph/0312071} \BibitemShut {NoStop}%
\bibitem [{\citenamefont {{Lyne}}\ \emph {et~al.}(2004)\citenamefont {{Lyne}},
  \citenamefont {{Burgay}}, \citenamefont {{Kramer}}, \citenamefont
  {{Possenti}}, \citenamefont {{Manchester}}, \citenamefont {{Camilo}},
  \citenamefont {{McLaughlin}}, \citenamefont {{Lorimer}}, \citenamefont
  {{D'Amico}}, \citenamefont {{Joshi}}, \citenamefont {{Reynolds}},\ and\
  \citenamefont {{Freire}}}]{lyne}%
  \BibitemOpen
  \bibfield  {author} {\bibinfo {author} {\bibfnamefont {A.~G.}\ \bibnamefont
  {{Lyne}}}, \bibinfo {author} {\bibfnamefont {M.}~\bibnamefont {{Burgay}}},
  \bibinfo {author} {\bibfnamefont {M.}~\bibnamefont {{Kramer}}}, \bibinfo
  {author} {\bibfnamefont {A.}~\bibnamefont {{Possenti}}}, \bibinfo {author}
  {\bibfnamefont {R.~N.}\ \bibnamefont {{Manchester}}}, \bibinfo {author}
  {\bibfnamefont {F.}~\bibnamefont {{Camilo}}}, \bibinfo {author}
  {\bibfnamefont {M.~A.}\ \bibnamefont {{McLaughlin}}}, \bibinfo {author}
  {\bibfnamefont {D.~R.}\ \bibnamefont {{Lorimer}}}, \bibinfo {author}
  {\bibfnamefont {N.}~\bibnamefont {{D'Amico}}}, \bibinfo {author}
  {\bibfnamefont {B.~C.}\ \bibnamefont {{Joshi}}}, \bibinfo {author}
  {\bibfnamefont {J.}~\bibnamefont {{Reynolds}}}, \ and\ \bibinfo {author}
  {\bibfnamefont {P.~C.~C.}\ \bibnamefont {{Freire}}},\ }\href {\doibase
  10.1126/science.1094645} {\bibfield  {journal} {\bibinfo  {journal}
  {Science}\ }\textbf {\bibinfo {volume} {303}},\ \bibinfo {pages} {1153}
  (\bibinfo {year} {2004})},\ \Eprint
  {http://arxiv.org/abs/arXiv:astro-ph/0401086} {arXiv:astro-ph/0401086}
  \BibitemShut {NoStop}%
\bibitem [{\citenamefont {Kramer}\ \emph {et~al.}(2006)\citenamefont {Kramer},
  \citenamefont {Stairs}, \citenamefont {Manchester}, \citenamefont
  {McLaughlin}, \citenamefont {Lyne} \emph {et~al.}}]{kramer-double-pulsar}%
  \BibitemOpen
  \bibfield  {author} {\bibinfo {author} {\bibfnamefont {M.}~\bibnamefont
  {Kramer}}, \bibinfo {author} {\bibfnamefont {I.~H.}\ \bibnamefont {Stairs}},
  \bibinfo {author} {\bibfnamefont {R.}~\bibnamefont {Manchester}}, \bibinfo
  {author} {\bibfnamefont {M.}~\bibnamefont {McLaughlin}}, \bibinfo {author}
  {\bibfnamefont {A.}~\bibnamefont {Lyne}},  \emph {et~al.},\ }\href {\doibase
  10.1126/science.1132305} {\bibfield  {journal} {\bibinfo  {journal}
  {Science}\ }\textbf {\bibinfo {volume} {314}},\ \bibinfo {pages} {97}
  (\bibinfo {year} {2006})},\ \Eprint {http://arxiv.org/abs/astro-ph/0609417}
  {arXiv:astro-ph/0609417 [astro-ph]} \BibitemShut {NoStop}%
%%CITATION = ASTRO-PH/0609417;%%
\bibitem [{\citenamefont {Gair}\ \emph {et~al.}(2012)\citenamefont {Gair},
  \citenamefont {Vallisneri}, \citenamefont {Larson},\ and\ \citenamefont
  {Baker}}]{gair-living}%
  \BibitemOpen
  \bibfield  {author} {\bibinfo {author} {\bibfnamefont {J.~R.}\ \bibnamefont
  {Gair}}, \bibinfo {author} {\bibfnamefont {M.}~\bibnamefont {Vallisneri}},
  \bibinfo {author} {\bibfnamefont {S.~L.}\ \bibnamefont {Larson}}, \ and\
  \bibinfo {author} {\bibfnamefont {J.~G.}\ \bibnamefont {Baker}},\ }\href@noop
  {} {\  (\bibinfo {year} {2012})},\ \Eprint {http://arxiv.org/abs/1212.5575}
  {arXiv:1212.5575 [gr-qc]} \BibitemShut {NoStop}%
%%CITATION = ARXIV:1212.5575;%%
\bibitem [{\citenamefont {Damour}\ and\ \citenamefont
  {Esposito-Farese}(1998)}]{damour-GW-vs-pulsar}%
  \BibitemOpen
  \bibfield  {author} {\bibinfo {author} {\bibfnamefont {T.}~\bibnamefont
  {Damour}}\ and\ \bibinfo {author} {\bibfnamefont {G.}~\bibnamefont
  {Esposito-Farese}},\ }\href {\doibase 10.1103/PhysRevD.58.042001} {\bibfield
  {journal} {\bibinfo  {journal} {Phys.Rev.}\ }\textbf {\bibinfo {volume}
  {D58}},\ \bibinfo {pages} {042001} (\bibinfo {year} {1998})},\ \Eprint
  {http://arxiv.org/abs/gr-qc/9803031} {arXiv:gr-qc/9803031 [gr-qc]}
  \BibitemShut {NoStop}%
%%CITATION = GR-QC/9803031;%%
\bibitem [{\citenamefont {Fujii}\ and\ \citenamefont {Maeda}(2003)}]{fujii}%
  \BibitemOpen
  \bibfield  {author} {\bibinfo {author} {\bibfnamefont {Y.}~\bibnamefont
  {Fujii}}\ and\ \bibinfo {author} {\bibfnamefont {K.}~\bibnamefont {Maeda}},\
  }\href@noop {} {\  (\bibinfo {year} {2003})}\BibitemShut {NoStop}%
%%CITATION = INSPIRE-618647;%%
\bibitem [{\citenamefont {Eardley}(1975)}]{eardley}%
  \BibitemOpen
  \bibfield  {author} {\bibinfo {author} {\bibfnamefont {D.~M.}\ \bibnamefont
  {Eardley}},\ }\href@noop {} {\bibfield  {journal} {\bibinfo  {journal}
  {Astrophys. J.}\ }\textbf {\bibinfo {volume} {196}},\ \bibinfo {pages} {L59}
  (\bibinfo {year} {1975})}\BibitemShut {NoStop}%
%%CITATION = ASTRO-PH/0206404;%%
\bibitem [{\citenamefont {Will}(1977)}]{will1977}%
  \BibitemOpen
  \bibfield  {author} {\bibinfo {author} {\bibfnamefont {C.~M.}\ \bibnamefont
  {Will}},\ }\href {\doibase 10.1086/155313} {\bibfield  {journal} {\bibinfo
  {journal} {Astrophys. J.}\ }\textbf {\bibinfo {volume} {214}},\ \bibinfo
  {pages} {826} (\bibinfo {year} {1977})}\BibitemShut {NoStop}%
%%CITATION = ASJOA,214,826;%%
\bibitem [{\citenamefont {Will}\ and\ \citenamefont
  {Zaglauer}(1989)}]{zaglauer}%
  \BibitemOpen
  \bibfield  {author} {\bibinfo {author} {\bibfnamefont {C.~M.}\ \bibnamefont
  {Will}}\ and\ \bibinfo {author} {\bibfnamefont {H.~W.}\ \bibnamefont
  {Zaglauer}},\ }\href {\doibase 10.1086/168016} {\bibfield  {journal}
  {\bibinfo  {journal} {Astrophys. J.}\ }\textbf {\bibinfo {volume} {346}},\
  \bibinfo {pages} {366} (\bibinfo {year} {1989})}\BibitemShut {NoStop}%
%%CITATION = ASJOA,346,366;%%
\bibitem [{\citenamefont {Will}(1994)}]{will1994}%
  \BibitemOpen
  \bibfield  {author} {\bibinfo {author} {\bibfnamefont {C.~M.}\ \bibnamefont
  {Will}},\ }\href {\doibase 10.1103/PhysRevD.50.6058} {\bibfield  {journal}
  {\bibinfo  {journal} {Phys. Rev.}\ }\textbf {\bibinfo {volume} {D50}},\
  \bibinfo {pages} {6058} (\bibinfo {year} {1994})},\ \Eprint
  {http://arxiv.org/abs/gr-qc/9406022} {arXiv:gr-qc/9406022} \BibitemShut
  {NoStop}%
%%CITATION = GR-QC/9406022;%%
\bibitem [{\citenamefont {Arun}(2012)}]{arun-dipole}%
  \BibitemOpen
  \bibfield  {author} {\bibinfo {author} {\bibfnamefont {K.}~\bibnamefont
  {Arun}},\ }\href {\doibase 10.1088/0264-9381/29/7/075011} {\bibfield
  {journal} {\bibinfo  {journal} {Class.Quant.Grav.}\ }\textbf {\bibinfo
  {volume} {29}},\ \bibinfo {pages} {075011} (\bibinfo {year} {2012})},\
  \Eprint {http://arxiv.org/abs/1202.5911} {arXiv:1202.5911 [gr-qc]}
  \BibitemShut {NoStop}%
%%CITATION = ARXIV:1202.5911;%%
\bibitem [{\citenamefont {Will}(1998)}]{will1998}%
  \BibitemOpen
  \bibfield  {author} {\bibinfo {author} {\bibfnamefont {C.~M.}\ \bibnamefont
  {Will}},\ }\href {\doibase 10.1103/PhysRevD.57.2061} {\bibfield  {journal}
  {\bibinfo  {journal} {Phys. Rev.}\ }\textbf {\bibinfo {volume} {D57}},\
  \bibinfo {pages} {2061} (\bibinfo {year} {1998})},\ \Eprint
  {http://arxiv.org/abs/gr-qc/9709011} {arXiv:gr-qc/9709011} \BibitemShut
  {NoStop}%
%%CITATION = GR-QC/9709011;%%
\bibitem [{\citenamefont {Arun}\ and\ \citenamefont {Will}(2009)}]{arunwill}%
  \BibitemOpen
  \bibfield  {author} {\bibinfo {author} {\bibfnamefont {K.~G.}\ \bibnamefont
  {Arun}}\ and\ \bibinfo {author} {\bibfnamefont {C.~M.}\ \bibnamefont
  {Will}},\ }\href {\doibase 10.1088/0264-9381/26/15/155002} {\bibfield
  {journal} {\bibinfo  {journal} {Class. Quant. Grav.}\ }\textbf {\bibinfo
  {volume} {26}},\ \bibinfo {pages} {155002} (\bibinfo {year} {2009})},\
  \Eprint {http://arxiv.org/abs/0904.1190} {arXiv:0904.1190 [gr-qc]}
  \BibitemShut {NoStop}%
%%CITATION = 0904.1190;%%
\bibitem [{\citenamefont {Keppel}\ and\ \citenamefont {Ajith}(2010)}]{keppel}%
  \BibitemOpen
  \bibfield  {author} {\bibinfo {author} {\bibfnamefont {D.}~\bibnamefont
  {Keppel}}\ and\ \bibinfo {author} {\bibfnamefont {P.}~\bibnamefont {Ajith}},\
  }\href {\doibase 10.1103/PhysRevD.82.122001} {\bibfield  {journal} {\bibinfo
  {journal} {Phys. Rev.}\ }\textbf {\bibinfo {volume} {D82}},\ \bibinfo {pages}
  {122001} (\bibinfo {year} {2010})},\ \Eprint {http://arxiv.org/abs/1004.0284}
  {arXiv:1004.0284 [gr-qc]} \BibitemShut {NoStop}%
%%CITATION = 1004.0284;%%
\bibitem [{\citenamefont {Del~Pozzo}\ \emph {et~al.}(2011)\citenamefont
  {Del~Pozzo}, \citenamefont {Veitch},\ and\ \citenamefont
  {Vecchio}}]{delpozzo}%
  \BibitemOpen
  \bibfield  {author} {\bibinfo {author} {\bibfnamefont {W.}~\bibnamefont
  {Del~Pozzo}}, \bibinfo {author} {\bibfnamefont {J.}~\bibnamefont {Veitch}}, \
  and\ \bibinfo {author} {\bibfnamefont {A.}~\bibnamefont {Vecchio}},\ }\href
  {\doibase 10.1103/PhysRevD.83.082002} {\bibfield  {journal} {\bibinfo
  {journal} {Phys. Rev.}\ }\textbf {\bibinfo {volume} {D83}},\ \bibinfo {pages}
  {082002} (\bibinfo {year} {2011})},\ \Eprint {http://arxiv.org/abs/1101.1391}
  {arXiv:1101.1391 [gr-qc]} \BibitemShut {NoStop}%
%%CITATION = 1101.1391;%%
\bibitem [{\citenamefont {Cornish}\ \emph {et~al.}(2011)\citenamefont
  {Cornish}, \citenamefont {Sampson}, \citenamefont {Yunes},\ and\
  \citenamefont {Pretorius}}]{cornishsampson}%
  \BibitemOpen
  \bibfield  {author} {\bibinfo {author} {\bibfnamefont {N.}~\bibnamefont
  {Cornish}}, \bibinfo {author} {\bibfnamefont {L.}~\bibnamefont {Sampson}},
  \bibinfo {author} {\bibfnamefont {N.}~\bibnamefont {Yunes}}, \ and\ \bibinfo
  {author} {\bibfnamefont {F.}~\bibnamefont {Pretorius}},\ }\href {\doibase
  10.1103/PhysRevD.84.062003} {\bibfield  {journal} {\bibinfo  {journal}
  {Phys.Rev.}\ }\textbf {\bibinfo {volume} {D84}},\ \bibinfo {pages} {062003}
  (\bibinfo {year} {2011})},\ \Eprint {http://arxiv.org/abs/1105.2088}
  {arXiv:1105.2088 [gr-qc]} \BibitemShut {NoStop}%
%%CITATION = ARXIV:1105.2088;%%
\bibitem [{\citenamefont {Mirshekari}\ \emph {et~al.}(2012)\citenamefont
  {Mirshekari}, \citenamefont {Yunes},\ and\ \citenamefont
  {Will}}]{mirshekari}%
  \BibitemOpen
  \bibfield  {author} {\bibinfo {author} {\bibfnamefont {S.}~\bibnamefont
  {Mirshekari}}, \bibinfo {author} {\bibfnamefont {N.}~\bibnamefont {Yunes}}, \
  and\ \bibinfo {author} {\bibfnamefont {C.~M.}\ \bibnamefont {Will}},\ }\href
  {\doibase 10.1103/PhysRevD.85.024041} {\bibfield  {journal} {\bibinfo
  {journal} {Phys.Rev.}\ }\textbf {\bibinfo {volume} {D85}},\ \bibinfo {pages}
  {024041} (\bibinfo {year} {2012})},\ \Eprint {http://arxiv.org/abs/1110.2720}
  {arXiv:1110.2720 [gr-qc]} \BibitemShut {NoStop}%
%%CITATION = ARXIV:1110.2720;%%
\bibitem [{\citenamefont {Finn}\ and\ \citenamefont {Sutton}(2002)}]{sutton}%
  \BibitemOpen
  \bibfield  {author} {\bibinfo {author} {\bibfnamefont {L.~S.}\ \bibnamefont
  {Finn}}\ and\ \bibinfo {author} {\bibfnamefont {P.~J.}\ \bibnamefont
  {Sutton}},\ }\href {\doibase 10.1103/PhysRevD.65.044022} {\bibfield
  {journal} {\bibinfo  {journal} {Phys. Rev.}\ }\textbf {\bibinfo {volume}
  {D65}},\ \bibinfo {pages} {044022} (\bibinfo {year} {2002})},\ \Eprint
  {http://arxiv.org/abs/gr-qc/0109049} {arXiv:gr-qc/0109049} \BibitemShut
  {NoStop}%
%%CITATION = GR-QC/0109049;%%
\bibitem [{\citenamefont {Alexander}\ and\ \citenamefont
  {Yunes}(2009)}]{CSreview}%
  \BibitemOpen
  \bibfield  {author} {\bibinfo {author} {\bibfnamefont {S.}~\bibnamefont
  {Alexander}}\ and\ \bibinfo {author} {\bibfnamefont {N.}~\bibnamefont
  {Yunes}},\ }\href {\doibase 10.1016/j.physrep.2009.07.002} {\bibfield
  {journal} {\bibinfo  {journal} {Phys. Rept.}\ }\textbf {\bibinfo {volume}
  {480}},\ \bibinfo {pages} {1} (\bibinfo {year} {2009})},\ \Eprint
  {http://arxiv.org/abs/0907.2562} {arXiv:0907.2562 [hep-th]} \BibitemShut
  {NoStop}%
%%CITATION = 0907.2562;%%
\bibitem [{\citenamefont {Ali-Haimoud}\ and\ \citenamefont
  {Chen}(2011)}]{alihaimoud-chen}%
  \BibitemOpen
  \bibfield  {author} {\bibinfo {author} {\bibfnamefont {Y.}~\bibnamefont
  {Ali-Haimoud}}\ and\ \bibinfo {author} {\bibfnamefont {Y.}~\bibnamefont
  {Chen}},\ }\href@noop {} {\bibfield  {journal} {\bibinfo  {journal}
  {Phys.Rev.}\ }\textbf {\bibinfo {volume} {D84}},\ \bibinfo {pages} {124033}
  (\bibinfo {year} {2011})},\ \Eprint {http://arxiv.org/abs/1110.5329}
  {arXiv:1110.5329 [astro-ph.HE]} \BibitemShut {NoStop}%
%%CITATION = ARXIV:1110.5329;%%
\bibitem [{\citenamefont {Yagi}\ \emph
  {et~al.}(2012{\natexlab{a}})\citenamefont {Yagi}, \citenamefont {Yunes},\
  and\ \citenamefont {Tanaka}}]{kent-CSBH}%
  \BibitemOpen
  \bibfield  {author} {\bibinfo {author} {\bibfnamefont {K.}~\bibnamefont
  {Yagi}}, \bibinfo {author} {\bibfnamefont {N.}~\bibnamefont {Yunes}}, \ and\
  \bibinfo {author} {\bibfnamefont {T.}~\bibnamefont {Tanaka}},\ }\href
  {\doibase 10.1103/PhysRevD.86.044037} {\bibfield  {journal} {\bibinfo
  {journal} {Phys.Rev.}\ }\textbf {\bibinfo {volume} {D86}},\ \bibinfo {pages}
  {044037} (\bibinfo {year} {2012}{\natexlab{a}})},\ \Eprint
  {http://arxiv.org/abs/1206.6130} {arXiv:1206.6130 [gr-qc]} \BibitemShut
  {NoStop}%
%%CITATION = ARXIV:1206.6130;%%
\bibitem [{\citenamefont {Sopuerta}\ and\ \citenamefont
  {Yunes}(2009)}]{sopuerta-yunes-DCS-EMRI}%
  \BibitemOpen
  \bibfield  {author} {\bibinfo {author} {\bibfnamefont {C.~F.}\ \bibnamefont
  {Sopuerta}}\ and\ \bibinfo {author} {\bibfnamefont {N.}~\bibnamefont
  {Yunes}},\ }\href {\doibase 10.1103/PhysRevD.80.064006} {\bibfield  {journal}
  {\bibinfo  {journal} {Phys.Rev.}\ }\textbf {\bibinfo {volume} {D80}},\
  \bibinfo {pages} {064006} (\bibinfo {year} {2009})},\ \Eprint
  {http://arxiv.org/abs/0904.4501} {arXiv:0904.4501 [gr-qc]} \BibitemShut
  {NoStop}%
\bibitem [{\citenamefont {Pani}\ \emph
  {et~al.}(2011{\natexlab{a}})\citenamefont {Pani}, \citenamefont {Cardoso},\
  and\ \citenamefont {Gualtieri}}]{pani-DCS-EMRI}%
  \BibitemOpen
  \bibfield  {author} {\bibinfo {author} {\bibfnamefont {P.}~\bibnamefont
  {Pani}}, \bibinfo {author} {\bibfnamefont {V.}~\bibnamefont {Cardoso}}, \
  and\ \bibinfo {author} {\bibfnamefont {L.}~\bibnamefont {Gualtieri}},\ }\href
  {\doibase 10.1103/PhysRevD.83.104048} {\bibfield  {journal} {\bibinfo
  {journal} {Phys. Rev.}\ }\textbf {\bibinfo {volume} {D83}},\ \bibinfo {pages}
  {104048} (\bibinfo {year} {2011}{\natexlab{a}})},\ \Eprint
  {http://arxiv.org/abs/1104.1183} {arXiv:1104.1183 [gr-qc]} \BibitemShut
  {NoStop}%
%%CITATION = 1104.1183;%%
\bibitem [{\citenamefont {Canizares}\ \emph {et~al.}(2012)\citenamefont
  {Canizares}, \citenamefont {Gair},\ and\ \citenamefont
  {Sopuerta}}]{canizares}%
  \BibitemOpen
  \bibfield  {author} {\bibinfo {author} {\bibfnamefont {P.}~\bibnamefont
  {Canizares}}, \bibinfo {author} {\bibfnamefont {J.~R.}\ \bibnamefont {Gair}},
  \ and\ \bibinfo {author} {\bibfnamefont {C.~F.}\ \bibnamefont {Sopuerta}},\
  }\href@noop {} {\  (\bibinfo {year} {2012})},\ \Eprint
  {http://arxiv.org/abs/1205.1253} {arXiv:1205.1253 [gr-qc]} \BibitemShut
  {NoStop}%
%%CITATION = ARXIV:1205.1253;%%
\bibitem [{\citenamefont {Yagi}\ \emph
  {et~al.}(2012{\natexlab{b}})\citenamefont {Yagi}, \citenamefont {Yunes},\
  and\ \citenamefont {Tanaka}}]{kent-CSGW}%
  \BibitemOpen
  \bibfield  {author} {\bibinfo {author} {\bibfnamefont {K.}~\bibnamefont
  {Yagi}}, \bibinfo {author} {\bibfnamefont {N.}~\bibnamefont {Yunes}}, \ and\
  \bibinfo {author} {\bibfnamefont {T.}~\bibnamefont {Tanaka}},\ }\href@noop {}
  {\  (\bibinfo {year} {2012}{\natexlab{b}})},\ \Eprint
  {http://arxiv.org/abs/1208.5102} {arXiv:1208.5102 [gr-qc]} \BibitemShut
  {NoStop}%
%%CITATION = ARXIV:1208.5102;%%
\bibitem [{\citenamefont {Polchinski}(1998{\natexlab{a}})}]{polchinski2}%
  \BibitemOpen
  \bibfield  {author} {\bibinfo {author} {\bibfnamefont {J.}~\bibnamefont
  {Polchinski}},\ }\href@noop {} {\emph {\bibinfo {title} {String theory. Vol.
  2: Superstring theory and beyond}}}\ (\bibinfo  {publisher} {Cambridge
  University Press},\ \bibinfo {address} {Cambridge, UK},\ \bibinfo {year}
  {1998})\BibitemShut {NoStop}%
\bibitem [{\citenamefont {Polchinski}(1998{\natexlab{b}})}]{polchinski1}%
  \BibitemOpen
  \bibfield  {author} {\bibinfo {author} {\bibfnamefont {J.}~\bibnamefont
  {Polchinski}},\ }\href@noop {} {\emph {\bibinfo {title} {String theory. Vol.
  1: An introduction to the bosonic string}}}\ (\bibinfo  {publisher}
  {Cambridge University Press},\ \bibinfo {address} {Cambridge, UK},\ \bibinfo
  {year} {1998})\BibitemShut {NoStop}%
\bibitem [{\citenamefont {Alexander}\ and\ \citenamefont
  {Gates}(2006)}]{alexandergates}%
  \BibitemOpen
  \bibfield  {author} {\bibinfo {author} {\bibfnamefont {S.~H.~S.}\
  \bibnamefont {Alexander}}\ and\ \bibinfo {author} {\bibfnamefont {S.~J.}\
  \bibnamefont {Gates}, \bibfnamefont {Jr.}},\ }\href {\doibase
  10.1088/1475-7516/2006/06/018} {\bibfield  {journal} {\bibinfo  {journal}
  {JCAP}\ }\textbf {\bibinfo {volume} {0606}},\ \bibinfo {pages} {018}
  (\bibinfo {year} {2006})},\ \Eprint {http://arxiv.org/abs/hep-th/0409014}
  {arXiv:hep-th/0409014} \BibitemShut {NoStop}%
%%CITATION = HEP-TH/0409014;%%
\bibitem [{\citenamefont {Taveras}\ and\ \citenamefont
  {Yunes}(2008)}]{taveras}%
  \BibitemOpen
  \bibfield  {author} {\bibinfo {author} {\bibfnamefont {V.}~\bibnamefont
  {Taveras}}\ and\ \bibinfo {author} {\bibfnamefont {N.}~\bibnamefont
  {Yunes}},\ }\href {\doibase 10.1103/PhysRevD.78.064070} {\bibfield  {journal}
  {\bibinfo  {journal} {Phys. Rev.}\ }\textbf {\bibinfo {volume} {D78}},\
  \bibinfo {pages} {064070} (\bibinfo {year} {2008})},\ \Eprint
  {http://arxiv.org/abs/0807.2652} {arXiv:0807.2652 [gr-qc]} \BibitemShut
  {NoStop}%
%%CITATION = 0807.2652;%%
\bibitem [{\citenamefont {Calcagni}\ and\ \citenamefont
  {Mercuri}(2009)}]{calcagni}%
  \BibitemOpen
  \bibfield  {author} {\bibinfo {author} {\bibfnamefont {G.}~\bibnamefont
  {Calcagni}}\ and\ \bibinfo {author} {\bibfnamefont {S.}~\bibnamefont
  {Mercuri}},\ }\href {\doibase 10.1103/PhysRevD.79.084004} {\bibfield
  {journal} {\bibinfo  {journal} {Phys. Rev.}\ }\textbf {\bibinfo {volume}
  {D79}},\ \bibinfo {pages} {084004} (\bibinfo {year} {2009})},\ \Eprint
  {http://arxiv.org/abs/0902.0957} {arXiv:0902.0957 [gr-qc]} \BibitemShut
  {NoStop}%
%%CITATION = 0902.0957;%%
\bibitem [{\citenamefont {Mercuri}\ and\ \citenamefont
  {Taveras}(2009)}]{Mercuri:2009zt}%
  \BibitemOpen
  \bibfield  {author} {\bibinfo {author} {\bibfnamefont {S.}~\bibnamefont
  {Mercuri}}\ and\ \bibinfo {author} {\bibfnamefont {V.}~\bibnamefont
  {Taveras}},\ }\href {\doibase 10.1103/PhysRevD.80.104007} {\bibfield
  {journal} {\bibinfo  {journal} {Phys. Rev.}\ }\textbf {\bibinfo {volume}
  {D80}},\ \bibinfo {pages} {104007} (\bibinfo {year} {2009})},\ \Eprint
  {http://arxiv.org/abs/0903.4407} {arXiv:0903.4407 [gr-qc]} \BibitemShut
  {NoStop}%
%%CITATION = 0903.4407;%%
\bibitem [{\citenamefont {Weinberg}(2008)}]{weinberg-CS}%
  \BibitemOpen
  \bibfield  {author} {\bibinfo {author} {\bibfnamefont {S.}~\bibnamefont
  {Weinberg}},\ }\href {\doibase 10.1103/PhysRevD.77.123541} {\bibfield
  {journal} {\bibinfo  {journal} {Phys. Rev.}\ }\textbf {\bibinfo {volume}
  {D77}},\ \bibinfo {pages} {123541} (\bibinfo {year} {2008})},\ \Eprint
  {http://arxiv.org/abs/0804.4291} {arXiv:0804.4291 [hep-th]} \BibitemShut
  {NoStop}%
%%CITATION = 0804.4291;%%
\bibitem [{\citenamefont {Jackiw}\ and\ \citenamefont {Pi}(2003)}]{jackiw}%
  \BibitemOpen
  \bibfield  {author} {\bibinfo {author} {\bibfnamefont {R.}~\bibnamefont
  {Jackiw}}\ and\ \bibinfo {author} {\bibfnamefont {S.~Y.}\ \bibnamefont
  {Pi}},\ }\href {\doibase 10.1103/PhysRevD.68.104012} {\bibfield  {journal}
  {\bibinfo  {journal} {Phys. Rev.}\ }\textbf {\bibinfo {volume} {D68}},\
  \bibinfo {pages} {104012} (\bibinfo {year} {2003})},\ \Eprint
  {http://arxiv.org/abs/gr-qc/0308071} {arXiv:gr-qc/0308071} \BibitemShut
  {NoStop}%
%%CITATION = GR-QC/0308071;%%
\bibitem [{\citenamefont {Alexander}\ \emph {et~al.}(2008)\citenamefont
  {Alexander}, \citenamefont {Finn},\ and\ \citenamefont
  {Yunes}}]{Alexander:2007kv}%
  \BibitemOpen
  \bibfield  {author} {\bibinfo {author} {\bibfnamefont {S.}~\bibnamefont
  {Alexander}}, \bibinfo {author} {\bibfnamefont {L.~S.}\ \bibnamefont {Finn}},
  \ and\ \bibinfo {author} {\bibfnamefont {N.}~\bibnamefont {Yunes}},\ }\href
  {\doibase 10.1103/PhysRevD.78.066005} {\bibfield  {journal} {\bibinfo
  {journal} {Phys. Rev.}\ }\textbf {\bibinfo {volume} {D78}},\ \bibinfo {pages}
  {066005} (\bibinfo {year} {2008})},\ \Eprint {http://arxiv.org/abs/0712.2542}
  {arXiv:0712.2542 [gr-qc]} \BibitemShut {NoStop}%
%%CITATION = 0712.2542;%%
\bibitem [{\citenamefont {Smith}\ \emph {et~al.}(2008)\citenamefont {Smith},
  \citenamefont {Erickcek}, \citenamefont {Caldwell},\ and\ \citenamefont
  {Kamionkowski}}]{Smith:2007jm}%
  \BibitemOpen
  \bibfield  {author} {\bibinfo {author} {\bibfnamefont {T.~L.}\ \bibnamefont
  {Smith}}, \bibinfo {author} {\bibfnamefont {A.~L.}\ \bibnamefont {Erickcek}},
  \bibinfo {author} {\bibfnamefont {R.~R.}\ \bibnamefont {Caldwell}}, \ and\
  \bibinfo {author} {\bibfnamefont {M.}~\bibnamefont {Kamionkowski}},\ }\href
  {\doibase 10.1103/PhysRevD.77.024015} {\bibfield  {journal} {\bibinfo
  {journal} {Phys. Rev.}\ }\textbf {\bibinfo {volume} {D77}},\ \bibinfo {pages}
  {024015} (\bibinfo {year} {2008})},\ \Eprint {http://arxiv.org/abs/0708.0001}
  {arXiv:0708.0001 [astro-ph]} \BibitemShut {NoStop}%
%%CITATION = 0708.0001;%%
\bibitem [{\citenamefont {Yunes}\ and\ \citenamefont
  {Pretorius}(2009)}]{yunespretorius}%
  \BibitemOpen
  \bibfield  {author} {\bibinfo {author} {\bibfnamefont {N.}~\bibnamefont
  {Yunes}}\ and\ \bibinfo {author} {\bibfnamefont {F.}~\bibnamefont
  {Pretorius}},\ }\href {\doibase 10.1103/PhysRevD.79.084043} {\bibfield
  {journal} {\bibinfo  {journal} {Phys. Rev.}\ }\textbf {\bibinfo {volume}
  {D79}},\ \bibinfo {pages} {084043} (\bibinfo {year} {2009})},\ \Eprint
  {http://arxiv.org/abs/0902.4669} {arXiv:0902.4669 [gr-qc]} \BibitemShut
  {NoStop}%
%%CITATION = 0902.4669;%%
\bibitem [{\citenamefont {Konno}\ \emph {et~al.}(2009)\citenamefont {Konno},
  \citenamefont {Matsuyama},\ and\ \citenamefont {Tanda}}]{konnoBH}%
  \BibitemOpen
  \bibfield  {author} {\bibinfo {author} {\bibfnamefont {K.}~\bibnamefont
  {Konno}}, \bibinfo {author} {\bibfnamefont {T.}~\bibnamefont {Matsuyama}}, \
  and\ \bibinfo {author} {\bibfnamefont {S.}~\bibnamefont {Tanda}},\ }\href
  {\doibase 10.1143/PTP.122.561} {\bibfield  {journal} {\bibinfo  {journal}
  {Prog.Theor.Phys.}\ }\textbf {\bibinfo {volume} {122}},\ \bibinfo {pages}
  {561} (\bibinfo {year} {2009})},\ \Eprint {http://arxiv.org/abs/0902.4767}
  {arXiv:0902.4767 [gr-qc]} \BibitemShut {NoStop}%
%%CITATION = ARXIV:0902.4767;%%
\bibitem [{\citenamefont {Yunes}\ \emph
  {et~al.}(2010{\natexlab{a}})\citenamefont {Yunes}, \citenamefont {Psaltis},
  \citenamefont {Ozel},\ and\ \citenamefont {Loeb}}]{yunes-CSNS}%
  \BibitemOpen
  \bibfield  {author} {\bibinfo {author} {\bibfnamefont {N.}~\bibnamefont
  {Yunes}}, \bibinfo {author} {\bibfnamefont {D.}~\bibnamefont {Psaltis}},
  \bibinfo {author} {\bibfnamefont {F.}~\bibnamefont {Ozel}}, \ and\ \bibinfo
  {author} {\bibfnamefont {A.}~\bibnamefont {Loeb}},\ }\href {\doibase
  10.1103/PhysRevD.81.064020} {\bibfield  {journal} {\bibinfo  {journal}
  {Phys.Rev.}\ }\textbf {\bibinfo {volume} {D81}},\ \bibinfo {pages} {064020}
  (\bibinfo {year} {2010}{\natexlab{a}})},\ \Eprint
  {http://arxiv.org/abs/0912.2736} {arXiv:0912.2736 [gr-qc]} \BibitemShut
  {NoStop}%
%%CITATION = ARXIV:0912.2736;%%
\bibitem [{\citenamefont {{Demorest}}\ \emph {et~al.}(2010)\citenamefont
  {{Demorest}}, \citenamefont {{Pennucci}}, \citenamefont {{Ransom}},
  \citenamefont {{Roberts}},\ and\ \citenamefont {{Hessels}}}]{1.97NS}%
  \BibitemOpen
  \bibfield  {author} {\bibinfo {author} {\bibfnamefont {P.~B.}\ \bibnamefont
  {{Demorest}}}, \bibinfo {author} {\bibfnamefont {T.}~\bibnamefont
  {{Pennucci}}}, \bibinfo {author} {\bibfnamefont {S.~M.}\ \bibnamefont
  {{Ransom}}}, \bibinfo {author} {\bibfnamefont {M.~S.~E.}\ \bibnamefont
  {{Roberts}}}, \ and\ \bibinfo {author} {\bibfnamefont {J.~W.~T.}\
  \bibnamefont {{Hessels}}},\ }\href {\doibase 10.1038/nature09466} {\bibfield
  {journal} {\bibinfo  {journal} {Nature}\ }\textbf {\bibinfo {volume} {467}},\
  \bibinfo {pages} {1081} (\bibinfo {year} {2010})},\ \Eprint
  {http://arxiv.org/abs/1010.5788} {arXiv:1010.5788 [astro-ph.HE]} \BibitemShut
  {NoStop}%
\bibitem [{\citenamefont {Hartle}(1967)}]{hartle1967}%
  \BibitemOpen
  \bibfield  {author} {\bibinfo {author} {\bibfnamefont {J.~B.}\ \bibnamefont
  {Hartle}},\ }\href@noop {} {\bibfield  {journal} {\bibinfo  {journal}
  {Astrophys.J.}\ }\textbf {\bibinfo {volume} {150}},\ \bibinfo {pages} {1005}
  (\bibinfo {year} {1967})}\BibitemShut {NoStop}%
%%CITATION = ASJOA,150,1005;%%
\bibitem [{\citenamefont {Akmal}\ \emph {et~al.}(1998)\citenamefont {Akmal},
  \citenamefont {Pandharipande},\ and\ \citenamefont {Ravenhall}}]{APR}%
  \BibitemOpen
  \bibfield  {author} {\bibinfo {author} {\bibfnamefont {A.}~\bibnamefont
  {Akmal}}, \bibinfo {author} {\bibfnamefont {V.}~\bibnamefont
  {Pandharipande}}, \ and\ \bibinfo {author} {\bibfnamefont {D.}~\bibnamefont
  {Ravenhall}},\ }\href {\doibase 10.1103/PhysRevC.58.1804} {\bibfield
  {journal} {\bibinfo  {journal} {Phys.Rev.}\ }\textbf {\bibinfo {volume}
  {C58}},\ \bibinfo {pages} {1804} (\bibinfo {year} {1998})},\ \Eprint
  {http://arxiv.org/abs/nucl-th/9804027} {arXiv:nucl-th/9804027 [nucl-th]}
  \BibitemShut {NoStop}%
%%CITATION = NUCL-TH/9804027;%%
\bibitem [{\citenamefont {{Douchin}}\ and\ \citenamefont
  {{Haensel}}(2001)}]{SLy}%
  \BibitemOpen
  \bibfield  {author} {\bibinfo {author} {\bibfnamefont {F.}~\bibnamefont
  {{Douchin}}}\ and\ \bibinfo {author} {\bibfnamefont {P.}~\bibnamefont
  {{Haensel}}},\ }\href {\doibase 10.1051/0004-6361:20011402} {\bibfield
  {journal} {\bibinfo  {journal} {Astron. Astrophys.}\ }\textbf {\bibinfo
  {volume} {380}},\ \bibinfo {pages} {151} (\bibinfo {year} {2001})},\ \Eprint
  {http://arxiv.org/abs/arXiv:astro-ph/0111092} {arXiv:astro-ph/0111092}
  \BibitemShut {NoStop}%
\bibitem [{\citenamefont {Shibata}\ \emph {et~al.}(2005)\citenamefont
  {Shibata}, \citenamefont {Taniguchi},\ and\ \citenamefont
  {Uryu}}]{shibata-fitting}%
  \BibitemOpen
  \bibfield  {author} {\bibinfo {author} {\bibfnamefont {M.}~\bibnamefont
  {Shibata}}, \bibinfo {author} {\bibfnamefont {K.}~\bibnamefont {Taniguchi}},
  \ and\ \bibinfo {author} {\bibfnamefont {K.}~\bibnamefont {Uryu}},\ }\href
  {\doibase 10.1103/PhysRevD.71.084021} {\bibfield  {journal} {\bibinfo
  {journal} {Phys.Rev.}\ }\textbf {\bibinfo {volume} {D71}},\ \bibinfo {pages}
  {084021} (\bibinfo {year} {2005})},\ \Eprint
  {http://arxiv.org/abs/gr-qc/0503119} {arXiv:gr-qc/0503119 [gr-qc]}
  \BibitemShut {NoStop}%
%%CITATION = GR-QC/0503119;%%
\bibitem [{\citenamefont {{Lattimer}}\ and\ \citenamefont {{Douglas
  Swesty}}(1991)}]{LS}%
  \BibitemOpen
  \bibfield  {author} {\bibinfo {author} {\bibfnamefont {J.~M.}\ \bibnamefont
  {{Lattimer}}}\ and\ \bibinfo {author} {\bibfnamefont {F.}~\bibnamefont
  {{Douglas Swesty}}},\ }\href {\doibase 10.1016/0375-9474(91)90452-C}
  {\bibfield  {journal} {\bibinfo  {journal} {Nuclear Physics A}\ }\textbf
  {\bibinfo {volume} {535}},\ \bibinfo {pages} {331} (\bibinfo {year}
  {1991})}\BibitemShut {NoStop}%
\bibitem [{\citenamefont {O'Connor}\ and\ \citenamefont {Ott}(2010)}]{ott-EOS}%
  \BibitemOpen
  \bibfield  {author} {\bibinfo {author} {\bibfnamefont {E.}~\bibnamefont
  {O'Connor}}\ and\ \bibinfo {author} {\bibfnamefont {C.~D.}\ \bibnamefont
  {Ott}},\ }\href {\doibase 10.1088/0264-9381/27/11/114103} {\bibfield
  {journal} {\bibinfo  {journal} {Class.Quant.Grav.}\ }\textbf {\bibinfo
  {volume} {27}},\ \bibinfo {pages} {114103} (\bibinfo {year} {2010})},\
  \Eprint {http://arxiv.org/abs/0912.2393} {arXiv:0912.2393 [astro-ph.HE]}
  \BibitemShut {NoStop}%
%%CITATION = ARXIV:0912.2393;%%
\bibitem [{\citenamefont {{Shen}}\ \emph
  {et~al.}(1998{\natexlab{a}})\citenamefont {{Shen}}, \citenamefont {{Toki}},
  \citenamefont {{Oyamatsu}},\ and\ \citenamefont {{Sumiyoshi}}}]{Shen1}%
  \BibitemOpen
  \bibfield  {author} {\bibinfo {author} {\bibfnamefont {H.}~\bibnamefont
  {{Shen}}}, \bibinfo {author} {\bibfnamefont {H.}~\bibnamefont {{Toki}}},
  \bibinfo {author} {\bibfnamefont {K.}~\bibnamefont {{Oyamatsu}}}, \ and\
  \bibinfo {author} {\bibfnamefont {K.}~\bibnamefont {{Sumiyoshi}}},\ }\href
  {\doibase 10.1016/S0375-9474(98)00236-X} {\bibfield  {journal} {\bibinfo
  {journal} {Nuclear Physics A}\ }\textbf {\bibinfo {volume} {637}},\ \bibinfo
  {pages} {435} (\bibinfo {year} {1998}{\natexlab{a}})},\ \Eprint
  {http://arxiv.org/abs/arXiv:nucl-th/9805035} {arXiv:nucl-th/9805035}
  \BibitemShut {NoStop}%
\bibitem [{\citenamefont {{Shen}}\ \emph
  {et~al.}(1998{\natexlab{b}})\citenamefont {{Shen}}, \citenamefont {{Toki}},
  \citenamefont {{Oyamatsu}},\ and\ \citenamefont {{Sumiyoshi}}}]{Shen2}%
  \BibitemOpen
  \bibfield  {author} {\bibinfo {author} {\bibfnamefont {H.}~\bibnamefont
  {{Shen}}}, \bibinfo {author} {\bibfnamefont {H.}~\bibnamefont {{Toki}}},
  \bibinfo {author} {\bibfnamefont {K.}~\bibnamefont {{Oyamatsu}}}, \ and\
  \bibinfo {author} {\bibfnamefont {K.}~\bibnamefont {{Sumiyoshi}}},\ }\href
  {\doibase 10.1143/PTP.100.1013} {\bibfield  {journal} {\bibinfo  {journal}
  {Progress of Theoretical Physics}\ }\textbf {\bibinfo {volume} {100}},\
  \bibinfo {pages} {1013} (\bibinfo {year} {1998}{\natexlab{b}})},\ \Eprint
  {http://arxiv.org/abs/arXiv:nucl-th/9806095} {arXiv:nucl-th/9806095}
  \BibitemShut {NoStop}%
\bibitem [{\citenamefont {Damour}\ and\ \citenamefont
  {Taylor}(1992)}]{damour-taylor}%
  \BibitemOpen
  \bibfield  {author} {\bibinfo {author} {\bibfnamefont {T.}~\bibnamefont
  {Damour}}\ and\ \bibinfo {author} {\bibfnamefont {J.~H.}\ \bibnamefont
  {Taylor}},\ }\href {\doibase 10.1103/PhysRevD.45.1840} {\bibfield  {journal}
  {\bibinfo  {journal} {Phys.Rev.}\ }\textbf {\bibinfo {volume} {D45}},\
  \bibinfo {pages} {1840} (\bibinfo {year} {1992})}\BibitemShut {NoStop}%
%%CITATION = PHRVA,D45,1840;%%
\bibitem [{\citenamefont {Misner}\ \emph {et~al.}(1973)\citenamefont {Misner},
  \citenamefont {Thorne},\ and\ \citenamefont {Wheeler}}]{MTW}%
  \BibitemOpen
  \bibfield  {author} {\bibinfo {author} {\bibfnamefont {C.~W.}\ \bibnamefont
  {Misner}}, \bibinfo {author} {\bibfnamefont {K.}~\bibnamefont {Thorne}}, \
  and\ \bibinfo {author} {\bibfnamefont {J.~A.}\ \bibnamefont {Wheeler}},\
  }\href@noop {} {\emph {\bibinfo {title} {Gravitation}}}\ (\bibinfo
  {publisher} {W. H. Freeman \& Co.},\ \bibinfo {address} {San Francisco},\
  \bibinfo {year} {1973})\BibitemShut {NoStop}%
\bibitem [{\citenamefont {{Yagi}}\ \emph {et~al.}(2012)\citenamefont {{Yagi}},
  \citenamefont {{Stein}}, \citenamefont {{Yunes}},\ and\ \citenamefont
  {{Tanaka}}}]{quadratic}%
  \BibitemOpen
  \bibfield  {author} {\bibinfo {author} {\bibfnamefont {K.}~\bibnamefont
  {{Yagi}}}, \bibinfo {author} {\bibfnamefont {L.~C.}\ \bibnamefont {{Stein}}},
  \bibinfo {author} {\bibfnamefont {N.}~\bibnamefont {{Yunes}}}, \ and\
  \bibinfo {author} {\bibfnamefont {T.}~\bibnamefont {{Tanaka}}},\ }\href
  {\doibase 10.1103/PhysRevD.85.064022} {\bibfield  {journal} {\bibinfo
  {journal} {Phys. Rev.}\ }\textbf {\bibinfo {volume} {D85}},\ \bibinfo {eid}
  {064022} (\bibinfo {year} {2012})},\ \Eprint {http://arxiv.org/abs/1110.5950}
  {arXiv:1110.5950 [gr-qc]} \BibitemShut {NoStop}%
\bibitem [{\citenamefont {Dyda}\ \emph {et~al.}(2012)\citenamefont {Dyda},
  \citenamefont {Flanagan},\ and\ \citenamefont {Kamionkowski}}]{Dyda:2012rj}%
  \BibitemOpen
  \bibfield  {author} {\bibinfo {author} {\bibfnamefont {S.}~\bibnamefont
  {Dyda}}, \bibinfo {author} {\bibfnamefont {E.~E.}\ \bibnamefont {Flanagan}},
  \ and\ \bibinfo {author} {\bibfnamefont {M.}~\bibnamefont {Kamionkowski}},\
  }\href {\doibase 10.1103/PhysRevD.86.124031} {\bibfield  {journal} {\bibinfo
  {journal} {Phys.Rev.}\ }\textbf {\bibinfo {volume} {D86}},\ \bibinfo {pages}
  {124031} (\bibinfo {year} {2012})},\ \Eprint {http://arxiv.org/abs/1208.4871}
  {arXiv:1208.4871 [gr-qc]} \BibitemShut {NoStop}%
%%CITATION = ARXIV:1208.4871;%%
\bibitem [{\citenamefont {Poisson}(1998)}]{poisson-quadrupole}%
  \BibitemOpen
  \bibfield  {author} {\bibinfo {author} {\bibfnamefont {E.}~\bibnamefont
  {Poisson}},\ }\href {\doibase 10.1103/PhysRevD.57.5287} {\bibfield  {journal}
  {\bibinfo  {journal} {Phys.Rev.}\ }\textbf {\bibinfo {volume} {D57}},\
  \bibinfo {pages} {5287} (\bibinfo {year} {1998})},\ \Eprint
  {http://arxiv.org/abs/gr-qc/9709032} {arXiv:gr-qc/9709032 [gr-qc]}
  \BibitemShut {NoStop}%
%%CITATION = GR-QC/9709032;%%
\bibitem [{\citenamefont {Laarakkers}\ and\ \citenamefont
  {Poisson}(1999)}]{laarakkers}%
  \BibitemOpen
  \bibfield  {author} {\bibinfo {author} {\bibfnamefont {W.~G.}\ \bibnamefont
  {Laarakkers}}\ and\ \bibinfo {author} {\bibfnamefont {E.}~\bibnamefont
  {Poisson}},\ }\href {\doibase 10.1086/306732} {\bibfield  {journal} {\bibinfo
   {journal} {Astrophys.J.}\ }\textbf {\bibinfo {volume} {512}},\ \bibinfo
  {pages} {282} (\bibinfo {year} {1999})},\ \Eprint
  {http://arxiv.org/abs/gr-qc/9709033} {arXiv:gr-qc/9709033 [gr-qc]}
  \BibitemShut {NoStop}%
%%CITATION = GR-QC/9709033;%%
\bibitem [{\citenamefont {Kalogera}\ and\ \citenamefont
  {Psaltis}(2000)}]{kalogera-psaltis}%
  \BibitemOpen
  \bibfield  {author} {\bibinfo {author} {\bibfnamefont {V.}~\bibnamefont
  {Kalogera}}\ and\ \bibinfo {author} {\bibfnamefont {D.}~\bibnamefont
  {Psaltis}},\ }\href {\doibase 10.1103/PhysRevD.61.024009} {\bibfield
  {journal} {\bibinfo  {journal} {Phys.Rev.}\ }\textbf {\bibinfo {volume}
  {D61}},\ \bibinfo {pages} {024009} (\bibinfo {year} {2000})},\ \Eprint
  {http://arxiv.org/abs/astro-ph/9903415} {arXiv:astro-ph/9903415 [astro-ph]}
  \BibitemShut {NoStop}%
%%CITATION = ASTRO-PH/9903415;%%
\bibitem [{\citenamefont {Galassi}\ \emph {et~al.}(2009)\citenamefont {Galassi}
  \emph {et~al.}}]{gsl}%
  \BibitemOpen
  \bibfield  {author} {\bibinfo {author} {\bibfnamefont {M.}~\bibnamefont
  {Galassi}} \emph {et~al.},\ }\href@noop {} {\emph {\bibinfo {title} {GNU
  Scientific Library Reference Manual}}}\ (\bibinfo  {publisher} {Network
  Theory Ltd.; 3rd Revised edition},\ \bibinfo {year} {2009})\BibitemShut
  {NoStop}%
\bibitem [{\citenamefont {Yunes}\ \emph
  {et~al.}(2010{\natexlab{b}})\citenamefont {Yunes}, \citenamefont {Psaltis},
  \citenamefont {Ozel},\ and\ \citenamefont {Loeb}}]{yunespsaltis}%
  \BibitemOpen
  \bibfield  {author} {\bibinfo {author} {\bibfnamefont {N.}~\bibnamefont
  {Yunes}}, \bibinfo {author} {\bibfnamefont {D.}~\bibnamefont {Psaltis}},
  \bibinfo {author} {\bibfnamefont {F.}~\bibnamefont {Ozel}}, \ and\ \bibinfo
  {author} {\bibfnamefont {A.}~\bibnamefont {Loeb}},\ }\href {\doibase
  10.1103/PhysRevD.81.064020} {\bibfield  {journal} {\bibinfo  {journal} {Phys.
  Rev.}\ }\textbf {\bibinfo {volume} {D81}},\ \bibinfo {pages} {064020}
  (\bibinfo {year} {2010}{\natexlab{b}})},\ \Eprint
  {http://arxiv.org/abs/0912.2736} {arXiv:0912.2736 [gr-qc]} \BibitemShut
  {NoStop}%
%%CITATION = 0912.2736;%%
\bibitem [{\citenamefont {{Dieci}}\ \emph
  {et~al.}(1988{\natexlab{a}})\citenamefont {{Dieci}}, \citenamefont
  {{Osborne}},\ and\ \citenamefont {{Russell}}}]{dieci1}%
  \BibitemOpen
  \bibfield  {author} {\bibinfo {author} {\bibfnamefont {L.}~\bibnamefont
  {{Dieci}}}, \bibinfo {author} {\bibfnamefont {M.~R.}\ \bibnamefont
  {{Osborne}}}, \ and\ \bibinfo {author} {\bibfnamefont {R.~D.}\ \bibnamefont
  {{Russell}}},\ }\href {\doibase 10.1137/0725061} {\bibfield  {journal}
  {\bibinfo  {journal} {SIAM Journal on Numerical Analysis}\ }\textbf {\bibinfo
  {volume} {25}},\ \bibinfo {pages} {1055} (\bibinfo {year}
  {1988}{\natexlab{a}})}\BibitemShut {NoStop}%
\bibitem [{\citenamefont {{Dieci}}\ \emph
  {et~al.}(1988{\natexlab{b}})\citenamefont {{Dieci}}, \citenamefont
  {{Osborne}},\ and\ \citenamefont {{Russell}}}]{dieci2}%
  \BibitemOpen
  \bibfield  {author} {\bibinfo {author} {\bibfnamefont {L.}~\bibnamefont
  {{Dieci}}}, \bibinfo {author} {\bibfnamefont {M.~R.}\ \bibnamefont
  {{Osborne}}}, \ and\ \bibinfo {author} {\bibfnamefont {R.~D.}\ \bibnamefont
  {{Russell}}},\ }\href {\doibase 10.1137/0725062} {\bibfield  {journal}
  {\bibinfo  {journal} {SIAM Journal on Numerical Analysis}\ }\textbf {\bibinfo
  {volume} {25}},\ \bibinfo {pages} {1074} (\bibinfo {year}
  {1988}{\natexlab{b}})}\BibitemShut {NoStop}%
\bibitem [{\citenamefont {{Takata}}\ and\ \citenamefont
  {{L{\"o}ffler}}(2004)}]{takata}%
  \BibitemOpen
  \bibfield  {author} {\bibinfo {author} {\bibfnamefont {M.}~\bibnamefont
  {{Takata}}}\ and\ \bibinfo {author} {\bibfnamefont {W.}~\bibnamefont
  {{L{\"o}ffler}}},\ }\href@noop {} {\bibfield  {journal} {\bibinfo  {journal}
  {Pabl. Astron. Soc. Japan}\ }\textbf {\bibinfo {volume} {56}},\ \bibinfo
  {pages} {645} (\bibinfo {year} {2004})}\BibitemShut {NoStop}%
\bibitem [{\citenamefont {Pani}\ \emph
  {et~al.}(2011{\natexlab{b}})\citenamefont {Pani}, \citenamefont {Berti},
  \citenamefont {Cardoso},\ and\ \citenamefont {Read}}]{pani-NS-EDGB}%
  \BibitemOpen
  \bibfield  {author} {\bibinfo {author} {\bibfnamefont {P.}~\bibnamefont
  {Pani}}, \bibinfo {author} {\bibfnamefont {E.}~\bibnamefont {Berti}},
  \bibinfo {author} {\bibfnamefont {V.}~\bibnamefont {Cardoso}}, \ and\
  \bibinfo {author} {\bibfnamefont {J.}~\bibnamefont {Read}},\ }\href {\doibase
  10.1103/PhysRevD.84.104035} {\bibfield  {journal} {\bibinfo  {journal}
  {Phys.Rev.}\ }\textbf {\bibinfo {volume} {D84}},\ \bibinfo {pages} {104035}
  (\bibinfo {year} {2011}{\natexlab{b}})},\ \Eprint
  {http://arxiv.org/abs/1109.0928} {arXiv:1109.0928 [gr-qc]} \BibitemShut
  {NoStop}%
%%CITATION = ARXIV:1109.0928;%%
\bibitem [{\citenamefont {{Smart}}(1953)}]{smart}%
  \BibitemOpen
  \bibfield  {author} {\bibinfo {author} {\bibfnamefont {W.~M.}\ \bibnamefont
  {{Smart}}},\ }\href@noop {} {\emph {\bibinfo {title} {Celestial Mechanics,
  London, UK: Longman Green}}}\ (\bibinfo {year} {1953})\BibitemShut {NoStop}%
\bibitem [{\citenamefont {{Robertson}}\ and\ \citenamefont
  {{Noonan}}(1968)}]{robertson-noonan}%
  \BibitemOpen
  \bibfield  {author} {\bibinfo {author} {\bibfnamefont {H.~P.}\ \bibnamefont
  {{Robertson}}}\ and\ \bibinfo {author} {\bibfnamefont {T.~W.}\ \bibnamefont
  {{Noonan}}},\ }\href@noop {} {\emph {\bibinfo {title} {Relativity and
  Cosmology, Philadelphia, USA: Saunders}}}\ (\bibinfo {year}
  {1968})\BibitemShut {NoStop}%
\bibitem [{\citenamefont {Peters}\ and\ \citenamefont
  {Mathews}(1963)}]{PetersMathews}%
  \BibitemOpen
  \bibfield  {author} {\bibinfo {author} {\bibfnamefont {P.~C.}\ \bibnamefont
  {Peters}}\ and\ \bibinfo {author} {\bibfnamefont {J.}~\bibnamefont
  {Mathews}},\ }\href@noop {} {\bibfield  {journal} {\bibinfo  {journal}
  {Physical Review}\ }\textbf {\bibinfo {volume} {131}},\ \bibinfo {pages}
  {435} (\bibinfo {year} {1963})}\BibitemShut {NoStop}%
\bibitem [{\citenamefont {Eling}\ \emph {et~al.}(2007)\citenamefont {Eling},
  \citenamefont {Jacobson},\ and\ \citenamefont
  {Coleman~Miller}}]{eling-AE-NS}%
  \BibitemOpen
  \bibfield  {author} {\bibinfo {author} {\bibfnamefont {C.}~\bibnamefont
  {Eling}}, \bibinfo {author} {\bibfnamefont {T.}~\bibnamefont {Jacobson}}, \
  and\ \bibinfo {author} {\bibfnamefont {M.}~\bibnamefont {Coleman~Miller}},\
  }\href {\doibase 10.1103/PhysRevD.76.042003, 10.1103/PhysRevD.80.129906}
  {\bibfield  {journal} {\bibinfo  {journal} {Phys.Rev.}\ }\textbf {\bibinfo
  {volume} {D76}},\ \bibinfo {pages} {042003} (\bibinfo {year} {2007})},\
  \Eprint {http://arxiv.org/abs/0705.1565} {arXiv:0705.1565 [gr-qc]}
  \BibitemShut {NoStop}%
%%CITATION = ARXIV:0705.1565;%%
\bibitem [{\citenamefont {Pani}\ \emph
  {et~al.}(2011{\natexlab{c}})\citenamefont {Pani}, \citenamefont {Berti},
  \citenamefont {Cardoso},\ and\ \citenamefont {Read}}]{pani-EDGB-NS}%
  \BibitemOpen
  \bibfield  {author} {\bibinfo {author} {\bibfnamefont {P.}~\bibnamefont
  {Pani}}, \bibinfo {author} {\bibfnamefont {E.}~\bibnamefont {Berti}},
  \bibinfo {author} {\bibfnamefont {V.}~\bibnamefont {Cardoso}}, \ and\
  \bibinfo {author} {\bibfnamefont {J.}~\bibnamefont {Read}},\ }\href {\doibase
  10.1103/PhysRevD.84.104035} {\bibfield  {journal} {\bibinfo  {journal}
  {Phys.Rev.}\ }\textbf {\bibinfo {volume} {D84}},\ \bibinfo {pages} {104035}
  (\bibinfo {year} {2011}{\natexlab{c}})},\ \Eprint
  {http://arxiv.org/abs/1109.0928} {arXiv:1109.0928 [gr-qc]} \BibitemShut
  {NoStop}%
%%CITATION = ARXIV:1109.0928;%%
\bibitem [{\citenamefont {{Jenet}}\ and\ \citenamefont
  {{Ransom}}(2004)}]{jenet-ransom}%
  \BibitemOpen
  \bibfield  {author} {\bibinfo {author} {\bibfnamefont {F.~A.}\ \bibnamefont
  {{Jenet}}}\ and\ \bibinfo {author} {\bibfnamefont {S.~M.}\ \bibnamefont
  {{Ransom}}},\ }\href {\doibase 10.1038/nature02509} {\bibfield  {journal}
  {\bibinfo  {journal} {Nature}\ }\textbf {\bibinfo {volume} {428}},\ \bibinfo
  {pages} {919} (\bibinfo {year} {2004})},\ \Eprint
  {http://arxiv.org/abs/arXiv:astro-ph/0404569} {arXiv:astro-ph/0404569}
  \BibitemShut {NoStop}%
\bibitem [{\citenamefont {{Manchester}}\ \emph {et~al.}(2005)\citenamefont
  {{Manchester}}, \citenamefont {{Kramer}}, \citenamefont {{Possenti}},
  \citenamefont {{Lyne}}, \citenamefont {{Burgay}}, \citenamefont {{Stairs}},
  \citenamefont {{Hotan}}, \citenamefont {{McLaughlin}}, \citenamefont
  {{Lorimer}}, \citenamefont {{Hobbs}}, \citenamefont {{Sarkissian}},
  \citenamefont {{D'Amico}}, \citenamefont {{Camilo}}, \citenamefont
  {{Joshi}},\ and\ \citenamefont {{Freire}}}]{manchester}%
  \BibitemOpen
  \bibfield  {author} {\bibinfo {author} {\bibfnamefont {R.~N.}\ \bibnamefont
  {{Manchester}}}, \bibinfo {author} {\bibfnamefont {M.}~\bibnamefont
  {{Kramer}}}, \bibinfo {author} {\bibfnamefont {A.}~\bibnamefont
  {{Possenti}}}, \bibinfo {author} {\bibfnamefont {A.~G.}\ \bibnamefont
  {{Lyne}}}, \bibinfo {author} {\bibfnamefont {M.}~\bibnamefont {{Burgay}}},
  \bibinfo {author} {\bibfnamefont {I.~H.}\ \bibnamefont {{Stairs}}}, \bibinfo
  {author} {\bibfnamefont {A.~W.}\ \bibnamefont {{Hotan}}}, \bibinfo {author}
  {\bibfnamefont {M.~A.}\ \bibnamefont {{McLaughlin}}}, \bibinfo {author}
  {\bibfnamefont {D.~R.}\ \bibnamefont {{Lorimer}}}, \bibinfo {author}
  {\bibfnamefont {G.~B.}\ \bibnamefont {{Hobbs}}}, \bibinfo {author}
  {\bibfnamefont {J.~M.}\ \bibnamefont {{Sarkissian}}}, \bibinfo {author}
  {\bibfnamefont {N.}~\bibnamefont {{D'Amico}}}, \bibinfo {author}
  {\bibfnamefont {F.}~\bibnamefont {{Camilo}}}, \bibinfo {author}
  {\bibfnamefont {B.~C.}\ \bibnamefont {{Joshi}}}, \ and\ \bibinfo {author}
  {\bibfnamefont {P.~C.~C.}\ \bibnamefont {{Freire}}},\ }\href {\doibase
  10.1086/429128} {\bibfield  {journal} {\bibinfo  {journal} {Astrophys. J.
  Lett.}\ }\textbf {\bibinfo {volume} {621}},\ \bibinfo {pages} {L49} (\bibinfo
  {year} {2005})},\ \Eprint {http://arxiv.org/abs/arXiv:astro-ph/0501665}
  {arXiv:astro-ph/0501665} \BibitemShut {NoStop}%
\bibitem [{\citenamefont {Lattimer}\ and\ \citenamefont
  {Schutz}(2005)}]{lattimer-schutz}%
  \BibitemOpen
  \bibfield  {author} {\bibinfo {author} {\bibfnamefont {J.~M.}\ \bibnamefont
  {Lattimer}}\ and\ \bibinfo {author} {\bibfnamefont {B.~F.}\ \bibnamefont
  {Schutz}},\ }\href {\doibase 10.1086/431543} {\bibfield  {journal} {\bibinfo
  {journal} {Astrophys.J.}\ }\textbf {\bibinfo {volume} {629}},\ \bibinfo
  {pages} {979} (\bibinfo {year} {2005})},\ \Eprint
  {http://arxiv.org/abs/astro-ph/0411470} {arXiv:astro-ph/0411470 [astro-ph]}
  \BibitemShut {NoStop}%
%%CITATION = ASTRO-PH/0411470;%%
\bibitem [{\citenamefont {{Damour}}\ and\ \citenamefont
  {{Schafer}}(1988)}]{damour-schaefer}%
  \BibitemOpen
  \bibfield  {author} {\bibinfo {author} {\bibfnamefont {T.}~\bibnamefont
  {{Damour}}}\ and\ \bibinfo {author} {\bibfnamefont {G.}~\bibnamefont
  {{Schafer}}},\ }\href {\doibase 10.1007/BF02828697} {\bibfield  {journal}
  {\bibinfo  {journal} {Nuovo Cimento B Serie}\ }\textbf {\bibinfo {volume}
  {101}},\ \bibinfo {pages} {127} (\bibinfo {year} {1988})}\BibitemShut
  {NoStop}%
\bibitem [{\citenamefont {Flanagan}\ and\ \citenamefont
  {Hinderer}(2008)}]{flanagan-hinderer}%
  \BibitemOpen
  \bibfield  {author} {\bibinfo {author} {\bibfnamefont {E.~E.}\ \bibnamefont
  {Flanagan}}\ and\ \bibinfo {author} {\bibfnamefont {T.}~\bibnamefont
  {Hinderer}},\ }\href {\doibase 10.1103/PhysRevD.77.021502} {\bibfield
  {journal} {\bibinfo  {journal} {Phys.Rev.}\ }\textbf {\bibinfo {volume}
  {D77}},\ \bibinfo {pages} {021502} (\bibinfo {year} {2008})},\ \Eprint
  {http://arxiv.org/abs/0709.1915} {arXiv:0709.1915 [astro-ph]} \BibitemShut
  {NoStop}%
%%CITATION = ARXIV:0709.1915;%%
\bibitem [{\citenamefont {Read}\ \emph {et~al.}(2009)\citenamefont {Read},
  \citenamefont {Markakis}, \citenamefont {Shibata}, \citenamefont {Uryu},
  \citenamefont {Creighton} \emph {et~al.}}]{read-markakis-shibata}%
  \BibitemOpen
  \bibfield  {author} {\bibinfo {author} {\bibfnamefont {J.~S.}\ \bibnamefont
  {Read}}, \bibinfo {author} {\bibfnamefont {C.}~\bibnamefont {Markakis}},
  \bibinfo {author} {\bibfnamefont {M.}~\bibnamefont {Shibata}}, \bibinfo
  {author} {\bibfnamefont {K.}~\bibnamefont {Uryu}}, \bibinfo {author}
  {\bibfnamefont {J.~D.}\ \bibnamefont {Creighton}},  \emph {et~al.},\ }\href
  {\doibase 10.1103/PhysRevD.79.124033} {\bibfield  {journal} {\bibinfo
  {journal} {Phys.Rev.}\ }\textbf {\bibinfo {volume} {D79}},\ \bibinfo {pages}
  {124033} (\bibinfo {year} {2009})},\ \Eprint {http://arxiv.org/abs/0901.3258}
  {arXiv:0901.3258 [gr-qc]} \BibitemShut {NoStop}%
%%CITATION = ARXIV:0901.3258;%%
\bibitem [{\citenamefont {Hinderer}\ \emph {et~al.}(2010)\citenamefont
  {Hinderer}, \citenamefont {Lackey}, \citenamefont {Lang},\ and\ \citenamefont
  {Read}}]{hinderer-lackey-lang-read}%
  \BibitemOpen
  \bibfield  {author} {\bibinfo {author} {\bibfnamefont {T.}~\bibnamefont
  {Hinderer}}, \bibinfo {author} {\bibfnamefont {B.~D.}\ \bibnamefont
  {Lackey}}, \bibinfo {author} {\bibfnamefont {R.~N.}\ \bibnamefont {Lang}}, \
  and\ \bibinfo {author} {\bibfnamefont {J.~S.}\ \bibnamefont {Read}},\ }\href
  {\doibase 10.1103/PhysRevD.81.123016} {\bibfield  {journal} {\bibinfo
  {journal} {Phys.Rev.}\ }\textbf {\bibinfo {volume} {D81}},\ \bibinfo {pages}
  {123016} (\bibinfo {year} {2010})},\ \Eprint {http://arxiv.org/abs/0911.3535}
  {arXiv:0911.3535 [astro-ph.HE]} \BibitemShut {NoStop}%
%%CITATION = ARXIV:0911.3535;%%
\bibitem [{\citenamefont {Lackey}\ \emph {et~al.}(2012)\citenamefont {Lackey},
  \citenamefont {Kyutoku}, \citenamefont {Shibata}, \citenamefont {Brady},\
  and\ \citenamefont {Friedman}}]{lackey-kyutoku}%
  \BibitemOpen
  \bibfield  {author} {\bibinfo {author} {\bibfnamefont {B.~D.}\ \bibnamefont
  {Lackey}}, \bibinfo {author} {\bibfnamefont {K.}~\bibnamefont {Kyutoku}},
  \bibinfo {author} {\bibfnamefont {M.}~\bibnamefont {Shibata}}, \bibinfo
  {author} {\bibfnamefont {P.~R.}\ \bibnamefont {Brady}}, \ and\ \bibinfo
  {author} {\bibfnamefont {J.~L.}\ \bibnamefont {Friedman}},\ }\href {\doibase
  10.1103/PhysRevD.85.044061} {\bibfield  {journal} {\bibinfo  {journal}
  {Phys.Rev.}\ }\textbf {\bibinfo {volume} {D85}},\ \bibinfo {pages} {044061}
  (\bibinfo {year} {2012})},\ \Eprint {http://arxiv.org/abs/1109.3402}
  {arXiv:1109.3402 [astro-ph.HE]} \BibitemShut {NoStop}%
%%CITATION = ARXIV:1109.3402;%%
\bibitem [{\citenamefont {Damour}\ \emph {et~al.}(2012)\citenamefont {Damour},
  \citenamefont {Nagar},\ and\ \citenamefont {Villain}}]{damour-nagar-villain}%
  \BibitemOpen
  \bibfield  {author} {\bibinfo {author} {\bibfnamefont {T.}~\bibnamefont
  {Damour}}, \bibinfo {author} {\bibfnamefont {A.}~\bibnamefont {Nagar}}, \
  and\ \bibinfo {author} {\bibfnamefont {L.}~\bibnamefont {Villain}},\ }\href
  {\doibase 10.1103/PhysRevD.85.123007} {\bibfield  {journal} {\bibinfo
  {journal} {Phys.Rev.}\ }\textbf {\bibinfo {volume} {D85}},\ \bibinfo {pages}
  {123007} (\bibinfo {year} {2012})},\ \Eprint {http://arxiv.org/abs/1203.4352}
  {arXiv:1203.4352 [gr-qc]} \BibitemShut {NoStop}%
%%CITATION = ARXIV:1203.4352;%%
\bibitem [{\citenamefont {Bildsten}\ and\ \citenamefont
  {Cutler}(1992)}]{bildsten-cutler}%
  \BibitemOpen
  \bibfield  {author} {\bibinfo {author} {\bibfnamefont {L.}~\bibnamefont
  {Bildsten}}\ and\ \bibinfo {author} {\bibfnamefont {C.}~\bibnamefont
  {Cutler}},\ }\href {\doibase 10.1086/171983} {\bibfield  {journal} {\bibinfo
  {journal} {Astrophys.J.}\ }\textbf {\bibinfo {volume} {400}},\ \bibinfo
  {pages} {175} (\bibinfo {year} {1992})}\BibitemShut {NoStop}%
%%CITATION = ASJOA,400,175;%%
\bibitem [{\citenamefont {{Kokkotas}}\ and\ \citenamefont
  {{Schmidt}}(1999)}]{kokkotas-living}%
  \BibitemOpen
  \bibfield  {author} {\bibinfo {author} {\bibfnamefont {K.}~\bibnamefont
  {{Kokkotas}}}\ and\ \bibinfo {author} {\bibfnamefont {B.}~\bibnamefont
  {{Schmidt}}},\ }\href@noop {} {\bibfield  {journal} {\bibinfo  {journal}
  {Living Reviews in Relativity}\ }\textbf {\bibinfo {volume} {2}},\ \bibinfo
  {pages} {2} (\bibinfo {year} {1999})}\BibitemShut {NoStop}%
\bibitem [{\citenamefont {Sotani}\ and\ \citenamefont
  {Kokkotas}(2004)}]{sotani-fp-ST}%
  \BibitemOpen
  \bibfield  {author} {\bibinfo {author} {\bibfnamefont {H.}~\bibnamefont
  {Sotani}}\ and\ \bibinfo {author} {\bibfnamefont {K.~D.}\ \bibnamefont
  {Kokkotas}},\ }\href {\doibase 10.1103/PhysRevD.70.084026} {\bibfield
  {journal} {\bibinfo  {journal} {Phys.Rev.}\ }\textbf {\bibinfo {volume}
  {D70}},\ \bibinfo {pages} {084026} (\bibinfo {year} {2004})},\ \Eprint
  {http://arxiv.org/abs/gr-qc/0409066} {arXiv:gr-qc/0409066 [gr-qc]}
  \BibitemShut {NoStop}%
%%CITATION = GR-QC/0409066;%%
\bibitem [{\citenamefont {Sotani}(2009{\natexlab{a}})}]{sotani-fp-TeVeS}%
  \BibitemOpen
  \bibfield  {author} {\bibinfo {author} {\bibfnamefont {H.}~\bibnamefont
  {Sotani}},\ }\href {\doibase 10.1103/PhysRevD.79.064033} {\bibfield
  {journal} {\bibinfo  {journal} {Phys.Rev.}\ }\textbf {\bibinfo {volume}
  {D79}},\ \bibinfo {pages} {064033} (\bibinfo {year} {2009}{\natexlab{a}})},\
  \Eprint {http://arxiv.org/abs/0903.2424} {arXiv:0903.2424 [gr-qc]}
  \BibitemShut {NoStop}%
%%CITATION = ARXIV:0903.2424;%%
\bibitem [{\citenamefont {Sotani}\ and\ \citenamefont
  {Kokkotas}(2005)}]{sotani-w-ST}%
  \BibitemOpen
  \bibfield  {author} {\bibinfo {author} {\bibfnamefont {H.}~\bibnamefont
  {Sotani}}\ and\ \bibinfo {author} {\bibfnamefont {K.~D.}\ \bibnamefont
  {Kokkotas}},\ }\href {\doibase 10.1103/PhysRevD.71.124038} {\bibfield
  {journal} {\bibinfo  {journal} {Phys.Rev.}\ }\textbf {\bibinfo {volume}
  {D71}},\ \bibinfo {pages} {124038} (\bibinfo {year} {2005})},\ \Eprint
  {http://arxiv.org/abs/gr-qc/0506060} {arXiv:gr-qc/0506060 [gr-qc]}
  \BibitemShut {NoStop}%
%%CITATION = GR-QC/0506060;%%
\bibitem [{\citenamefont {Sotani}(2009{\natexlab{b}})}]{sotani-w-TeVeS}%
  \BibitemOpen
  \bibfield  {author} {\bibinfo {author} {\bibfnamefont {H.}~\bibnamefont
  {Sotani}},\ }\href {\doibase 10.1103/PhysRevD.80.064035} {\bibfield
  {journal} {\bibinfo  {journal} {Phys.Rev.}\ }\textbf {\bibinfo {volume}
  {D80}},\ \bibinfo {pages} {064035} (\bibinfo {year} {2009}{\natexlab{b}})},\
  \Eprint {http://arxiv.org/abs/0909.2411} {arXiv:0909.2411 [gr-qc]}
  \BibitemShut {NoStop}%
%%CITATION = ARXIV:0909.2411;%%
\bibitem [{\citenamefont {Cardoso}\ and\ \citenamefont
  {Gualtieri}(2009)}]{cardoso-gualtieri}%
  \BibitemOpen
  \bibfield  {author} {\bibinfo {author} {\bibfnamefont {V.}~\bibnamefont
  {Cardoso}}\ and\ \bibinfo {author} {\bibfnamefont {L.}~\bibnamefont
  {Gualtieri}},\ }\href {\doibase 10.1103/PhysRevD.81.089903,
  10.1103/PhysRevD.80.064008} {\bibfield  {journal} {\bibinfo  {journal}
  {Phys.Rev.}\ }\textbf {\bibinfo {volume} {D80}},\ \bibinfo {pages} {064008}
  (\bibinfo {year} {2009})},\ \Eprint {http://arxiv.org/abs/0907.5008}
  {arXiv:0907.5008 [gr-qc]} \BibitemShut {NoStop}%
%%CITATION = ARXIV:0907.5008;%%
\bibitem [{\citenamefont {Garfinkle}\ \emph {et~al.}(2010)\citenamefont
  {Garfinkle}, \citenamefont {Pretorius},\ and\ \citenamefont
  {Yunes}}]{garfinkle}%
  \BibitemOpen
  \bibfield  {author} {\bibinfo {author} {\bibfnamefont {D.}~\bibnamefont
  {Garfinkle}}, \bibinfo {author} {\bibfnamefont {F.}~\bibnamefont
  {Pretorius}}, \ and\ \bibinfo {author} {\bibfnamefont {N.}~\bibnamefont
  {Yunes}},\ }\href {\doibase 10.1103/PhysRevD.82.041501} {\bibfield  {journal}
  {\bibinfo  {journal} {Phys.Rev.}\ }\textbf {\bibinfo {volume} {D82}},\
  \bibinfo {pages} {041501} (\bibinfo {year} {2010})},\ \Eprint
  {http://arxiv.org/abs/1007.2429} {arXiv:1007.2429 [gr-qc]} \BibitemShut
  {NoStop}%
%%CITATION = ARXIV:1007.2429;%%
\bibitem [{\citenamefont {Molina}\ \emph {et~al.}(2010)\citenamefont {Molina},
  \citenamefont {Pani}, \citenamefont {Cardoso},\ and\ \citenamefont
  {Gualtieri}}]{molina}%
  \BibitemOpen
  \bibfield  {author} {\bibinfo {author} {\bibfnamefont {C.}~\bibnamefont
  {Molina}}, \bibinfo {author} {\bibfnamefont {P.}~\bibnamefont {Pani}},
  \bibinfo {author} {\bibfnamefont {V.}~\bibnamefont {Cardoso}}, \ and\
  \bibinfo {author} {\bibfnamefont {L.}~\bibnamefont {Gualtieri}},\ }\href
  {\doibase 10.1103/PhysRevD.81.124021} {\bibfield  {journal} {\bibinfo
  {journal} {Phys.Rev.}\ }\textbf {\bibinfo {volume} {D81}},\ \bibinfo {pages}
  {124021} (\bibinfo {year} {2010})},\ \Eprint {http://arxiv.org/abs/1004.4007}
  {arXiv:1004.4007 [gr-qc]} \BibitemShut {NoStop}%
%%CITATION = ARXIV:1004.4007;%%
\bibitem [{\citenamefont {Kojima}(1998)}]{kojima}%
  \BibitemOpen
  \bibfield  {author} {\bibinfo {author} {\bibfnamefont {Y.}~\bibnamefont
  {Kojima}},\ }\href@noop {} {\bibfield  {journal} {\bibinfo  {journal}
  {Mon.Not.Roy.Astron.Soc.}\ }\textbf {\bibinfo {volume} {293}},\ \bibinfo
  {pages} {49} (\bibinfo {year} {1998})},\ \Eprint
  {http://arxiv.org/abs/gr-qc/9709003} {arXiv:gr-qc/9709003 [gr-qc]}
  \BibitemShut {NoStop}%
%%CITATION = GR-QC/9709003;%%
\bibitem [{\citenamefont {Beyer}\ and\ \citenamefont {Kokkotas}(1999)}]{beyer}%
  \BibitemOpen
  \bibfield  {author} {\bibinfo {author} {\bibfnamefont {H.~R.}\ \bibnamefont
  {Beyer}}\ and\ \bibinfo {author} {\bibfnamefont {K.~D.}\ \bibnamefont
  {Kokkotas}},\ }\href {\doibase 10.1046/j.1365-8711.1999.02739.x} {\bibfield
  {journal} {\bibinfo  {journal} {Mon.Not.Roy.Astron.Soc.}\ }\textbf {\bibinfo
  {volume} {308}},\ \bibinfo {pages} {745} (\bibinfo {year} {1999})},\ \Eprint
  {http://arxiv.org/abs/gr-qc/9903019} {arXiv:gr-qc/9903019 [gr-qc]}
  \BibitemShut {NoStop}%
%%CITATION = GR-QC/9903019;%%
\bibitem [{\citenamefont {Sotani}(2010)}]{sotani-toroidal}%
  \BibitemOpen
  \bibfield  {author} {\bibinfo {author} {\bibfnamefont {H.}~\bibnamefont
  {Sotani}},\ }\href {\doibase 10.1103/PhysRevD.82.124061} {\bibfield
  {journal} {\bibinfo  {journal} {Phys.Rev.}\ }\textbf {\bibinfo {volume}
  {D82}},\ \bibinfo {pages} {124061} (\bibinfo {year} {2010})},\ \Eprint
  {http://arxiv.org/abs/1012.2143} {arXiv:1012.2143 [astro-ph.HE]} \BibitemShut
  {NoStop}%
%%CITATION = ARXIV:1012.2143;%%
\bibitem [{\citenamefont {Ruoff}\ \emph {et~al.}(2003)\citenamefont {Ruoff},
  \citenamefont {Stavridis},\ and\ \citenamefont {Kokkotas}}]{ruoff}%
  \BibitemOpen
  \bibfield  {author} {\bibinfo {author} {\bibfnamefont {J.}~\bibnamefont
  {Ruoff}}, \bibinfo {author} {\bibfnamefont {A.}~\bibnamefont {Stavridis}}, \
  and\ \bibinfo {author} {\bibfnamefont {K.~D.}\ \bibnamefont {Kokkotas}},\
  }\href {\doibase 10.1046/j.1365-8711.2003.06267.x} {\bibfield  {journal}
  {\bibinfo  {journal} {Mon.Not.Roy.Astron.Soc.}\ }\textbf {\bibinfo {volume}
  {339}},\ \bibinfo {pages} {1170} (\bibinfo {year} {2003})},\ \Eprint
  {http://arxiv.org/abs/gr-qc/0203052} {arXiv:gr-qc/0203052 [gr-qc]}
  \BibitemShut {NoStop}%
%%CITATION = GR-QC/0203052;%%
\bibitem [{\citenamefont {Gaertig}\ and\ \citenamefont
  {Kokkotas}(2009)}]{gaertig}%
  \BibitemOpen
  \bibfield  {author} {\bibinfo {author} {\bibfnamefont {E.}~\bibnamefont
  {Gaertig}}\ and\ \bibinfo {author} {\bibfnamefont {K.~D.}\ \bibnamefont
  {Kokkotas}},\ }\href {\doibase 10.1103/PhysRevD.80.064026} {\bibfield
  {journal} {\bibinfo  {journal} {Phys.Rev.}\ }\textbf {\bibinfo {volume}
  {D80}},\ \bibinfo {pages} {064026} (\bibinfo {year} {2009})},\ \Eprint
  {http://arxiv.org/abs/0905.0821} {arXiv:0905.0821 [astro-ph.SR]} \BibitemShut
  {NoStop}%
%%CITATION = ARXIV:0905.0821;%%
\bibitem [{\citenamefont {Musgrave}\ \emph {et~al.}(1996)\citenamefont
  {Musgrave}, \citenamefont {Pollney},\ and\ \citenamefont {Lake}}]{grtensor}%
  \BibitemOpen
  \bibfield  {author} {\bibinfo {author} {\bibfnamefont {P.}~\bibnamefont
  {Musgrave}}, \bibinfo {author} {\bibfnamefont {D.}~\bibnamefont {Pollney}}, \
  and\ \bibinfo {author} {\bibfnamefont {K.}~\bibnamefont {Lake}},\ }\href@noop
  {} {\enquote {\bibinfo {title} {{GRTensorII software}},}\ }\bibinfo
  {howpublished} {\url{http://grtensor.org/}} (\bibinfo {year} {1996}),\
  \bibinfo {note} {{Queen's University, Kingston, Ontario,
  Canada.}}\BibitemShut {Stop}%
\bibitem [{\citenamefont {{Mart{\'{\i}}n-Garc{\'{\i}}a}}\ \emph
  {et~al.}(2008)\citenamefont {{Mart{\'{\i}}n-Garc{\'{\i}}a}}, \citenamefont
  {{Yllanes}},\ and\ \citenamefont {{Portugal}}}]{2008CoPhC.179..586M}%
  \BibitemOpen
  \bibfield  {author} {\bibinfo {author} {\bibfnamefont {J.~M.}\ \bibnamefont
  {{Mart{\'{\i}}n-Garc{\'{\i}}a}}}, \bibinfo {author} {\bibfnamefont
  {D.}~\bibnamefont {{Yllanes}}}, \ and\ \bibinfo {author} {\bibfnamefont
  {R.}~\bibnamefont {{Portugal}}},\ }\href {\doibase 10.1016/j.cpc.2008.04.018}
  {\bibfield  {journal} {\bibinfo  {journal} {Computer Physics Communications}\
  }\textbf {\bibinfo {volume} {179}},\ \bibinfo {pages} {586} (\bibinfo {year}
  {2008})},\ \Eprint {http://arxiv.org/abs/0802.1274} {arXiv:0802.1274 [cs.SC]}
  \BibitemShut {NoStop}%
\bibitem [{\citenamefont {{Brizuela}}\ \emph {et~al.}(2009)\citenamefont
  {{Brizuela}}, \citenamefont {{Mart{\'{\i}}n-Garc{\'{\i}}a}},\ and\
  \citenamefont {{Mena Marug{\'a}n}}}]{2009GReGr..41.2415B}%
  \BibitemOpen
  \bibfield  {author} {\bibinfo {author} {\bibfnamefont {D.}~\bibnamefont
  {{Brizuela}}}, \bibinfo {author} {\bibfnamefont {J.~M.}\ \bibnamefont
  {{Mart{\'{\i}}n-Garc{\'{\i}}a}}}, \ and\ \bibinfo {author} {\bibfnamefont
  {G.~A.}\ \bibnamefont {{Mena Marug{\'a}n}}},\ }\href {\doibase
  10.1007/s10714-009-0773-2} {\bibfield  {journal} {\bibinfo  {journal}
  {General Relativity and Gravitation}\ }\textbf {\bibinfo {volume} {41}},\
  \bibinfo {pages} {2415} (\bibinfo {year} {2009})},\ \Eprint
  {http://arxiv.org/abs/0807.0824} {arXiv:0807.0824 [gr-qc]} \BibitemShut
  {NoStop}%
\bibitem [{\citenamefont {{Stein}}\ and\ \citenamefont
  {{Yunes}}(2011)}]{Stein:2010pn}%
  \BibitemOpen
  \bibfield  {author} {\bibinfo {author} {\bibfnamefont {L.~C.}\ \bibnamefont
  {{Stein}}}\ and\ \bibinfo {author} {\bibfnamefont {N.}~\bibnamefont
  {{Yunes}}},\ }\href {\doibase 10.1103/PhysRevD.83.064038} {\bibfield
  {journal} {\bibinfo  {journal} {\prd}\ }\textbf {\bibinfo {volume} {83}},\
  \bibinfo {pages} {064038} (\bibinfo {year} {2011})},\ \Eprint
  {http://arxiv.org/abs/1012.3144} {arXiv:1012.3144 [gr-qc]} \BibitemShut
  {NoStop}%
\bibitem [{\citenamefont {Thorne}(1980)}]{Thorne:1980rm}%
  \BibitemOpen
  \bibfield  {author} {\bibinfo {author} {\bibfnamefont {K.~S.}\ \bibnamefont
  {Thorne}},\ }\href@noop {} {\bibfield  {journal} {\bibinfo  {journal} {Rev.
  Mod. Phys.}\ }\textbf {\bibinfo {volume} {52}},\ \bibinfo {pages} {299}
  (\bibinfo {year} {1980})}\BibitemShut {NoStop}%
\bibitem [{\citenamefont {Peters}(1964)}]{Peters:1964zz}%
  \BibitemOpen
  \bibfield  {author} {\bibinfo {author} {\bibfnamefont {P.}~\bibnamefont
  {Peters}},\ }\href {\doibase 10.1103/PhysRev.136.B1224} {\bibfield  {journal}
  {\bibinfo  {journal} {Phys.Rev.}\ }\textbf {\bibinfo {volume} {136}},\
  \bibinfo {pages} {B1224} (\bibinfo {year} {1964})}\BibitemShut {NoStop}%
%%CITATION = PHRVA,136,B1224;%%
\bibitem [{\citenamefont {Kidder}\ \emph {et~al.}(1993)\citenamefont {Kidder},
  \citenamefont {Will},\ and\ \citenamefont {Wiseman}}]{Kidder:1993do}%
  \BibitemOpen
  \bibfield  {author} {\bibinfo {author} {\bibfnamefont {L.~E.}\ \bibnamefont
  {Kidder}}, \bibinfo {author} {\bibfnamefont {C.~M.}\ \bibnamefont {Will}}, \
  and\ \bibinfo {author} {\bibfnamefont {A.~G.}\ \bibnamefont {Wiseman}},\
  }\href {\doibase 10.1103/PhysRevD.47.R4183} {\bibfield  {journal} {\bibinfo
  {journal} {Phys.Rev.}\ }\textbf {\bibinfo {volume} {D47}},\ \bibinfo {pages}
  {4183} (\bibinfo {year} {1993})},\ \Eprint
  {http://arxiv.org/abs/gr-qc/9211025} {arXiv:gr-qc/9211025 [gr-qc]}
  \BibitemShut {NoStop}%
%%CITATION = GR-QC/9211025;%%
\bibitem [{\citenamefont {Clifton}\ and\ \citenamefont
  {Weisberg}(2008)}]{Clifton:2008gr}%
  \BibitemOpen
  \bibfield  {author} {\bibinfo {author} {\bibfnamefont {T.}~\bibnamefont
  {Clifton}}\ and\ \bibinfo {author} {\bibfnamefont {J.~M.}\ \bibnamefont
  {Weisberg}},\ }\href@noop {} {\bibfield  {journal} {\bibinfo  {journal} {The
  Astrophysical Journal}\ }\textbf {\bibinfo {volume} {679}},\ \bibinfo {pages}
  {687} (\bibinfo {year} {2008})}\BibitemShut {NoStop}%
\bibitem [{\citenamefont {Goldstein}\ \emph {et~al.}(2002)\citenamefont
  {Goldstein}, \citenamefont {Poole},\ and\ \citenamefont {Safko}}]{goldstein}%
  \BibitemOpen
  \bibfield  {author} {\bibinfo {author} {\bibfnamefont {H.}~\bibnamefont
  {Goldstein}}, \bibinfo {author} {\bibfnamefont {C.}~\bibnamefont {Poole}}, \
  and\ \bibinfo {author} {\bibfnamefont {J.}~\bibnamefont {Safko}},\
  }\href@noop {} {\emph {\bibinfo {title} {Classical mechanics}}}\ (\bibinfo
  {publisher} {Addison-Wesley},\ \bibinfo {address} {San Francisco},\ \bibinfo
  {year} {2002})\BibitemShut {NoStop}%
\bibitem [{\citenamefont {{Lai}}\ and\ \citenamefont
  {{Shapiro}}(1995)}]{lai-shapiro}%
  \BibitemOpen
  \bibfield  {author} {\bibinfo {author} {\bibfnamefont {D.}~\bibnamefont
  {{Lai}}}\ and\ \bibinfo {author} {\bibfnamefont {S.~L.}\ \bibnamefont
  {{Shapiro}}},\ }\href {\doibase 10.1086/175562} {\bibfield  {journal}
  {\bibinfo  {journal} {Astrophys. J.}\ }\textbf {\bibinfo {volume} {443}},\
  \bibinfo {pages} {705} (\bibinfo {year} {1995})},\ \Eprint
  {http://arxiv.org/abs/arXiv:astro-ph/9408054} {arXiv:astro-ph/9408054}
  \BibitemShut {NoStop}%
\end{thebibliography}%
\end{document}

\title{Erratum: Isolated and Binary Neutron Stars\\* in Dynamical Chern-Simons Gravity}

\author{Kent Yagi}
%\email{kyagi@physics.montana.edu}
\affiliation{Department of Physics, Montana State University, Bozeman, MT 59717, USA.}
\affiliation{Department of Physics, Princeton University, Princeton, NJ 08544, USA.}

\author{Leo C. Stein}
\affiliation{Cornell Center for Astrophysics and
Planetary Science (CCAPS), Cornell University, Ithaca, NY 14853, USA}
\affiliation{Theoretical Astrophysics,
Walter Burke Institute for Theoretical Physics,
California Institute of Technology, Pasadena, CA
91125, USA}

\author{Nicol\'as Yunes}
\affiliation{Department of Physics, Montana State University, Bozeman, MT 59717, USA.}

\author{Takahiro Tanaka}
\affiliation{Yukawa Institute for Theoretical Physics, Kyoto University, Kyoto, 606-8502, Japan.}
\affiliation{Department of Physics, Kyoto University, Kyoto, 606-8502, Japan.}

\date{\today}

\pacs{04.30.-w,04.50.Kd,04.25.-g,97.60.Jd}

%04.30.Db Wave generation and sources
% 04.50.Kd Modified theories of gravity
% 04.25.-g Approximation methods; equations of motion
%04.25.Nx Post-Newtonian approximation; perturbation theory; related approximations

\maketitle

%%%%%%%%%%%%%%%%%%%%%%%%%%%%%%%%%%%%%%%%%%%%%%%%%%%%%

In Ref.~\cite{Yagi:2013mbt}, the Chern-Simons correction to the osculating orbital elements due to the dipole-dipole interaction for the scalar field was derived in Eq.~(130) by integrating out the interaction Lagrangian, which was taken to be the kinetic term for the scalar field [Eq.~(122) in~\cite{Yagi:2013mbt}], using the scalar field solution. However, such kinetic term only leads to the left-hand side of the evolution equation for the scalar field, i.e.
\be
\square \vartheta = J\,,
\ee
where $J$ is the effective source term of the scalar field equation~\cite{quadratic}. In order to obtain the correct interaction Lagrangian, one also needs to include the source term ($L_\mrm{source}$) that reproduces the right-hand side of the above evolution equation. The Lagrangian density for such an additional term is given by $- \vartheta J$. Therefore, Eq.~(126) needs to be changed to 
\ba
L_\mrm{int} &=& \int \left( \mathcal{L}_\mrm{kin,int} + \mathcal{L}_\mrm{source,int} \right) d^3x\,, \nn \\ 
& = & \beta \mu_1^j \mu_2^k \partial^{(1)}_{ij} \partial^{(2)}_{ik} \int \frac{1}{r_1} \frac{1}{r_2} d^3x\,,
\ea
where $\mathcal{L}_\mrm{source,int}$ is the interaction part of the source term Lagrangian density. Notice that, by accounting for such additional term, the sign of the interaction Lagrangian now flips. This means that the sign in Eqs.~(127),~(129) and~(130) needs to be reversed. The results in Sec. IX~B are unaffected since we have neglected contribution from the dipole-dipole interaction because the spin of the secondary pulsar in the double binary pulsar is much smaller than that of the primary.

We also report errors in App.~B of~\cite{Yagi:2013mbt}. In this appendix, we calculated dissipative corrections to the energy and angular momentum energy flux emitted from neutron star binaries based on calculations in~\cite{quadratic}. However, some terms in the perturbed scalar and metric field equations were not properly accounted for in~\cite{quadratic}. Taking these terms into account, we found that corrections to the gravitational energy and angular momentum flux due to metric perturbations enter at third post-Newtonian order relative to general relativity, which is of 1 post-Newtonian order higher than the effect due to the scalar field. Therefore, Apps.~B.1.b and~B.2.b in~\cite{Yagi:2013mbt} are now irrelevant. We discuss this in more detail in~\cite{quadratic-erratum}.

On the other hand, the scalar energy flux in App.~B.1.a is affected as follows. First, the far-zone solution for the scalar field in Eq.~(B3) becomes
\be
\vartheta^{\FZ} = \frac{1}{r} \ddot{\mu}_{ij} \left( n^{ij} + \delta^{ij} \right)\,.
\ee
Then, the scalar energy flux in Eqs.~(B5) and~(B6) becomes
\begin{align}
\delta  \dot{E}^{(\vartheta)} &=\, -\frac{8\pi}{15}\beta
\left\langle \left[\dddot{\mu}_{ij}\dddot{\mu}^{ij} + 13
    \left(\dddot{\mu}^i{}_i\right)^2 \right]
\right\rangle_{\omega}\,, \\
\delta\dot{E}^{(\vartheta)} &= -\frac{5}{768} \frac{\xi_\CS}{m^4} \left<
  \frac{m^{6}}{r_{12}^{6}} \left(
 \Delta^{2} v_{12}^{2}+ 27 (\Delta \cdot v_{12})^{2}
\right.\right.
\nonumber \\
&+ \left. \left.
3 \Delta^{2}(n_{12}\cdot v_{12})^{2}  - 162 (n_{12}\cdot v_{12}) (\Delta\cdot n_{12})(\Delta\cdot v_{12})
\right.\right.
\nonumber \\
&+ \left. \left.
243 (\Delta \cdot n_{12})^{2} (n_{12}\cdot v_{12})^{2}
\right)  \right>_{\omega}\,,
\end{align}
respectively. Finally, Eq.~(B8) is corrected to
\begin{equation}
\delta\dot{E}^{(\vartheta)} = -\frac{5}{1536}\zeta\left(\frac{m}{a}
\right)^{7} \frac{ 29 [ \Delta_{1}^{2}f_{1}(e) + \Delta_{2}^{2}
  f_{2}(e) ] +  2 \Delta_{3}^{2}f_{3}(e)}{(1-e^{2})^{11/2}}\,,
  \label{eq:dotEtheta}
\end{equation}
with Eqs.~(B9) and~(B10) now become
\begin{align}
f_{1}(e) &= 1+\frac{1129}{116}e^{2} + \frac{2271}{232}e^{4}+\frac{1287}{1856}e^{6}\,, \\
f_{2}(e) &= 1+\frac{1075}{116}e^{2} + \frac{1731}{232}e^{4}+\frac{801}{1856}e^{6}\,, 
\end{align}
respectively.
We checked that the scalar angular momentum flux [Eq.~(B24)] is not modified.
 
Regarding other typos in~\cite{Yagi:2013mbt}, $\bar \mu$ in Fig.~1 and Table~1 should be corrected to $\bar \mu/4$. 
\if0%%%%%%%%%%%%%%%%%
During the publication process, a typo was made in Eq.~(B.24), which needs to be corrected as
%
\be
\delta \dot{L}^{(\vartheta)}_{i} = \frac{16\pi}{15}\beta \epsilon_{ijk}
\left< \dddot{\mu}_{jp} \ddot{\mu}_{kp}
\right>_{\omega}\,.
\ee
%
References~[93-102] were also incorrectly processed during the publication process. They should be corrected as follows: [93]$\to$[96], [94]$\to$[97], [95]$\to$[98], [96]$\to$[99], [97]$\to$[100], [98]$\to$[101], [99]$\to$[102], [100]$\to$[93], [101]$\to$[94] and [102]$\to$[95]. 
\fi%%%%%%%%%%%%%%%%%

%%%%%%%%%%%%%%%%%%%%%%%%%%%%%%%%%%%%%%%%%%%%
%%%%%%%%%%%%%%%%%%%%%%%%%%%%%%%%%%%%%%%%%%%%
\bibliography{master}